\documentclass[fleqn,usenatbib]{mnras}
\usepackage{mathptmx}
\usepackage[T1]{fontenc}
\DeclareRobustCommand{\VAN}[3]{#2}
\let\VANthebibliography\thebibliography
\def\thebibliography{\DeclareRobustCommand{\VAN}[3]{##3}\VANthebibliography}

\usepackage{graphicx}
\graphicspath{{./plots/}}
\usepackage{amsmath}
\usepackage{cleveref}
\usepackage{upgreek}
\usepackage{tipa}
\usepackage{amssymb}
\usepackage[T1]{fontenc}
\newcommand*{\dittostraight}{$\,\,\,\,\,\,\,\,\,\,\,\,\,\,\,\,\,\,\,\,\,\,\,\,\,\,\,\,\,\,\,$\textquotedbl}
\def \aap{A\&A}

\def \apj{ApJ}
\def \apjl{ApJ Lett.}
\def \apjs{ApJS}

\def \mnras{MNRAS}

\def \prl{Phys. Rev. Lett.}

\newcommand{\eagle}{\textsc{eagle}{ }}
\newcommand{\galics}{\textsc{GalICs}{ }}
\newcommand{\gadget}{\textsc{Gadget2}{ }}

\newcommand{\be}{\begin{equation}}
\newcommand{\ee}{\end{equation}}

\def\ltsima{$\; \buildrel < \over \sim \;$}
\def\simlt{\lower.5ex\hbox{\ltsima}}
\def\gtsima{$\; \buildrel > \over \sim \;$}
\def\simgt{\lower.5ex\hbox{\gtsima}}
\def\la{$\langle$}

\DeclareMathOperator{\sech}{sech}

\title[Spurious collisional disk heating]{Spurious heating of stellar motions in simulated galactic disks by dark matter halo particles}

\author[Ludlow et al.] {\parbox{18cm}{
    Aaron D. Ludlow$^{1,\star}$,
    S. Michael Fall$^{2}$,
    Joop Schaye$^{3}$ and
    Danail Obreschkow$^{1}$
  }\vspace{0.3cm}\\
  $^{1}${International Centre for Radio Astronomy Research, University of Western Australia, 35 Stirling Highway, Crawley,}\\
  {Western Australia, 6009, Australia}\\
  $^{2}${Space Telescope Science Institute, 3700 San Martin Drive, Baltimore, MD 21218, USA.
  }\\
  $^{3}${Leiden Observatory, Leiden University, PO Box 9513, 2300 RA Leiden, the Netherlands}\\
}

\date{Accepted XXX. Received YYY; in original form ZZZ}

\pubyear{2021}

\hypersetup{draft}
\begin{document}
\label{firstpage}
\pagerange{\pageref{firstpage}--\pageref{lastpage}}
\maketitle

\begin{abstract}
  We use idealized N-body simulations of equilibrium stellar disks embedded within course-grained dark matter haloes
  to study the effects of spurious collisional heating on disk structure and kinematics. Collisional heating artificially
  increases the vertical and radial velocity dispersions of disk stars, as well as the thickness and size of
  disks; the effects are felt at all galacto-centric radii. The integrated effects of collisional heating are determined
  by the mass of dark matter halo particles (or equivalently, by the number of particles at fixed halo mass), their local
  density and characteristic velocity dispersion, but are largely insensitive to the stellar particle mass. The effects 
  can therefore be reduced by increasing the mass resolution of dark matter in cosmological simulations, with
  limited benefits from increasing the baryonic (or stellar) mass resolution. We provide a simple empirical model that
  accurately captures the effects of spurious collisional heating on the structure and kinematics of simulated disks,
  and use it to assess the importance of disk heating for simulations of galaxy formation. We find that the majority of state-of-the-art
  zoom simulations, and a few of the highest-resolution, smallest-volume cosmological runs, are in principle able to
  resolve thin stellar disks in Milky Way-mass haloes, but most large-volume cosmological simulations cannot. For example,
  dark matter haloes resolved with fewer than $\approx 10^6$ particles
  will collisionally heat stars near the stellar half-mass radius such that their vertical velocity dispersion increases
  by $\gtrsim 10$ per cent of the halo's virial velocity in approximately one Hubble time.
\end{abstract}

\begin{keywords}
Galaxy: kinematics and dynamics -- Galaxy: evolution -- Galaxy: structure -- Galaxy: disc -- methods: numerical
\end{keywords}

\renewcommand{\thefootnote}{\fnsymbol{footnote}}
\footnotetext[1]{E-mail: aaron.ludlow@icrar.org}

\section{Introduction}
\label{SecIntro}

Large-volume cosmological hydrodynamical simulations are now commonplace, and published runs span a considerable range in
both force and mass resolution \citep[e.g.][]{Schaye2010,Dubois2014,Vogelsberger2014,Schaye2015,Dolag2016,McCarthy2017,Pillepich2019}.
By self-consistently co-evolving the dark matter (DM), gas and stellar particles, these simulations can in principle be used to
inspect the morphological and kinematic properties of galaxies, the relation between them and their cosmological
origins \citep[e.g.][]{Dubois2016,Correa2017,Clauwens2018,Trayford2019,Thob2019,Du2020}. Recent simulations have gained traction in both the
theoretical and observational communities, in large part due to their ability to reproduce a variety of observed galaxy scaling
relations, often interpreted as a testament to their credibility. The results of these studies are reassuring: in many respects,
simulated galaxies resemble observed ones \citep[e.g.][]{Furlong2017,Ludlow2017,Trayford2017,Nelson2018,vandeSande2019}.

The initial conditions for smoothed particle hydrodynamical simulations typically sample the linear density field with an equal
number of DM and gas particles such that the particle mass ratio is
$\mu_{\rm bar}\equiv m_{\rm DM}/m_{\rm bar}=\Omega_{\rm DM}/\Omega_{\rm bar}\approx 5.4$ ($\Omega_i$ is the cosmological density parameter of
species $i$). Cosmological adaptive mesh refinement (AMR) or moving-mesh simulations also adopt DM particle masses that exceed those
of baryonic fluid elements. Under the right conditions stellar particles form from these fluid elements, inheriting their masses. While not strictly
necessary, this set-up is practical for a number of reasons: it simplifies the creation of initial conditions, it reduces compute
times which in turn permits larger volumes to be simulated, and may appear to offer better resolution
in the stellar components of galaxies\footnote{Several simulations artificially increase the stellar mass resolution by creating
  multiple stellar particles per gas fluid
  element such that the resulting stellar-to-DM particle mass ratio is $\mu\equiv m_{\rm DM}/m_\star > 5$
  \citep[e.g.][]{Dubois2014,Revaz2018,Dubois2020}.}
by shifting the computational effort away from their larger, rounder DM haloes. 

Recently, \citet[][see also \citealt{Revaz2018}]{Ludlow2019,Ludlow2020} showed that long-range gravitational interactions between stellar and DM particles
(a scattering process often referred to as ``collisions'', a vernacular we adopt throughout the paper) heat the stellar particles within galaxies,
resulting in spurious growth of their sizes. The effect is numerical and arises due to momentum exchange between the two particle
species as galaxies and their DM haloes progress toward energy equipartition. This undesirable outcome affects both cosmological
and idealized\footnote{Idealized simulations, such as the ones used in this paper, refer to non-cosmological simulations of isolated
  galaxies and their dark matter haloes
  whose properties (e.g. characteristic sizes, masses, shapes, etc) can be specified a priori. They are sometimes refereed to as ``controlled''
  simulations.} simulations; it can be mitigated (but not eliminated) by adopting baryonic and dark matter particles of approximately
equal mass, i.e. by increasing the resolution of DM particles relative to baryonic ones such that $\mu\approx 1$. These results likely have broader implications
for the kinematics and internal structure of galaxies in hydrodynamical simulations. For example, collisional heating will modify the
structure of simulated disks, thickening them and making them appear more spheroidal than real disks, for which this heating effect
is negligible.

A simple, order-of-magnitude calculation relates the collisional heating timescale $t_{\rm heat}$ of an initially-cold
stellar system (i.e. the timescale over which the stellar velocity dispersion increases by of order $\Delta\sigma$) to the
mass ($m$), local density ($\rho$) and velocity ($v$) of perturbers \citep[e.g.][]{Chandrasekhar1960,LO1985}:
\begin{equation}
  t_{\rm heat}\approx \frac{\Delta\sigma^2 \, v}{8\,\pi\,G^2\,\,\rho\, m\,\ln\Lambda},
  \label{eq1}
\end{equation}
where $G$ and $\ln\Lambda$ are Newton's constant and the Coulomb logarithm, respectively. Assuming a heating time of
$t_{\rm heat}\approx 10^{10}\, {\rm yr}$ and typical properties of a Milky Way-like
galaxy -- a virial mass\footnote{Throughout the paper we quote halo masses as $M_{200}$, the mass contained within a sphere of radius $r_{200}$
  that encloses a mean density of $\rho_{200}\equiv 200\times \rho_{\rm crit}$ (where $\rho_{\rm crit}=3\,H^2/8\,\pi\, G$ is the critical density for a closed
  universe and $H$ is the Hubble constant); the circular velocity at $r_{200}$ is $V_{200}=(G\,{\rm M_{200}}/r_{200})^{1/2}$.}
${\rm M_{200}}\approx 10^{12}\,{\rm M_\odot}$ and characteristic velocity $v\approx 220\,{\rm km\,s^{-1}}$, a local
density at $r=8\,{\rm kpc}$ of $\rho\approx 10^7\, {\rm M_\odot/kpc^3}$, and a Coulomb logarithm $\ln\Lambda\approx 10$ -- 
eq.~\ref{eq1} yields a perturber mass of $m\approx 10^6\,{\rm M_{\odot}}$ for $\Delta\sigma\approx 0.25\times v=55\,{\rm km\,s^{-1}}$,
which is similar to the observed vertical velocity dispersion of disk stars.
This mass is comparable to (or smaller than) the DM particle masses adopted for the majority of recent large-volume cosmological
hydrodynamical simulations, suggesting that simulated DM haloes and the galaxies that they contain may be affected by numerical heating.

\citet[][hereafter, LO85]{LO1985} developed a more rigorous theoretical framework for modelling the collisional heating of disk stars by DM
halo objects in a different context. These authors derived the (vertical, radial and azimuthal)
heating rates caused by a halo composed of massive black holes (at the time viable DM candidates).
Based on observations of the vertical velocity dispersion of stars in the Milky Way's disk, they concluded that black holes with a critical
mass $\gtrsim 10^6{\rm M_\odot}$ cannot make a significant contribution to the DM; the integrated effects of black hole-stellar
scattering would otherwise heat the disk to levels above observational constraints.

The extent to which spurious heating affects various estimates of the morphology and kinematics of
simulated galaxies has not been established, yet the results above suggest that it should be given careful
consideration. The purpose of this paper is to address a particular
problem: the spurious collisional heating of initially thin stellar disks embedded in spherically-symmetric,
coarse-grained haloes of dark matter. We target the issue using N-body simulations of idealized equilibrium systems, and focus
on how DM-star and star-star collisions affect the vertical and horizontal kinematics of disk stars, and
their vertical scale heights. Implications of numerical heating for the morphological transformation
of simulated disk galaxies will be addressed in follow-up work.

Our paper proceeds as follows. In Section~\ref{SecPrelim}, we lay out the analytic groundwork that helps
guide the interpretation of our simulations, which are described in Section~\ref{SecSims}.
Section~\ref{SecResults} details our main results: We provide an initial assessment of the possible importance
of collisional disk heating in Section~\ref{sSecGlimpse}; we present an empirical model for disk heating in Section~\ref{sSecModel}
and apply it to vertical and radial heating rates of stars measured in our simulations in Section~\ref{sSecSigma}; we use
those results to interpret the evolution of disk scale heights in
Section~\ref{sSecScaleHeight}. We discuss the implementation of our model and its implications for current and future cosmological
hydrodynamical simulations in Section~\ref{SecCosmo} before providing some concluding remarks in Section~\ref{SecSummary}. The main
body of the paper contains the most important results; we remit a full discussion of several technical but important points to
the Appendix.

\section{Preliminaries}
\label{SecPrelim}

This section provides a brief overview of the theoretical background required to interpret and model our simulation
results. Throughout the paper, we adopt a cylindrical coordinate system with the $z-$axis normal to the disk
plane (i.e. aligned with its angular momentum vector), and $R=(r^2-z^2)^{1/2}$ is the distance from the $z$-axis,
where $r$ is the radial coordinate in three dimensions.

\subsection{Parameterizing the impact of collisional heating on the velocity dispersion profiles of disk stars}
\label{sSec_prelim1}

As mentioned above, LO85 provided an analytic description of the collisional heating of cold stellar disks.
Their detailed calculations implicitly assume a {\em cold} disk embedded in a {\em hot} isotropic halo, such that
$\sigma_i\ll \sigma_{\rm DM}$ at all times ($\sigma_i$ is the stellar velocity dispersion in the $i$ direction, and $\sigma_{\rm DM}$ the
1D velocity dispersion of halo particles; we adopt this nomenclature throughout the paper). Their model also assumes that
the mass of DM particles is much greater than that of disk stars, i.e. $m_{\rm DM}\gg m_\star$.

For perturbers of mass $m_{\rm DM}$ with local density $\rho_{\rm DM}$, the heating rate is given by
\begin{equation}
  \frac{\Delta\sigma_i^2}{\Delta t}= \sqrt{2}\,\pi\,\ln\Lambda\, f_i(...) \, \frac{G^2\, \rho_{\rm DM} \, m_{\rm DM}}{\sigma_{\rm DM}},
  \label{eq:LO}
\end{equation}
where $\Delta\sigma_i^2(t)=\sigma_i^2(t)-\sigma_{i,0}^2$ ($\sigma_{i,0}$ is the {\em initial} velocity dispersion
of stars in the $i$ direction). The functions $f_i(...)$
(see LO85 for details), which are dimensionless and of order unity
for the cases discussed in this paper, are derived from epicyclic theory \citep{Chandrasekhar1960} assuming a
Gaussian velocity distribution for halo particles; their values differ for different velocity components. 

Equation~\ref{eq:LO} has clear implications. First, it suggests that the stellar velocity variance, $\sigma_i^2(t)$,
grows linearly with time as a result of scattering. And second, that collisional heating
depends linearly on the density and mass of perturbers, $\rho_{\rm DM}$ and $m_{\rm DM}$, respectively, and inversely on their 
characteristic velocity dispersion, $\sigma_{\rm DM}$. One of the goals of this paper is to compare the heating rates predicted by
eq.~\ref{eq:LO} to the results of numerical simulations and, if necessary, to empirically calibrate its dependence on the density
and characteristic velocity dispersion of DM particles.

In the LO85 model, $\rho_{\rm DM}$ and $\sigma_{\rm DM}$ are {\em local} quantities; it is indeed sensible to assume that the heating
rate at a given location in the disk should depend on the density and velocity of perturbers there. Their model, however, neglects the
strong density gradients of DM particles across the radial extent of galactic disks, as well as the weak gradients in their characteristic
velocity dispersions. However, gravitational scattering is {\em not} a local phenomenon: it is an integrated effect resulting from a small number
of short-range interactions, and a large number of distant ones \citep{BinneyTremaine2008,LSB2019}. It is therefore plausible that suitable
halo-averaged values of $\rho_{\rm DM}$ and $\sigma_{\rm DM}$ may be more appropriate than local ones, or that simulation results will
be best described by a suitable balance between the two extremes.

To accommodate a possible non-linear density and velocity dependence on the rate of disk heating, we generalize eq.~\ref{eq:LO} as
\begin{equation}
  \frac{\Delta\sigma_i^2}{\Delta t}= \sqrt{2}\,\pi\, \ln\Lambda\, \frac{G^2\,\rho_{\rm DM} \, m_{\rm DM}}{\sigma_{\rm DM}} \times \biggr[k_i\,\biggr(\frac{\rho_{\rm DM}}{\rho_{200}}\biggl)^\alpha\biggr(\frac{V_{200}}{\sigma_{\rm DM}}\biggl)^\beta\biggl],
  \label{eq:LOII}
\end{equation}
where $\rho_{200}$ and $V_{200}$ are the mean enclosed density and circular velocity at the halo's virial radius, $r_{200}$, respectively.
Note that we have absorbed the functions $f_i(...)$ into the dimensionless constant $k_i$; its value will depend on the velocity component,
and on the exponents $\alpha$ and $\beta$; the latter two parameters determine the degree to which heating rates depend on local values of
the density and velocity dispersion of dark matter.  

For some purposes it is convenient to express eq.~\ref{eq:LOII} in dimensionless form. To do so, we define the
normalized components of the stellar velocity dispersion,
\begin{equation}
  \overline{\sigma}_i\equiv \frac{\sigma_i(r)}{\sigma_{\rm DM}}
  \label{eq:vchar}
\end{equation}
and a characteristic timescale
\begin{equation}
  t_{\rm c}\equiv \biggr(\frac{G^2\rho_{\rm DM} m_{\rm DM}}{\sigma_{\rm DM}^3}\biggl)^{-1}.
  \label{eq:tchar}
\end{equation}
This allows us to rewrite eq.~\ref{eq:LOII} more compactly, as
\begin{equation}
  \frac{\Delta\overline{\sigma}_i^2}{\Delta \tau}= \sqrt{2}\,\pi\, \ln\Lambda\, \times \Big[k_i\,\delta_{\rm DM}^\alpha\,\upsilon_{\rm DM}^{-\beta}\Big],
  \label{eq:LOred}
\end{equation}
where $\tau\equiv t/t_{\rm c}$, and we have defined $\delta_{\rm DM}\equiv\rho_{\rm DM}/\rho_{200}$ and $\upsilon_{\rm DM}\equiv\sigma_{\rm DM}/V_{200}$.

We consider $\alpha$, $\beta$ and $k_i\,\ln\Lambda$ to be free parameters, and calibrate their values using
results from idealized simulations of secularly-evolving, equilibrium disks embedded within ``live'' DM haloes.
Although $\ln\Lambda$ can, in principle, be calculated explicitly from the density and velocity structure of the halo
and disk, fitting the combined term $k_i\,\ln\Lambda$ allows us to rectify the
limitations of the binary scattering model (see discussion in \citealt{Chandrasekhar1960}, p. 55-57) and to neglect density
gradients across the disk when calculating the Coulomb logarithm.
Note that the values of the free parameters are expected to differ for $\sigma_z$ and $\sigma_R$.

The primary goal of this paper is to devise criteria for the importance of spurious DM heating in galactic disks in
cosmological simulations. For that reason, we have normalized the DM density and velocity dispersion in eq.~\ref{eq:LOII}
by the mean enclosed density and circular velocity at the halo's virial radius, $r_{200}$. Our DM heating criteria are,
however, are equally valid for galactic disks in non-cosmological simulations, for which different choices of reference
density and velocity dispersion may be more appropriate. In the latter case, the value of the free parameter $k_i$ will differ from
the one described in Section~\ref{sSecModel}, but can be easily obtained by rescaling our best-fit model. For example,
one can express the DM density and velocity dispersion in eq.~\ref{eq:LOII} in terms of the enclosed density
$\rho_0=\langle\rho(r_0)\rangle$ and circular velocity $V_0=V_c(r_0)$ at an arbitrary radius $r_0$ provided $k_i$ 
is replaced by $k_i'=k_i\,(\rho_0/\rho_{200})^\alpha(V_{200}/V_0)^\beta$.

\subsection{The asymptotic velocity dispersion of disk stars}
\label{sSec_prelim2}

Stellar and DM particles in any simulated galaxy will progress toward energy equipartition. As discussed above, disk galaxies will
be heated by this process, reaching an asymptotic end state (which may not occur in a Hubble time) for which
$\sigma_i\approx \sqrt{\mu}\times \sigma_{\rm DM}$, at which point the net exchange of energy between the two particle
species ceases. It is useful to compare this asymptotic value to the halo's escape speed which, according to the
virial theorem\footnote{Note that we have normalized the escape speed by a factor of $\sqrt{3}$ to account for the fact that
    in our case $\sigma_{\rm DM}$ corresponds to the one dimensional velocity dispersion of an isotropic DM halo.}, is
$v_{\rm esc}/\sqrt{3}\approx \sqrt{2}\, \sigma_{\rm DM}$. This suggests that energy equipartition
cannot be reached for $\mu > 2$, and that the asymptotic velocity dispersion of disk stars will be
\begin{equation}
  \sigma_{i,{\rm max}}=
  \begin{cases}
    \sqrt{\mu}\times \sigma_{\rm DM}, & \text{if}\ \mu\leq 2 \\
    \sqrt{2}  \times \sigma_{\rm DM}, & \text{if}\ \mu > 2
  \end{cases}
\label{eq:asympv}.
\end{equation}
In practice (see Section~\ref{sSecSigma}), we find that $\sigma_{i,{\rm max}}(R)\approx \sigma_{\rm DM}(R)$ provides an adequate
description of our numerical simulations on timescales $\lesssim 10\,\, {\rm Gyr}$. We therefore assume
$\sigma_{i,{\rm max}}\approx \sigma_{\rm DM}$ throughout the paper. In Appendix~\ref{sec:A2} we provide a detailed assessment of
the {\em long-term} evolution of the vertical and radial velocity dispersion of disk stars for a sub-set of galaxy models with
different $\mu$, which validates eq.~\ref{eq:asympv}.

In the limiting case of an initially cold disk (i.e. $\sigma_{0,i}=0$) the dimensionless timescale at which
$\sigma_i=\sigma_{\rm DM}$ is given by
\begin{equation}
  \tau_{\rm vir}\equiv \frac{t_{\rm vir}}{t_{\rm c}}\equiv \biggr(\sqrt{2}\,\pi\,\ln\Lambda \, k_i\,\delta_{\rm DM}^\alpha\,\upsilon_{\rm DM}^{-\beta}\biggl)^{-1}.
  \label{eq:tsigmax}
\end{equation}
We will see in Section~\ref{sSecSigma} that $t_{\rm vir}$ can be considerably shorter than
a Hubble time in poorly-resolved systems.

\subsection{Collisional disk heating and the evolution of disk scale heights}
\label{sSec_prelim3}

The vertical density structure of galaxy disks, $\rho_\star(z)$, is determined by the balance between the kinetic energy stored in
vertical stellar motions and the combined vertical compressive forces of the disk and DM halo
\citep[see][for a recent discussion in the context of gaseous equilibrium disks]{Ale2018}.
For a vertically-isothermal disk (i.e. $\sigma_z$ independent of $z$), the structure is described
by the hydrostatic equilibrium equation:
\begin{equation}
  \frac{\sigma_z^2}{\rho_\star}\frac{\partial\rho_\star}{\partial z}=-\frac{\partial}{\partial z}(\Phi_{\rm DM}+\Phi_{\star}),
  \label{eq:hydro}
\end{equation}
where $\Phi_{\rm DM}$ and $\Phi_{\star}$ are the gravitational potential due to DM and stars, respectively.

For an equilibrium disk with stellar surface density $\Sigma_\star(R)$ embedded within a DM halo with circular velocity
profile $V_{\rm c}(r)$, eq.~\ref{eq:hydro} admits analytic solutions for the disk's vertical density distribution in
limiting cases where the gravitational potential
is dominated by either the DM halo (the ``non-self-gravitating'' case, NSG) or by the disk (the ``self-gravitating'', SG).
In the thin disk limit ($z/R\ll 1$), a NSG disk has a vertical density profile that declines exponentially from the midplane,
i.e. $\rho_\star(z)\propto\exp[-(z/z_{\rm d,NSG})^2]$, with a characteristic scale height of
\begin{equation}
  z_{\rm d,NSG}=\frac{\sqrt{2}\, \sigma_z}{V_{\rm c}(R)}\, R.
  \label{eq:zNSG}
\end{equation}
For the SG case, $\rho_\star(z)\propto\sech^2[-(z/z_{\rm d,SG})]$, where
\begin{equation}
  z_{\rm d,SG}=\frac{\sigma_z^2}{\pi\, G\,\Sigma_\star(R)}.
  \label{eq:zSG}
\end{equation}
To meaningfully compare the vertical scale heights in the SG and NSG cases, which are profile-dependent,
-- and to facilitate a comparison with our simulation results -- we will recast
these characteristic heights into their corresponding half-mass heights, denoted simply as $z_{\rm SG}$ and
$z_{\rm NSG}$ for the SG and NSG cases, respectively (throughout the paper we will typically denote
the half-mass height of a disk as $z_{1/2}$, but for the SG and NSG cases above, we drop the subscript ``1/2''
for simplicity). The conversion factors are given by $z_{\rm SG}\approx 0.55\, z_{\rm d,SG}$
and $z_{\rm NSG}\approx 0.47\, z_{\rm d,NSG}$, respectively.

As mentioned above, the characteristic timescale $\tau_{\rm max}$ on which the components of the disk velocity
dispersion grow to of order $\sigma_{\rm DM}$ can be significantly shorter than the Hubble time in
poorly-resolved systems. In such instances, the scale heights of disks can grow to of order their
scale-{\em radii} $R_{\rm d}$, and we must abandon the thin-disk approximation used to arrive at eqs.~\ref{eq:zNSG} and \ref{eq:zSG}
(see Section~\ref{sSecScaleHeight}). The structure of these ``thick''
disks can be calculated in the NSG limit by solving the hydrostatic equilibrium equation,
\begin{equation}
  \frac{\sigma_z^2}{\rho_\star}\frac{\partial\rho_\star}{\partial z}=-\frac{V_{\rm c}^2(r)}{r}\biggr(\frac{z}{r}\biggl),
  \label{eq:hydro_thick}
\end{equation}
where $r=(R^2+z^2)^{1/2}$ has an explicit $z$-dependence (neglected in the thin disk limit, for which $r\rightarrow R$).
We discuss the vertical structure of thick NSG disks in more detail in Section~\ref{sSecScaleHeight}.

\section{Simulations and Analysis}
\label{SecSims}

\subsection{Initial conditions}
\label{sSecICs}

Initial conditions (ICs) for all of our simulations were created using \galics \citep{Yurin2014} and are solutions
of the collisionless Boltzmann equation. In the absence of diffusion, scattering, or genuine
disk instabilities, these collisionless structures should remain stable for many dynamical times.
Axially-symmetric models that satisfy these criteria
can be used to assess the heating rates of disk stars due to collisions with massive DM halo particles.

We model the galaxy/halo pair as a \citet{Hernquist1990} sphere in equilibrium with a thin, rotationally-supported stellar
disk. Specifically, for the DM halo we adopt a radial density distribution given by
\begin{equation}
  \rho_{\rm DM}(r)=\frac{{\rm M_{DM}}}{2\pi}\frac{a}{r\, (r+a)^3},
  \label{eq:rhohern}
\end{equation}
where $a$ is the halo's scale radius and ${\rm M_{DM}}$ is its total mass. For an isotropic DM velocity distribution, the
(one dimensional) velocity dispersion profile of the halo is given by
\begin{equation}
  \begin{split}
    \sigma_{\rm DM}^2(r)&=\frac{G\,{\rm M_{DM}}}{12\, a}\Bigg\{ \frac{12\, r\, (r+a)^3}{a^4}\,\ln\biggr(1+\frac{a}{r}\biggl) - \\
      & \frac{r}{r+a}\Biggr[25 + 52\,\frac{r}{a} + 42\,\biggr(\frac{r}{a}\biggl)^2 + 12\,\biggr(\frac{r}{a}\biggl)^3\Biggl] \Bigg\}.
    \label{eq:sighern}
  \end{split}
\end{equation}

The density profile implied by eq.~\ref{eq:rhohern} scales as
$\rho\propto r^{-1}$ in the inner regions ($r\ll a$) and as $\rho\propto r^{-4}$ at large radii ($r\gg a$). Over the radial
extent of the stellar disk it resembles the Navarro-Frenk-White profile \citep[][hereafter, NFW]{Navarro1996} that is often used to
parameterize the density profiles of CDM haloes, but has a finite mass (for NFW, $\rho\propto r^{-1}$ at
small radii but as $\rho\propto r^{-3}$ at large $r$, resulting in a divergent mass profile). For that reason, we will
characterise the structural properties of haloes in terms of the circular virial velocity, ${\rm V_{200}}$,
and concentration, $c$, of an NFW halo whose density distribution matches that of eq.~\ref{eq:rhohern}
in the inner regions \citep[see][ for details]{Springel2005c}. In this case, ${\rm M_{DM}}=(1-f_\star)\, {\rm M_{200}}$,
where $f_\star$ is the disk's stellar mass fraction, and $a=(r_{200}/c)\, \sqrt{2\, f(c)}$, where
$f(c)=\ln(1+c)-c/(1+c)$.

The radial and vertical structure of the stellar disk are initially described by
\begin{equation}
  \rho_\star(R,z)=\frac{{\rm M_\star}}{4\,\pi \,z_{\rm d}\, R_{\rm d}^2}\sech^2\biggr(\frac{z}{z_{\rm d}}\biggl)\,\exp\biggr(-\frac{R}{R_{\rm d}}\biggl),
  \label{eq:rhodisk}
\end{equation}
where $R_{\rm d}$ and $z_{\rm d}$ are the scale radius and scale height of the disk, respectively. Most models considered in this
paper adopt a radially-constant initial disk scale height (i.e. constant $z_{\rm d}$) with $z_{\rm d}/R_{\rm d}\approx 0.05$.
The scale radius, $R_{\rm d}$, is calculated assuming that the specific angular momentum of the disk, $j_{\rm d}\equiv J_{\rm d}/M_\star$,
is a fraction $f_j\equiv j_{\rm d}/j_{\rm DM}$ of the halo's specific angular momentum, which we prescribe by means of the traditional
spin parameter, $\lambda_{\rm DM}$.

We will often describe the radial structure of our simulated disks in terms of their
half-mass radii, $R_{1/2}$, or analogous cylindrical radii, $R_f$, enclosing a fraction $f$ of all disk stars.
Similarly, $z_{1/2}$ is the vertical half-mass height of the disk (note that initially the half-mass heights of
our disks are given by $z_{1/2}\approx 0.55\,z_{\rm d}$).

As discussed above, analytic estimates of the collisional heating rates of stellar disks depend
only on properties of the DM halo: the density of dark matter, the mass of the DM
particles, and their characteristic velocity dispersions. To test these expectations,
we have carried out a fiducial suite of simulations that vary each parameter individually.
Thus, we initially hold a number
of the structural properties of the halo and disk constant: we adopt $c=10$ and $\lambda_{\rm DM}=0.03$
for the concentration and spin parameter of the halo, respectively; $f_\star=0.01$ is the
disk mass fraction\footnote{We acknowledge that this value is likely too low for Milky Way-like galaxies, perhaps
  considerably \citep{Posti2021}. This choice, however, simplifies our analysis for two important reasons: 1) it ensures that the
  disks are not massive enough to modify the distribution of dark matter during the simulation; and 2) results in disks that are
  Toomre stable and do not develop unwanted substructures or perturbations that may contribute to disk heating (see Appendix~\ref{sec:A1}
  for details).}
(relative to the total mass of the system; see Appendix~\ref{sec:A1} for a detailed
discussion of this choice), and $f_j=1$, i.e. the disk and halo have the same specific angular momentum.

\begin{center}
  \begin{table*}
    \caption{The main properties of disks and haloes in our suite of simulations. The first five rows list properties of our
      fiducial set of runs. The first column lists ${\rm V_{200}}$, the circular velocity at the virial radius, $r_{200}$,
      of a Navarro-Frenk-White halo whose density profile matches our adopted Hernquist density distribution in the central
      regions; $c$ is the concentration of the NFW halo; $a/r_{200}$ is the corresponding Hernquist scale radius in units of $r_{200}$.
      The disk-to-halo ratio of specific angular momentum is $f_j\equiv j_{\rm d}/j_{\rm DM}$; the characteristic radius and scale height of the disk are respectively
      $R_{\rm d}$ and $z_{\rm d}$ (see eq.~\ref{eq:rhodisk}), which have been normalized by $a$; $\sigma_{z,0}(R_{\rm d})$
      and $\sigma_{R,0}(R_{\rm d})$ are the initial velocity dispersion of disk stars at $R_{\rm d}$ in the vertical and
      radial directions, respectively (expressed in units of ${\rm V_{200}}$). $N_{\rm DM}$ is the number of DM particles
      and $\mu$ the ratio of the DM-to-stellar particle masses (the latter determines the stellar particle mass since all of our
      models used a stellar mass fraction of $f_\star=0.01$). The right-most column lists the gravitational softening lengths, which
      are expressed in units of the disk scale height, i.e. $\epsilon/z_{\rm d}$. }
    \begin{tabular}{r r l r c r r r r c c c c}\hline \hline
      & ${\rm V_{200}}$ & $c$ & $f_j$ & $a/r_{200}$ & $R_{\rm d}/a$  &  $z_{\rm d}/a$    & $\sigma_{z,0}(R_{\rm d})$  & $\sigma_{R,0}(R_{\rm d})$ & $N_{\rm DM}$            &  $\mu$                  & $\epsilon/z_{\rm d}$ &  \\
      & $[{\rm km\,s^{-1}}]$  &     &             &             &          & $[10^{-3}]$ &       $[V_{200}]$      &     $[V_{200}]$       &   $[10^5]$                &  $[m_{\rm DM}/m_\star]$ &  &   \\\hline
      &  50             & 10  &    1.0     &  0.173      &   0.120  &  6.0        &    0.0686              &     0.0687            & $0.091,0.29,0.91,2.9,9.1$ &  1, 5, 25               & 1 &   \\
      &  100            & 10  &    1.0     &  0.173      &   0.120  &  6.0        &    0.0716              &     0.0716            & $0.23,0.73,2.30,7.28,23.0$&  1, 5, 25               & 1 &   \\
      &  200            & 10  &    1.0     &  0.173      &   0.120  &  6.0        &    0.0724              &     0.0724            & $0.18,0.58,1.84,5.82,18.4$&  1, 5, 25               & 1 &   \\
      &  300            & 10  &    1.0     &  0.173      &   0.120  &  6.0        &    0.0728              &     0.0732            & $0.20,0.62,1.96,6.21,19.6$&  1, 5, 25               & 1 &   \\ \vspace{0.1cm}
      &  400            & 10  &    1.0     &  0.173      &   0.120  &  6.0        &    0.0732              &     0.0731            & $0.15,0.47,1.47,4.66,14.7$&  1, 5, 25               & 1 &   \\
      &  200            & 7   &    0.81   &  0.222      &   0.093  &  4.7        &    0.0708              &     0.0692            &  $0.18,1.84$              &  5                      & 1 &   \\
      &  200            & 15  &    1.03    &  0.128      &   0.162  &  8.1        &    0.0777              &     0.0784            &  $0.18,1.84$              &  5                      & 1 &   \\

      &  200            & 10  &    1.0     &  0.173      &   0.120  &  12.0       &    0.111               &     0.110             &  $0.18,1.84$              &  5                      & 1 &   \\
      &  200            & 10  &    1.0     &  0.173      &   0.120  &  24.0       &    0.179               &     0.173             &  $0.18,1.84$              &  5                      & 1 &   \\

      &  200            & 10  &    1.0     &  0.173      &   0.120  &  6.0        &    0.0724              &     0.0724            &  $0.18,1.84$              &  5                      & 0.25, 0.5, 2, 4 &   \\\hline
    \end{tabular}
    \label{TabSimParam}
  \end{table*}
\end{center}

We generate a series of models with ${\rm V_{200}}=50$, 100, 200, and 400 ${\rm km\,s^{-1}}$, and for each
of these systematically vary the DM particle mass by fixed factors of
$\Delta\log m_{\rm DM}=0.5$, ranging from $m_{\rm DM}={\rm 10^4\,{\rm M}_\odot}$ to
${\rm 10^9\,{\rm M}_\odot}$. For any particular ${\rm V_{200}}$ we simulate a subset of $m_{\rm DM}$
values such that the {\em least}-resolved haloes have $N_{\rm DM}\approx 10^4$ particles, and the {\em most}-resolved
have $N_{\rm DM}\approx\,{\rm a\, few}\times 10^6$. Note that for these values of $c$, $\lambda_{\rm DM}$, $f_\star$ and $f_j$,
the ratios $R_{\rm d}/a = 0.12$ and $R_{\rm d}/r_{200} = 0.02$ (which agree well with observational data; see, e.g. \citealt{Kravtsov2013,Huang2017})
are independent of ${\rm V_{200}}$ and the local DM density is constant at fixed $R_f$. We exploit this
self-similarity when testing the dependence of the collisional heating rate on the characteristic velocity and
local density of halo particles.

Although the dependence of heating rates on $\rho_{\rm DM}$ can
be assessed by comparing the rates measured at different disk radii $R_f$, we also carried out an additional subset of runs
that varied the halo's concentration parameter, carefully adjusting the angular momentum of disk stars so that
the {\em stellar} mass distribution remained fixed (each of these used ${\rm V_{200}}=200\,{\rm km\,s^{-1}}$).

For each ${\rm V_{200}}$ and $m_{\rm DM}$, disks are sampled with $N_\star$ stellar
particles of equal mass $m_\star$ such that $\mu\equiv m_{\rm DM}/m_\star=5$, which we take as our
default value. For a Planck cosmology, this is close to the DM-to-baryon particle mass ratio used for cosmological simulations that
adopt equal numbers of baryonic and DM particles (i.e. $\mu_{\rm bar}\equiv m_{\rm DM}/m_{\rm bar}=\Omega_{\rm DM}/\Omega_{\rm bar}\approx 5.4$).

The accumulation of DM-star scattering events -- if the masses are unequal, as is the case for
our fiducial runs -- may eventually
lead to a spatial segregation of the two components, with the heavier species becoming increasingly
concentrated and the lighter species more extended. This ``mass segregation'' is one way the system
tends toward energy equipartition, and has a measurable impact on the structural evolution of
galaxies in both cosmological simulations and in idealised numerical experiments
\citep[see][for a discussion]{BinneyKnebe2002,Ludlow2019}. For that reason, we have repeated our fiducial
runs for $\mu=1$ (which suppresses mass segregation but not energy equipartition) and $\mu=25$.

Finally, we stress that all of our disks are bar-mode and Toomre-stable, which precludes heating due to dynamical instabilities.
A summary of pertinent aspects of our runs is provided in Table~\ref{TabSimParam}.

%_________________________________________                                                                                                                                                                                 
\begin{figure*}
  \includegraphics[width=0.8\textwidth]{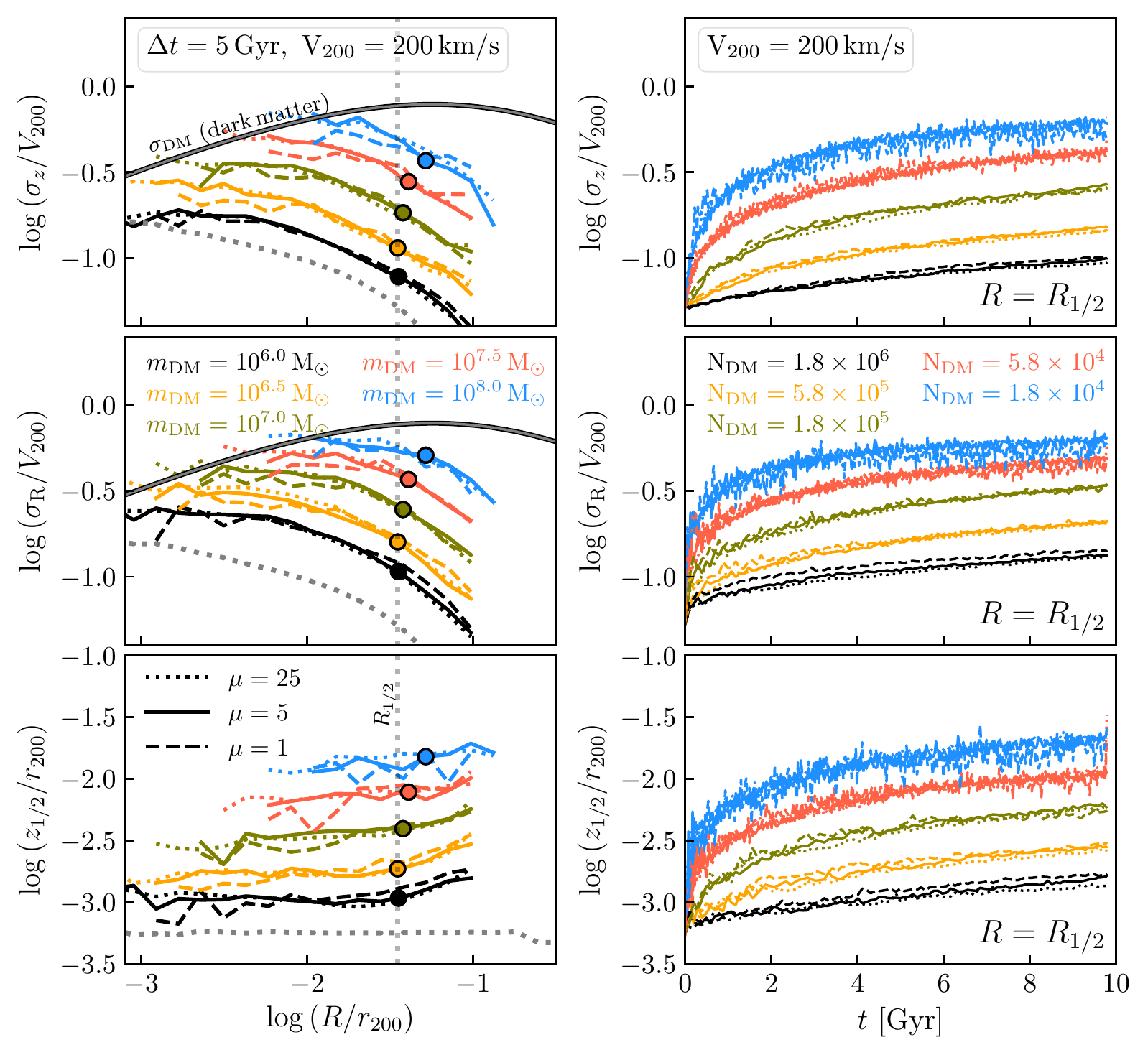}
  \caption{{\it Left-hand panels:} Radial profiles of the vertical (top) and radial (middle) dimensionless
    velocity dispersions (i.e. normalized to $V_{200}$) of stellar particles for our fiducial stellar disk models
    (${\rm V_{200}=200\, km\,s^{-1}}$, $c=10$) after $\Delta t=5\,{\rm Gyr}$; the bottom left panel shows the radial dependence of the
    vertical half-mass height of the stellar disk. The initial profiles for the highest-resolution run
    ($m_{\rm DM}=10^6\,{\rm M_\odot}$ and $\mu=25$) are shown for comparison using a grey dotted line in each of the left-hand panels. In all cases,
    profiles are plotted to the innermost radius that encloses at least 10 stellar particles.
    Vertical dotted lines in panels on the left mark the initial stellar half-mass radius of the disk, $R_{1/2}$; for comparison,
    the {\it measured} values of $R_{1/2}$ are shown using outsized circles (for $\mu=5$ runs). {\it Right-hand panels:}
    Evolution of the vertical (top) and radial (middle) velocity dispersions and the half-mass height, $z_{1/2}$ (bottom), measured
    at the initial $R_{1/2}$. In all panels, different colours correspond to different DM
    particle masses and different line-styles to different values of $\mu\equiv m_{\rm DM}/m_\star$. Note
    that the collisional heating of disk stars by DM particles is evident at {\it all} radii, even those that
    well-exceed the convergence radius of either DM or stars (not shown; see Appendix~\ref{A1a2}). Note also that the velocity dispersion
    of stars rarely, if ever,  exceeds the local 1-dimensional velocity dispersion of the surrounding DM halo (shown as a thick,
    solid grey line in the upper-left and middle-left panels; see eq.~\ref{eq:sighern}).}
  \label{fig1}
\end{figure*}
%_________________________________________                                                                                                                                                                                 

\subsection{The simulation code}
\label{sSecCode}

All runs were carried out for a total of $t=9.6\,{\rm Gyr}$. Particle orbits were integrated using the
N-body/SPH code \gadget \citep{Springel2005b} using a default value of the integration accuracy parameter
${\tt ErrTolIntAcc=0.01}$ (several runs were repeated using ${\tt ErrTolIntAcc=0.0025}$, which yields
similar results to those presented below). For all models, we adopt fixed (Plummer equivalent) softening
lengths for both DM and stellar particles that initially marginally resolve the disk scale heights, i.e. $\epsilon/z_{\rm d}=1$ (note that this also preserves
the self-similarity of our runs, since it ensures that $\epsilon$ scales self-consistently with the characteristic size
of the halo and disk). In Appendix~\ref{sec:A5} we show that our numerical results are largely insensitive
to gravitational softening provided $\epsilon\lesssim z_{\rm d}$; larger values, for which gravitational forces are not
properly modelled across the disk, suppress collisional heating, but do not eliminate it \citep[see also][]{LSB2019,Ludlow2020}.

For most runs, we output snapshots every $\Delta t=96\,{\rm Myr}$ (corresponding to a total of 100 snapshots), but for
those in which DM haloes are resolved with fewer than $10^5$ particles we adopt a higher cadence, $\Delta t=24\,{\rm Myr}$
(400 snapshots).

\subsection{Analysis}
\label{sSecAnalysis}

%_________________________________________                                                                                                                                                                                 
\begin{figure*}
  \includegraphics[width=0.8\textwidth]{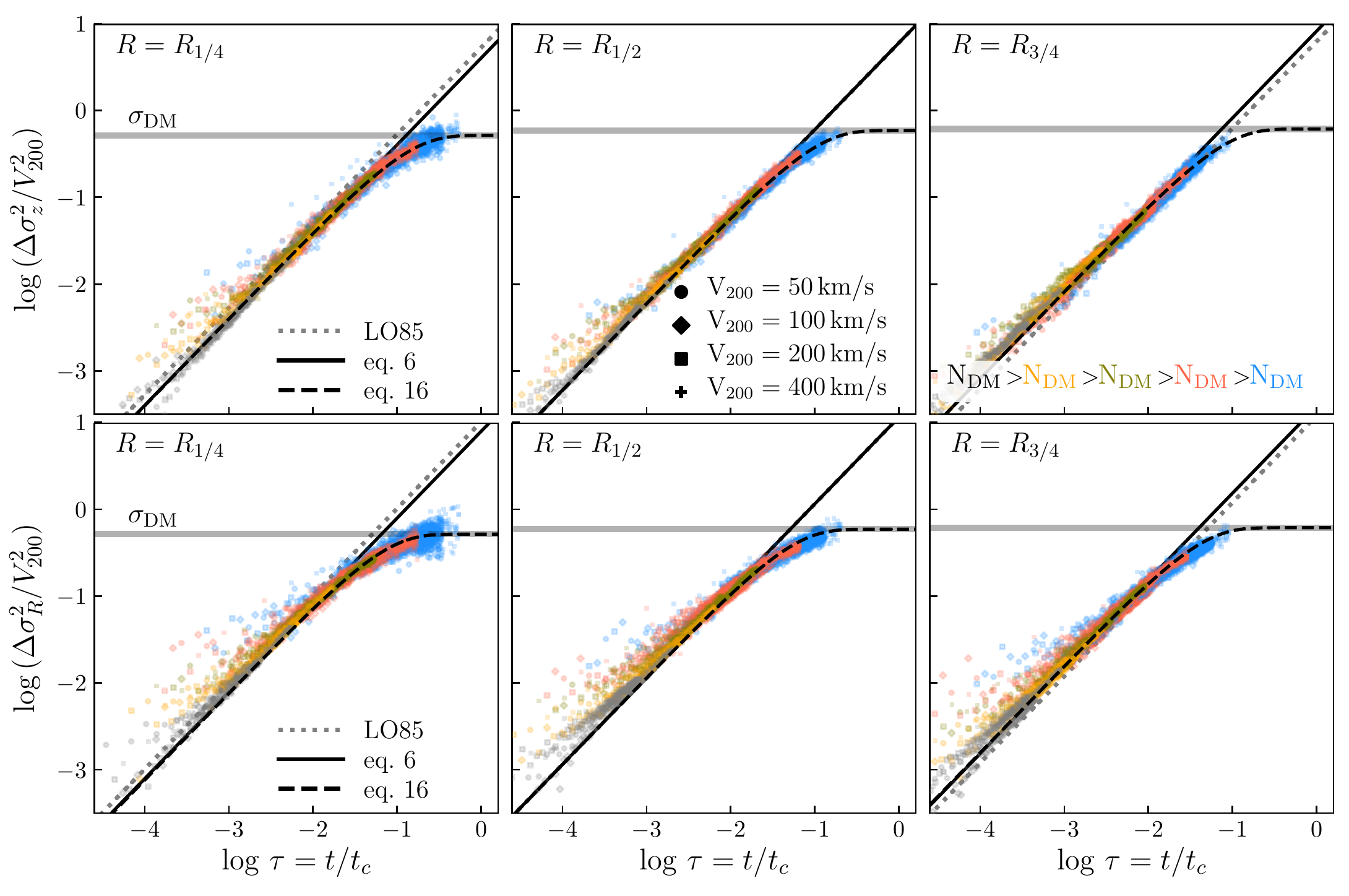}
  \caption{The evolution of the vertical (upper panels) and radial (lower panels) velocity dispersion
    of disk stars in our fiducial models (see section~\ref{sSecICs} for details). In all panels we plot the
    dimensionless velocity dispersion $\sigma_i^2/V_{200}^2$ versus $\tau=t/t_{\rm c}$ (see
    eq.~\ref{eq:tchar}) measured at the radii $R_{1/4}$ (left), $R_{1/2}$ (middle) and $R_{3/4}$ (right)
    that enclose one quarter, one half and three quarters of the {\em initial} stellar mass, respectively. Different colour
    points correspond to different $m_{\rm DM}$ (as in Figure~\ref{fig1}) and different symbols to different ${\rm V_{200}}$.
    At each radius, the evolution of $\sigma_i^2$ can be approximated by a single power-law (solid black lines)
    provided $\sigma_i\ll\sigma_{\rm DM}$ ($\sigma_{\rm DM}$ at each $R_f$ is indicated
    using a horizontal grey line in each panel). At $\tau \gtrsim 0.01$, departures from power-law growth occur
    when $\sigma_i$ approaches the local 1-dimensional DM velocity dispersion, a behaviour that can be
    approximated by eq.~\ref{eq:empheat} (black dashed lines).}
  \label{fig2}
\end{figure*}
%_________________________________________                                                                                                                                                                                 

We focus our analysis on the evolution of the vertical and radial velocity dispersions of stellar particles, and on
their half-mass height. In practice,
we measure these quantities in cylindrical annuli whose midpoints enclose specific fractions of
the total {\em initial} stellar mass of the galaxy. We typically adopt $R_{1/4}$, $R_{1/2}$ and $R_{3/4}$ enclosing
$1/4$, $1/2$ and $3/4$ of the disk stars, respectively; we use a logarithmic radial bin of width $\Delta\log R=0.2$.
The local DM density at each of these radii is independent of ${\rm V_{200}}$ for our fiducial models and, as
explained in Appendix~\ref{A1a2}, remains unaffected by collisional relaxation\footnote{We use the term ``relaxation''
  to describe the diffusion of a collisional system toward thermal equilibrium in which there is no change in
  the system's total energy. Collisional ``heating'' occurs when the total energy of a component, in our case stellar
  particles, increases when approaching thermal equilibrium.} for the duration of the simulations.
Heating rates are measured after a time interval $\Delta t$ by $\Delta\sigma_i^2/\Delta t$, where
$\Delta\sigma_i^2=\sigma_i^2(t)-\sigma_{i,0}^2$ and $\Delta t=t-t_{\rm init}$; we adopt $t_{\rm init}=0$ in all that follows (i.e. we
measure the {\em average} heating rates since the initial time $t=t_{\rm init}=0$). The vertical half-mass height, $z_{1/2}$,
is defined as the median value of $|z|$ within cylindrical shells at $R_f$.

For aesthetic purposes, results presented in several plots (Figures~\ref{fig4}, \ref{fig7} and \ref{fig8}) have been
smoothed using a Savitzky-Golay filter.

\section{Results}
\label{SecResults}

\subsection{A glimpse at the importance of collisional disk heating}
\label{sSecGlimpse}

The left panels of Figure~\ref{fig1} compare the vertical (top row) and radial (middle row) dimensionless velocity dispersion profiles
of stellar particles after $\Delta t=5\,{\rm Gyr}$ for haloes of virial velocity ${\rm V_{200}}=200\,{\rm km\,s^{-1}}$ (and $c=10$).
Different colours show
results for the range of DM particle masses spanning $10^6\,{\rm M_\odot}$ to $10^8\,{\rm M_\odot}$ ($N_{\rm DM}$ indicates
the corresponding number of DM particles), and for three values of $\mu\equiv m_{\rm DM}/m_\star$ (different line styles).
Vertical dotted lines correspond to the {\em initial} half-stellar mass radius $R_{1/2}$; for comparison, outsized circles
mark the {\em measured} $R_{1/2}$ for the subset of simulations with $\mu=5$ (similar results, not shown for clarity,
are obtained for the other values of $\mu$). The grey dotted lines show the initial profiles for the highest-resolution
run (i.e. $m_{\rm DM}=10^6\,{\rm M}_\odot$ and $\mu=25$). Profiles are plotted down to the radius that encloses at least
10 stellar particles.

As anticipated from eq.~\ref{eq:LO}, the velocity dispersion profiles have
increased at a rate proportional to the DM particle mass. For the highest resolution run
(black lines; $m_{\rm DM}=10^6\,{\rm M_\odot}$, $N_{\rm DM}=1.8\times 10^6$) heating rates are relatively low, but
still substantial. Take the $\mu=5$
runs for example. At the galaxy's half-stellar mass radius (vertical dashed line), the vertical
(radial) velocity dispersion of stars has increased by a factor of $\approx 1.5$ (2.0) after $\Delta t=5\,{\rm Gyr}$.
For increasing particle masses the situation becomes progressively worse: for $m_{\rm DM}=10^{7.5}\,{\rm M_\odot}$
(salmon lines; equivalent to $N_{\rm DM}=5.8\times 10^4$) the corresponding factor is $\approx 5.3$ (6.8), and for
$m_{\rm DM}=10^8\,{\rm M_\odot}$ (blue lines; $N_{\rm DM}=1.8\times 10^4$) it is $\approx 7.0$ (7.7).

Another important point can be inferred from Figure~\ref{fig1}: the velocity dispersion of stellar
particles does not significantly exceed the local (1-dimensional) velocity dispersion of the DM halo (plotted using a
thick solid black line, i.e. eq.~\ref{eq:sighern}) at any resolved radius. This is not unexpected: as discussed in Section~\ref{sSec_prelim2},
an equilibrium distribution of DM particles is cannot ``heat'' a cold stellar component beyond
$\sigma_i\approx \sigma_{\rm DM}$. Indeed, one may expect the collisional heating rate to drop
significantly when $\sigma_i$ approaches this value: any further injection of kinetic energy
will unbind stellar particles from the halo's potential, or if not, propel them onto extended, loosely bound orbits
(see Appendix~\ref{sec:A2} for a detailed discussion). Note that eq.~\ref{eq:LO} is valid provided $\sigma_i\ll \sigma_{\rm DM}$.

The collisional heating of disk particles has additional consequences for the evolution of the
vertical scale height of thin disks. This is shown in the lower-left panel of Figure~\ref{fig1},
where we plot the radial dependence of the disk half-mass height, $z_{1/2}(R)$. All disks start with a
uniform vertical scale height of $z_{\rm d}=(2/\ln 3)\, z_{1/2}=0.05\, R_{\rm d}$ ($\approx 113\,{\rm pc}$ for the runs shown here;
thin, dotted grey line), but become progressively thicker with time. After $\Delta t=5\,{\rm Gyr}$, $z_{1/2}$ at $R_{1/2}$ has increased to
$\approx 216\,{\rm pc}$ for $m_{\rm DM}=10^6\,{\rm M_\odot}$ ($N_{\rm DM}=1.8\times 10^6$) but, as with $\sigma_i$, the relative
increase is a strong function of $N_{\rm DM}$. For example, $z_{1/2}(R_{1/2})\approx 1.6\,{\rm kpc}$ for
$m_{\rm DM}=10^{7.5}\,{\rm M_\odot}$ ($N_{\rm DM}=5.8\times 10^4$), and $\approx 3.0\,{\rm kpc}$ for
$m_{\rm DM}=10^8\,{\rm M_\odot}$ ($N_{\rm DM}=1.8\times 10^4$) after the same time interval (values are again quoted for $\mu=5$).

Note also that the increase in the velocity dispersions and scale height of stellar particles depends 
only weakly on $\mu$, at least over the timescale plotted in Figure~\ref{fig1} (but see Appendix~\ref{sec:A2}).
This suggests that the rate of
disk heating is driven primarily by incoherent potential fluctuations brought about by shot noise in the DM
particle distribution, which lead to orbital deflections that also depend only weakly on $\mu$ (provided $\mu\gtrsim 1$).
This is because a DM particle of a given mass, $m_{\rm DM}$, will deflect a stellar particle's orbit by the same angle provided  $\mu\gg 1$;
and the deflection angle is at most a factor of 2 larger for $\mu=1$ \citep{Henon1973}, the maximum value we consider. 
Indeed, the same incoherent fluctuations in potential cause the DM halo to relax, albeit at a slower rate than that of the disk (see
Appendix~\ref{A1a2}). The motions of stars may also be scattered by any coherent density or potential fluctuation (resulting from,
e.g., molecular clouds, globular clusters, or those excited by disk instabilities or spiral arms), which also result in deflected
stellar particle  motions that are independent of their mass, although we have deliberately suppressed these in our
simulations. 

In addition to these potential fluctuations -- which affect {\em all} particles -- mass segregation leads to a
slow divergence in the average energies of the heavy (DM) and light (stellar) particle species. Since energy equipartition
demands $m_{\rm DM}\,\sigma_{\rm DM}^2\approx m_\star\, \sigma_\star^2$, we can eliminate the mass segregation effect by
setting $\mu=m_{\rm DM}/m_\star=1$, but collisions between particles in a cold stellar component and a comparatively
hot DM halo will nevertheless result in $\sigma_\star \rightarrow \sigma_{\rm DM}$. The latter is the dominant effect
in our simulations.

It is worth stressing that collisional heating affects the scale height and velocity dispersion of disk stars
at {\em all} radii. Unlike the mass profiles of DM haloes of differing numerical resolution, which have well-defined
radii beyond which convergence is achieved \citep[e.g.][]{Power2003,LSB2019},
{\em the kinematics and vertical structure of disks do not converge at any radius}.

The right-hand panels of Figure~\ref{fig1} show the time evolution of the vertical (top) and
radial (middle) velocity dispersion, and the half-mass height (bottom) measured at the initial value of $R_{1/2}$.
Different lines-styles and colours have the same meaning as those used in the corresponding panels on the left.

\subsection{An empirical model for collisional disk heating}
\label{sSecModel}
Figure~\ref{fig2} plots the normalized (i.e. to $V_{200}$) velocity dispersions, $\Delta\sigma_z^2(\tau)$ and $\Delta\sigma_R^2(\tau)$
(top and bottom panels, respectively), as a function
of the dimensionless time variable, $\tau=t/t_{\rm c}$ (see eq.~\ref{eq:tchar}), for our fiducial models. Results are shown for
three galacto-centric radii: $R_{1/4}$, $R_{1/2}$ and $R_{3/4}$ (left to right, respectively). We use different symbols for
different $V_{200}$ (for clarity, results are limited to those obtained for $\mu=5$); the colour coding indicates $m_{\rm DM}$, and
is the same as that adopted in Figure~\ref{fig1}. When expressed in scaled units, all runs exhibit a remarkable self-similarity,
regardless of ${\rm V_{200}}$ or $m_{\rm DM}$. Note also that, provided $\sigma_i\ll\sigma_{\rm DM}$ ($\sigma_{\rm DM}$ at $R_f$
is indicated using a horizontal grey
line in each panel), $\Delta\sigma_i^2(\tau)$ exhibits an approximate  power-law dependence on time,
i.e. $\Delta\sigma^2_i\propto \tau$ (the solid black lines show the best-fit power-laws obtained as described below).

However, a comparison of the results measured at different radii suggests that the local heating rate depends on the local density
or velocity dispersion of DM, rather than on a single halo-averaged value (i.e. the {\em normalization} of $\Delta\sigma_i^2$
depends on the radius at which it is measured). This is indeed expected from the LO85 model, in which heating rates depend linearly on
the local density of perturbers, and inversely on their characteristic velocity dispersions, both of which vary smoothly with radius.
In practice, we determine the dependence on $\rho_{\rm DM}$ and $\sigma_{\rm DM}$ empirically, by fitting
eq.~\ref{eq:LOred} to the results of our fiducial runs. Specifically, we determine the values of $\alpha$, $\beta$ and
$k_i\ln\Lambda$ that minimize the variance in $\Delta\overline{\sigma}_i^2/\Delta\tau$ given by eq.~\ref{eq:LOred}
(recall that $k_i\,\ln\Lambda$ is a constant for each direction $i$). When doing so, we combine the velocity dispersions
measured at $R_{1/4}$, $R_{1/2}$ and $R_{3/4}$ in all of our fiducial runs and in {\em all} simulation outputs
for which $\sigma_i^2< 0.2\times \sigma_{\rm DM}^2$ is satisfied (recall that we expect departures from eq.~\ref{eq:LOred}
unless $\sigma_i\ll\sigma_{\rm DM}$). Note that the fitting is carried out separately for the vertical and radial velocity
components.

\begin{figure}
  \includegraphics[width=0.48\textwidth]{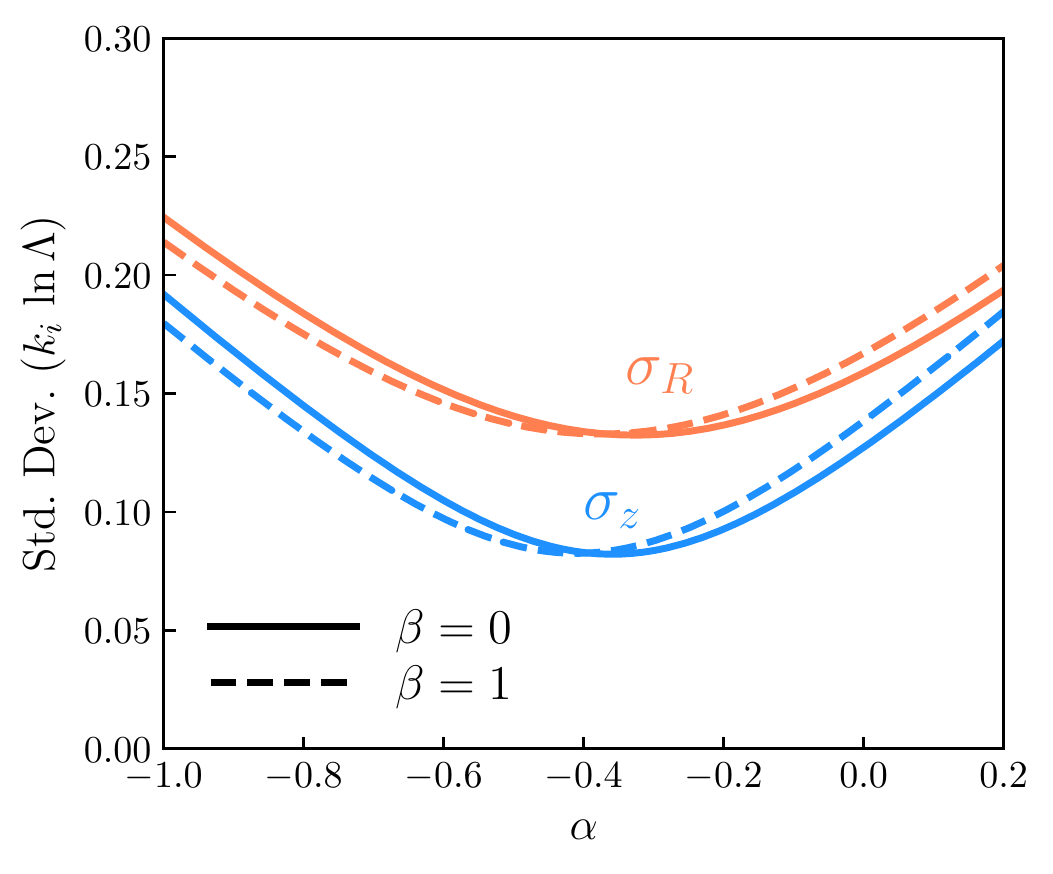}
  \caption{Standard deviation in $k_i\ln\Lambda$ obtained for our fiducial simulations plotted as
    a function of the local density exponent, $\alpha$. Different line styles correspond to different (fixed)
    values of $\beta$, as indicated in the legend. The parameters $\alpha$ and $\beta$ determine the degree to which
    the evolution of the stellar velocity dispersion due to collisional heating depends on the local value of the
    DM density and velocity dispersion, respectively (see eq.~\ref{eq:LOred}; $\alpha=\beta=0$ are the values anticipated
    by LO85). Blue lines corresponds to the vertical velocity dispersion, $\sigma_z$, and orange lines to the radial
    velocity dispersion, $\sigma_R$. }
  \label{fig3}
\end{figure}

Figure~\ref{fig3} plots the standard deviation in $k_i\,\ln \Lambda$ as a function of $\alpha$ for 
$\beta=0$ (solid lines; the value anticipated by LO) and $\beta=1$ (dashed lines). Note that the minimum variance is
largely independent of $\beta$ (we have
verified this result for the range $0\leq \beta \leq 3$), but depends significantly on $\alpha$. The best-fit values are
$\alpha_z=-0.356$ for $\sigma_z$ and $\alpha_R=-0.331$ for $\sigma_R$; the corresponding best-fit values of $k_i\,\ln\Lambda$ are
18.80 and 35.97 for $\sigma_z$ and $\sigma_R$, respectively (see Table~\ref{TabBestFit}). This suggests that, at fixed DM
particle mass, collisional heating depends on the local density of DM but not on its velocity dispersion. This is
likely because the density of DM exhibits strong radial gradients across the extent of the disk, whereas its velocity
dispersion does not. We henceforth adopt $\beta=0$.

For the particular case $\alpha=\beta=0$, the LO85 model makes precise predictions for the values of $k_i$,
or equivalently, for the vertical and radial heating rates that arise as a result of gravitational scattering (see
eq.~\ref{eq:LOred}). The predictions depend on the shape of the galaxy's rotation curve, on the Coulomb logarithm
and on the dimensionless relative velocity between stellar and DM particles, i.e. $\gamma\equiv V_c/\sqrt{2}\sigma_{\rm DM}$
(for a non-rotating DM halo);
we therefore compare our numerical results to their analytic predictions at the radius $R_{1/2}$ of our fiducial galaxy model,
where $\gamma(R_{1/2})\approx 0.82$ (we have verified that similar results are obtained at other characteristic radii, e.g.
$R_{1/4}$ and $R_{3/4}$). The Coulomb logarithm, assuming a maximum impact parameter $b_{\rm max}=r_{200}$ and
a minimum impact parameter $b_{\rm min}=\epsilon$
\citep[see, e.g.,][]{LSB2019}, is approximately $\ln\Lambda\approx \ln(r_{200}/\epsilon)\approx 7.24$ for our fiducial
runs. Using these values, LO85 predict a (dimensionless) vertical heating rate of
$\Delta\overline{\sigma}_z^2/\Delta \tau\approx 21.3$, and a radial heating rate of
$\Delta\overline{\sigma}_R^2/\Delta \tau\approx 45.9$. Our simulations have
$\Delta\overline{\sigma}_z^2/\Delta \tau\approx 10.2$ and $\Delta\overline{\sigma}_R^2/\Delta \tau\approx 22.6$
for the vertical and radial heating rates, respectively, about half of the analytically-predicted values. Whether
the assumption $b_{\rm max}=r_{200}$ is justified for our simulations is however doubtful. As discussed by \citet{Henon1958}
\citep[see also][]{Lacey1984,BinneyTremaine2008}, stellar particles will respond adiabatically to encounters with
dark matter particles when $b\gtrsim b_{\rm max}\sim R_{1/2}$ (such interactions will therefore not contribute to heating),
while those with $b\lesssim b_{\rm min}\sim \epsilon$ will
be suppressed by the gravitational softening. In this case $\ln\Lambda\approx \ln(R_{1/2}/\epsilon)\approx 3.51$, and LO85 predict
$\Delta\overline{\sigma}_z^2/\Delta \tau\approx 10.3$
and $\Delta\overline{\sigma}_R^2/\Delta \tau\approx 22.3$ for the (dimensionless) vertical and radial heating rates, respectively;
these values are in excellent agreement with our numerical results. 

The main uncertainty in this comparison therefore comes from the Coulomb logarithm, $\ln\Lambda$, which can only
be specified approximately.
A more meaningful comparison between our numerical results and the predictions of LO85 can therefore be obtained
by comparing corresponding ratios $\Delta\sigma_z^2/\Delta\sigma_R^2$, for which the value of $\ln\Lambda$
cancels out. Based on the results above, we find $\Delta\sigma_z^2/\Delta\sigma_R^2=k^{\rm LO85}_z/k^{\rm LO85}_R\approx 0.46$ for 
our fiducial galaxy-halo model.
To properly compare this ratio to our numerical results, we must account for the additional density
dependence of eq.~\ref{eq:LOII}, not present in LO85. Specifically, we find
$k_z/k_R\, \times [\rho(R_{1/2})/\rho_{200}]^{\alpha_z-\alpha_R}\approx 0.45$, which is within about 2 per cent
of the value predicted by LO85. (Note that the values quoted above are valid in the limit of
low stellar velocity dispersion, i.e. $\sigma_i\ll \sigma_{\rm DM}$.)

The solid black lines in Figure~\ref{fig2} show eq.~\ref{eq:LOred} plotted with these best-fit values of
$\alpha$ and $k_i\ln\Lambda$, and $\beta=0$. For comparison, the dotted lines in the left- and right-most panels show extrapolations
of the $R=R_{1/2}$ result (middle panels) assuming $\alpha=0$, i.e. the value anticipated by LO85. Although $\alpha=0$
describes our numerical simulations rather well, $\alpha\approx -0.3$ provides an even better description. 

Note too that our simple empirical model works well provided $\sigma_i\ll\sigma_{\rm DM}$. As $\sigma_i\rightarrow\sigma_{\rm DM}$,
however, a systematic departure from this relation is evident in Figure~\ref{fig2}. This is seen most
clearly for the two largest values of $m_{\rm DM}$ (i.e. $10^{7.5}\,{\rm M_\odot}$, red, and $10^8\,{\rm M_\odot}$, blue), for
which velocity dispersions are seen to saturate, approaching an asymptotic constant $\sigma_i\approx \sigma_{\rm DM}$
(horizontal grey lines). This behaviour can be captured quite accurately by the formula
\begin{equation}
  \sigma_i^2= \sigma_{\rm DM}^2\,\biggr[1-\exp\biggr(-\frac{(t+t_0)}{t_{\rm vir}}\biggl)\biggl],
  \label{eq:empheat}
\end{equation}
where $t_{\rm vir}$ is given by eq.~\ref{eq:tsigmax}, and $t_0$ is chosen so that $\sigma_i(t=0)=\sigma_{i,0}$, i.e.
\begin{equation}
  \frac{t_0}{t_{\rm vir}} = \ln\biggr(1+\frac{\sigma_{i,0}^2}{\sigma_{\rm DM}^2-\sigma_{i,0}^2}\biggl).
  \label{eq:t0}
\end{equation}

Equation~\ref{eq:empheat} has useful features: 1) it depends only on parameters already introduced in eq.~\ref{eq:LOred}; 2)
it predicts $\sigma_i\rightarrow\sigma_{\rm DM}$ as $\tau\rightarrow \infty$, and
3) in the limit $\sigma_{i,0}\rightarrow 0$ it returns the heating rates predicted by the LO85 model, 
eq.~\ref{eq:LOred} (this can be easily verified by taking the second-order Taylor expansion of eq.~\ref{eq:empheat} and
substituting in the definitions of $\tau_{\rm vir}$ and $t_0$, i.e. eq.~\ref{eq:tsigmax} and eq.~\ref{eq:t0}, respectively).
Note that eq.~\ref{eq:empheat} also ensures that the effects of collisional heating are the same
regardless of the initial velocity dispersion of stellar particles. The black dashed lines in Figure~\ref{fig2} correspond to
eq.~\ref{eq:empheat} for the same best-fit parameters described above (and provided in Table~\ref{TabBestFit}).

\begin{center}
  \begin{table}
    \caption{Best-fit parameters for eq.~\ref{eq:LOred} and \ref{eq:empheat}. The rms deviation
      from the best-fit model is quoted as $\sigma_{k_i\,\ln\Lambda}$}.
    \begin{tabular}{c c c c c r r}\hline \hline
      & Vel. Component          & $\sigma_{k_i\, \ln\Lambda}$ & $\alpha$ & $\beta$ & $k_i\,\ln\Lambda$         \\\hline
      & $\sigma_z$       & 0.082 & -0.356 & 0.0 & 18.80 \\
      & $\sigma_{\rm R}$ & 0.132 & -0.331 & 0.0 & 35.97 \\\hline
    \end{tabular}
    \label{TabBestFit}
  \end{table}
\end{center}

%_________________________________________                                                                                                                                                                                 
%_________________________________________     

%_________________________________________                                                                                                                                                                                 
\begin{figure*}
  \includegraphics[width=0.8\textwidth]{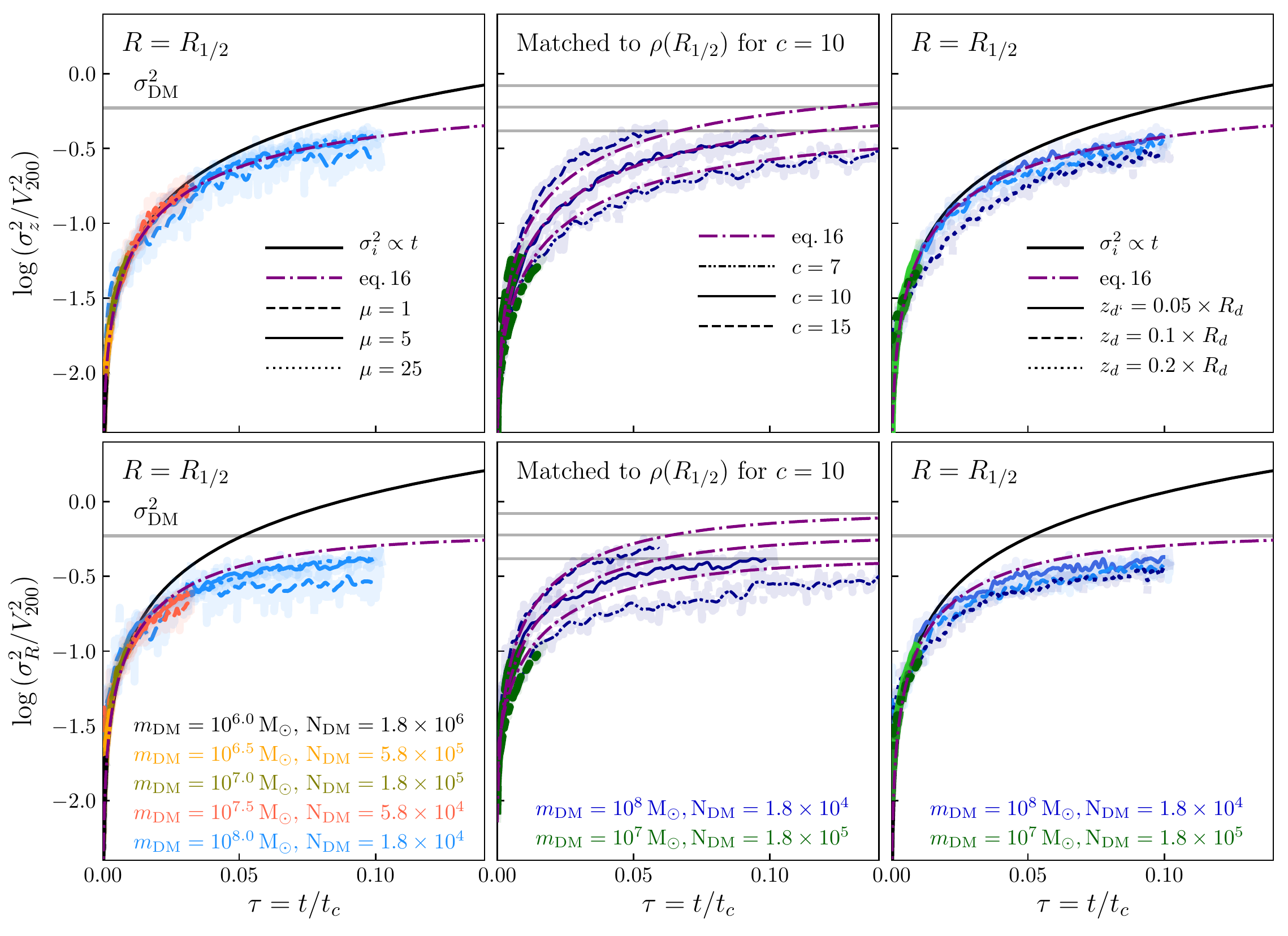}
  \caption{Vertical (upper panels) and radial (lower panels) dimensionless velocity dispersion, $\sigma_i^2/V^2_{200}$,
    as a function of (dimensionless) time, $\tau\equiv t/t_{\rm c}$ (see text for details), for variations of our fiducial
    model. Left panels adopt $f_j=1$, but vary the dark matter
    particle mass and $\mu=m_{\rm DM}/m_\star$ as indicated; results are plotted at the initial half-stellar mass radius,
    $R_{1/2}$, which is the same for all models. Middle panels vary the concentration of the dark
    matter halo, with $f_j$ adjusted in order to keep the {\em stellar} mass profile fixed (see Table~\ref{TabSimParam}).
    In this case, results for our fiducial model, i.e. $c=10$, are plotted at the initial radius $R_{1/2}$, but for the other
    values of $c$ we plot results at the radius corresponding to the same DM density (note that the DM velocity dispersion and
    stellar density differ at these radii for the three models). The right-hand panels vary the disk's {\em initial} scale
    height, $z_{\rm d}$, and plot the velocity dispersions at the initial value of $R_{1/2}$ (which is the same for all three values of $z_{\rm d}$).
    The horizontal grey lines in each panel mark the local one-dimensional velocity dispersion of DM particles at the corresponding
    radii. The solid black lines (shown only in the left- and right-hand panels, for clarity) show the best-fitting eq.~\ref{eq:LOred}, for
    which $\sigma_i^2\propto t$ (note that the fits are limited to $\sigma_i^2\leq\sigma_{\rm DM}^2$). Expressed in scaled units,
    the stellar velocity dispersions evolve approximately self-similarly, and
    are better described by eq.~\ref{eq:empheat} (purple dot-dashed lines).}
  \label{fig4}
\end{figure*}
%_________________________________________                                                                                                                                                                                 

\subsection{Evolution of the vertical and radial stellar velocity dispersion}
\label{sSecSigma}

The simple empirical model described above (eq.~\ref{eq:empheat}) successfully captures the collisional heating rates inferred
from all of our numerical simulations. This is summarized in Figure~\ref{fig4}, where we plot the evolution
of the vertical ($\sigma_z^2$, upper panels) and radial ($\sigma_R^2$, lower panels) stellar
velocity dispersion measured at several characteristic radii for a range of galaxy/halo models. All models correspond
to a DM halo with $V_{200}=200\,{\rm km\,s^{-1}}$ and $\lambda_{\rm DM}=0.03$ and a disk mass fraction of $f_\star=0.01$, but
other properties of the halo or disk are varied as described below.

The left-most panels show results for different $m_{\rm DM}$ (coloured lines) and $\mu$ (line
styles); in all cases, velocity dispersions have been measured at the initial half-stellar mass radius, $R_{1/2}$.
When expressed in scaled units, all models evolve approximately self-similarly, suggesting that
{\em heating rates are driven primarily by the mass resolution of the DM component, and are largely
  insensitive to the stellar particle mass}, although at arbitrarily late times the dispersion is significantly lower for
$\mu=1$ (see Appendix~\ref{sec:A2} and \citealt{Ludlow2019} for a more detailed discussion).

In the middle panels we plot results for three values of the halo's concentration parameter:
$c=7$ (dotted-dashed), $c=15$ (dashed) and our fiducial value $c=10$ (solid). Results are shown for $\mu=5$ and for
two different DM particle masses -- $m_{\rm DM}=10^7\,{\rm M_\odot}$ (green) and $m_{\rm DM}=10^8\,{\rm M_\odot}$ (blue).
Note that for $c=7$ and $15$ we have adjusted $f_j$ (see Table~\ref{TabSimParam} for details) to ensure that the
stellar mass profile remains unchanged -- $R_{1/2}=4.14\,{\rm kpc}$ is therefore independent of concentration, but
the local density and velocity dispersion of DM particles at that radius are not. For our fiducial model, with $c=10$,
we plot results at the initial value of $R_{1/2}$; for the other concentrations, we plot results at the radius that has the
same local DM density but {\em not} the same stellar density or DM velocity dispersion (the latter are indicated by
horizontal grey lines).

Finally, the right-most panels of Figure~\ref{fig4} compare results for three different {\em initial} disk scale heights.
Results are shown at the initial radius $R_{1/2}$.
Our fiducial models ($z_{\rm d}=0.05\, R_{\rm d}$) are shown using solid lines; initial scale heights two and four times larger
are shown using dashed and dotted lines, respectively. As with the middle panels, in this case we plot results for $\mu=5$, and
for two different DM particle masses. Note that varying the disk's initial scale height leads to different initial values
for the velocity dispersion at $R_{1/2}$ (Table~\ref{TabSimParam}), but the differences are small compared to the
integrated effects of collisional heating.

The solid black lines in the left- and right-hand panels of Figure~\ref{fig4} (not plotted in the middle panels for clarity) 
correspond to eq.~\ref{eq:LOred}, plotted using
the best-fit value of $\alpha$ (and $\beta=0$) determined above for our fiducial models (see Table~\ref{TabBestFit});
these curves describe the simulation
results reasonably well in the regime $\sigma_i\ll\sigma_{\rm DM}$ (horizontal lines indicate $\sigma^2_{\rm DM}$ at $R_{1/2}$).
However, as hinted at in Figure~\ref{fig2}, systematic departures are noticeable at late times, and are better captured by
eq.~\ref{eq:empheat}, shown as dot-dashed purple lines for the same parameters.

Overall, these results suggest that eq.~\ref{eq:empheat} provides an accurate account of the collisional heating rates
measured in all of our simulations, accommodating a wide variation in the masses of both DM and stellar particles, the
local density and velocity dispersion of DM particles, and the initial height (and velocity dispersion) of disk stars.
We next turn our attention to the evolution of the vertical scale height of disks.

\subsection{The impact of collisional heating on the vertical structure of thin disks}
\label{sSecScaleHeight}

\subsubsection{Thick versus thin disks}
\label{ssSecThickThin}

%_________________________________________                                                                                                                                                                                 
\begin{figure*}
  \includegraphics[width=0.9\textwidth]{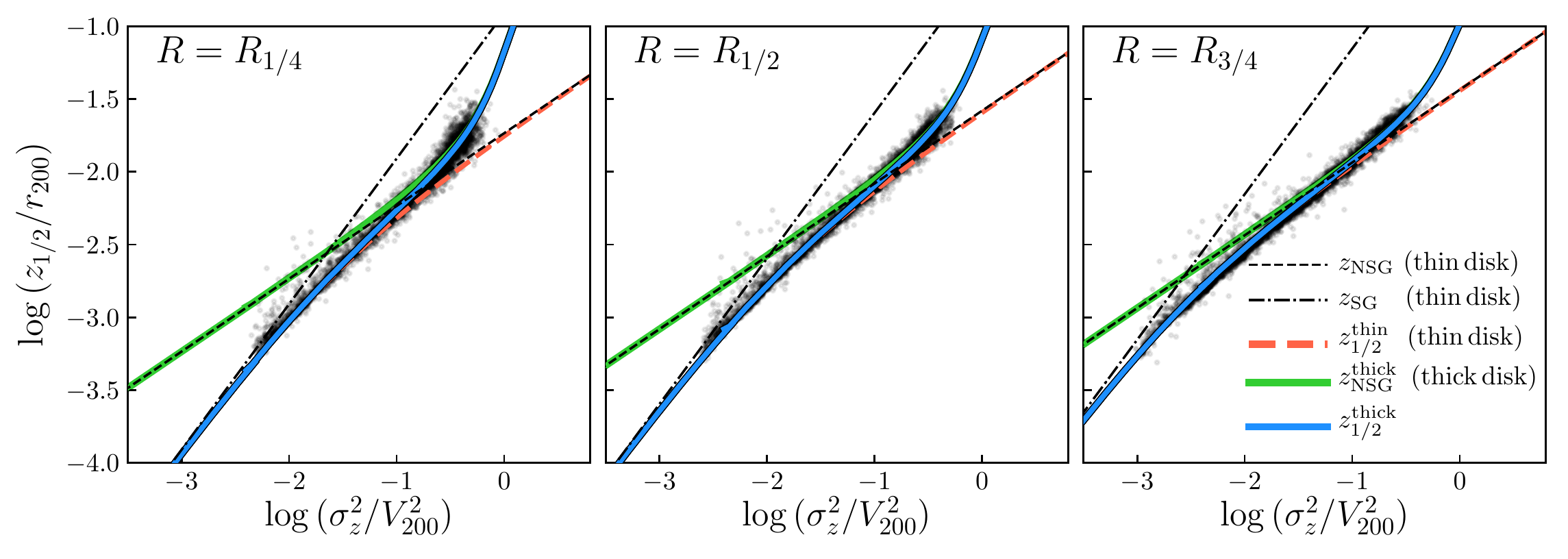}
  \caption{The vertical half-mass height of stellar particles (normalized by $r_{200}$) plotted as a function of
    their vertical velocity dispersion (normalized by $V_{200}$). Results are shown
    for the cylindrical radii corresponding to the initial values of $R_{1/4}$, $R_{1/2}$ and $R_{3/4}$ (left to right,
    respectively). Black dots show results obtained from our fiducial simulations (limited to $\mu=5$, but shown for all
    $V_{200}$ and $m_{\rm DM}$). The various curves correspond to different theoretical prescriptions for calculating the
    vertical half-mass heights of disks: $z_{\rm NSG}$ (dashed lines) and $z_{\rm SG}$ (dot-dashed) are the non-self-gravitating and
    self-gravitating half-mass heights calculated in the thin disk limit, respectively; the dashed orange line (labelled
    $z_{1/2}^{\rm thin}$) interpolates between $z_{\rm NSG}$ and $z_{\rm SG}$ (see eq.~\ref{eq:z50thin}). The thick green line corresponds to the
    half-mass height of a ``thick'' non-self-gravitating disk, $z_{\rm NSG}^{\rm thick}$ (see text for details), and
    the blue line is is an extrapolation between $z_{1/2}^{\rm thin}$ (at low $\sigma_z$) and $z_{\rm NSG}^{\rm thick}$
    (at high $\sigma_z$; see eq.~\ref{eq:zthick}).}
  \label{fig6}
\end{figure*}
%_________________________________________                                                                                                                                                                                 

In Section~\ref{sSec_prelim3} we outlined the relation between the vertical velocity dispersion of disk stars and their scale
height. The latter is determined by the balance between the vertical ``pressure'' gradients exerted by disk stars
and the combined vertical compressive forces of the DM halo and disk. In the thin
disk limit, where the latter are dominated by either the halo (non-self-gravitating, NSG) or by the 
disk (self-gravitating, SG), scales heights are given by eqs.~\ref{eq:zNSG} or \ref{eq:zSG},
respectively. The actual scale heights will be the smaller of the two, or in cases where disk and halo make comparable contributions
to the vertical acceleration profile, smaller than both. As in \citet{Ale2018}, we find that a useful approximation
for the thin-disk vertical half-mass height is given by 
\begin{equation}
  z_{1/2}^{\rm thin}=\biggr(\frac{1}{z_{\rm SG}^2}+\frac{1}{2\, z_{\rm SG}\,z_{\rm NSG}}+\frac{1}{z_{\rm NSG}^2}\biggl)^{-1/2},
  \label{eq:z50thin}
\end{equation}
where we have used the super-script 'thin' to indicate that the thin-disk approximation has been used. (Note that
$z_{\rm NSG}$ and $z_{\rm SG}$ are the half-mass heights of the NSG and SG disks, respectively.)

\begin{figure}
  \includegraphics[width=0.45\textwidth]{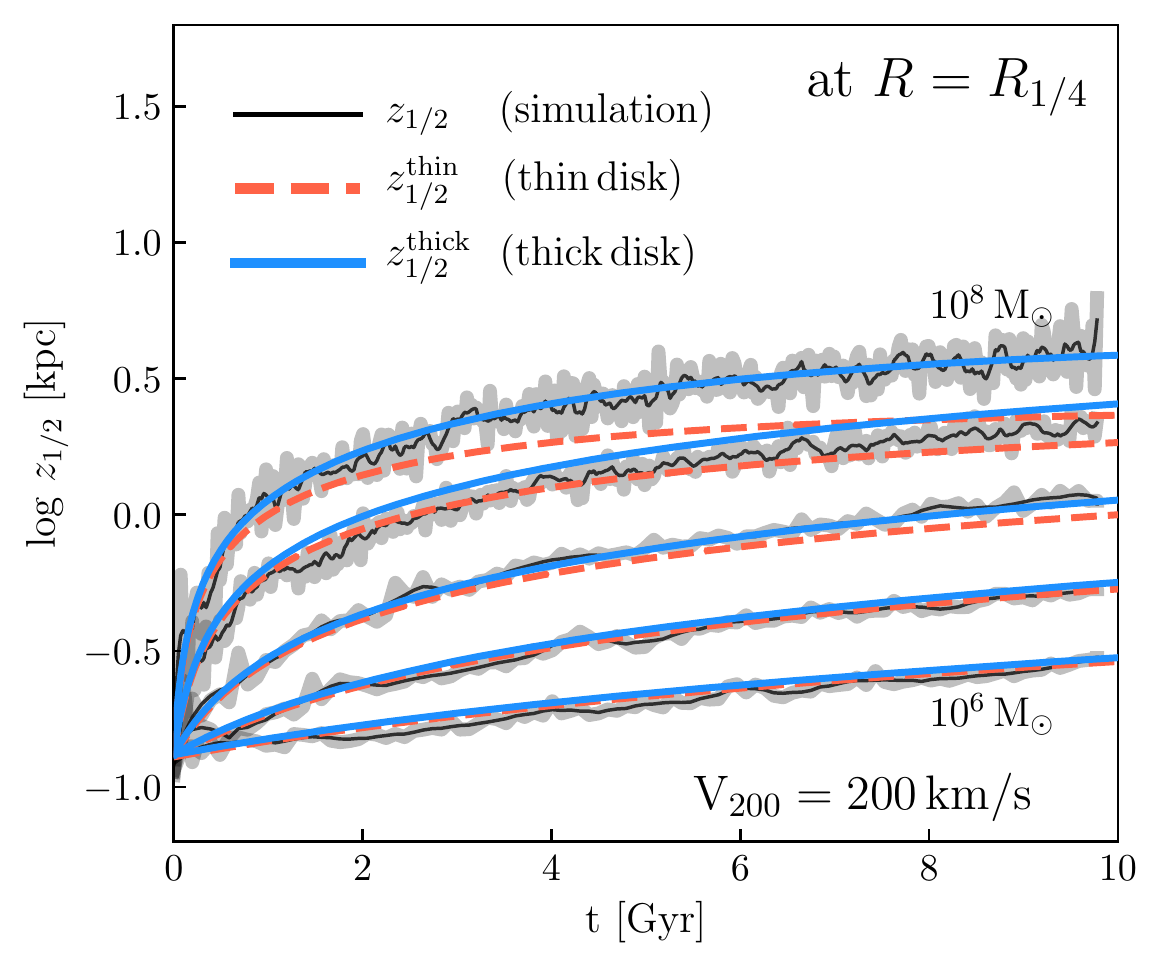}
  \caption{Evolution of the vertical half-mass height, $z_{1/2}$, measured at the radius $R_{1/4}$
    enclosing one-quarter of the initial stellar mass. Results are shown for one of our fiducial models ($V_{200}=200\,{\rm km\,s^{-1}}$, $\mu=5$)
    and for a range of dark matter particle masses (increasing from $m_{\rm DM}=10^6\,{\rm M_\odot}$ to $10^8\,{\rm M_\odot}$ in
    equally-spaced steps of $\Delta\log m_{\rm DM}=0.5$). Black lines show the simulation results; orange dashed lines show the
    evolution expected for {\em thin} stellar disks ($z_{1/2}^{\rm thin}$; eq.~\ref{eq:z50thin}) and blue lines correspond to
    $z_{1/2}^{\rm thick}$ (see text and eq.~\ref{eq:zthick} for details). In both cases,
    the vertical velocity dispersion is calculated from eq.~\ref{eq:empheat} using the best-fit parameters provided in
    Table~\ref{TabBestFit}.}
  \label{fig7}
\end{figure}

In Figure~\ref{fig6} we plot $z_{\rm SG}$ (dot-dashed lines), $z_{\rm NSG}$ (short dashed line) and $z_{1/2}^{\rm thin}$
(dashed orange lines; eq.~\ref{eq:z50thin}) as a function of the vertical stellar velocity dispersion at radii corresponding to the initial
values of $R_{1/4}$, $R_{1/2}$ and $R_{3/4}$ (left-to-right panels, respectively). Results
obtained from our fiducial set of simulations ($c=10$ and $\mu=5$, but for all values of $V_{200}$ and $m_{\rm DM}$) are underlaid as
black points (each simulation output is shown as a single point).
Note that, provided the vertical velocity dispersion remains small ($\sigma_z^2/V_{200}^2\lesssim 0.1$, or so), eq.~\ref{eq:z50thin}
accurately describes the $\sigma_z$-dependence of the measured half-mass heights. This is also shown in
Figure~\ref{fig7}, where we plot the time-dependence of $z_{1/2}$, measured at the initial value of $R_{1/4}$, for one of
our fiducial models ($V_{200}=200\,{\rm km\,s^{-1}}$, $\mu=5$), with $m_{\rm DM}$ spanning $10^6\,{\rm M_\odot}$ to
$10^8\,{\rm M_\odot}$ (different sets of curves). Results obtained when assuming the {\em thin} disk approximation,
$z_{1/2}^{\rm thin}$, are again shown using dashed orange lines.
This simple model works remarkably well {\em provided the disk remains thin}; If it does not, then the model does not,
which is most readily apparently for the lowest-resolution run in which disks are maximally heated; in this case $z_{1/2}^{\rm thin}$
systematically under-predicts the measured scale height.

%_________________________________________                                                                                                                                                                                 
%_________________________________________

These extreme instances of collisional disk thickening can nevertheless be accommodated by renouncing the thin
disk approximation and directly solving eq.~\ref{eq:hydro_thick}
in the NSG limit. In general, this must be done numerically but, for a Hernquist DM potential, the solution can be obtained analytically.
In this case, eq.~\ref{eq:hydro_thick} becomes
\begin{equation}
  \frac{1}{\rho_\star}\frac{\partial\rho_\star}{\partial z'}=-\frac{1}{(1-f_\star)\,\upsilon_z^2}\frac{z'}{(\sqrt{R'^2+z'^2}+a')^2\,\sqrt{R'^2+z'^2}},
  \label{eq:hydrohern}
\end{equation}
where $\upsilon_z=\sigma_z/V_{200}$ and the primes indicate quantities normalized by $r_{200}$ (i.e. $x'=x/r_{200}$).
This can be easily integrated, yielding
\begin{equation}
  \ln\biggr[\frac{\rho_\star(R',z')}{\rho_\star(R',0)}\biggl]=\frac{1}{(1-f_\star)\,\upsilon_z^2}\biggr[\frac{1}{\sqrt{R'^2+z'^2}+a'}-\frac{1}{R'+a'}\biggl].
  \label{eq:rhoDhern}
\end{equation}
Note that in the limit $z'\rightarrow \infty$, eq.~\ref{eq:rhoDhern} approaches a constant density, with a divergent total mass. This
is unphysical and requires a cut off in the density profile, for which we adopt $z'_{\rm max}=z_{\rm max}/r_{200}\approx 1$. We then find
$z_{\rm NSG}^{\rm thick}$ by equating the integrals of eq.~\ref{eq:rhoDhern} from 0 to $z_{\rm NSG}^{\rm thick}$, and from
$z_{\rm NSG}^{\rm thick}$ to $z_{\rm max}$. 
The results are shown in Figure~\ref{fig6} as thick
green lines\footnote{Note that, for the stellar disks and DM haloes considered in this paper, the vertical
  scale heights obtained this way are largely insensitive to the initial upper-limit of integration, $z'_{\rm max}$. For example,
  adopting $z'_{\rm max}=0.5,\,2.0$, or 10 yield scale heights that are virtually indistinguishable from those plotted in Figure~\ref{fig6}.}.
The thick blue lines labelled $z_{1/2}^{\rm thick}$ show a simple interpolation between $z_{1/2}^{\rm thin}$ (at low $\sigma_z$)
and $z_{\rm NSG}^{\rm thick}$ (at high $\sigma_z$), specifically,
\begin{equation}
  z_{1/2}^{\rm thick}=z_{1/2}^{\rm thin}+(z_{\rm NSG}^{\rm thick}-z_{\rm NSG}).
  \label{eq:zthick}
\end{equation}
Equation~\ref{eq:zthick} is also plotted in Figure~\ref{fig7} using thick blue lines, which in poorly resolved systems provides a clear
improvement over scale heights estimated using the thin disk approximation.

We show in Appendix~\ref{sec:A5} that the impact of spurious heating on the half-mass height of disks
is suppressed in simulations that adopt gravitational softening lengths $\epsilon\gtrsim z_{\rm d}$ (in which cases gravitational forces
are not properly modelled across the disk). Nevertheless, the relation between $z_{1/2}$
and $\sigma_z$ depicted in Figure~\ref{fig6} remains the same, suggesting that, regardless of $\epsilon$, vertical scale heights
of simulated disks can be inferred from their vertical velocity dispersion profiles as described above.

%_________________________________________                                                                                                                                                                                 
\begin{figure*}
  \includegraphics[width=0.8\textwidth]{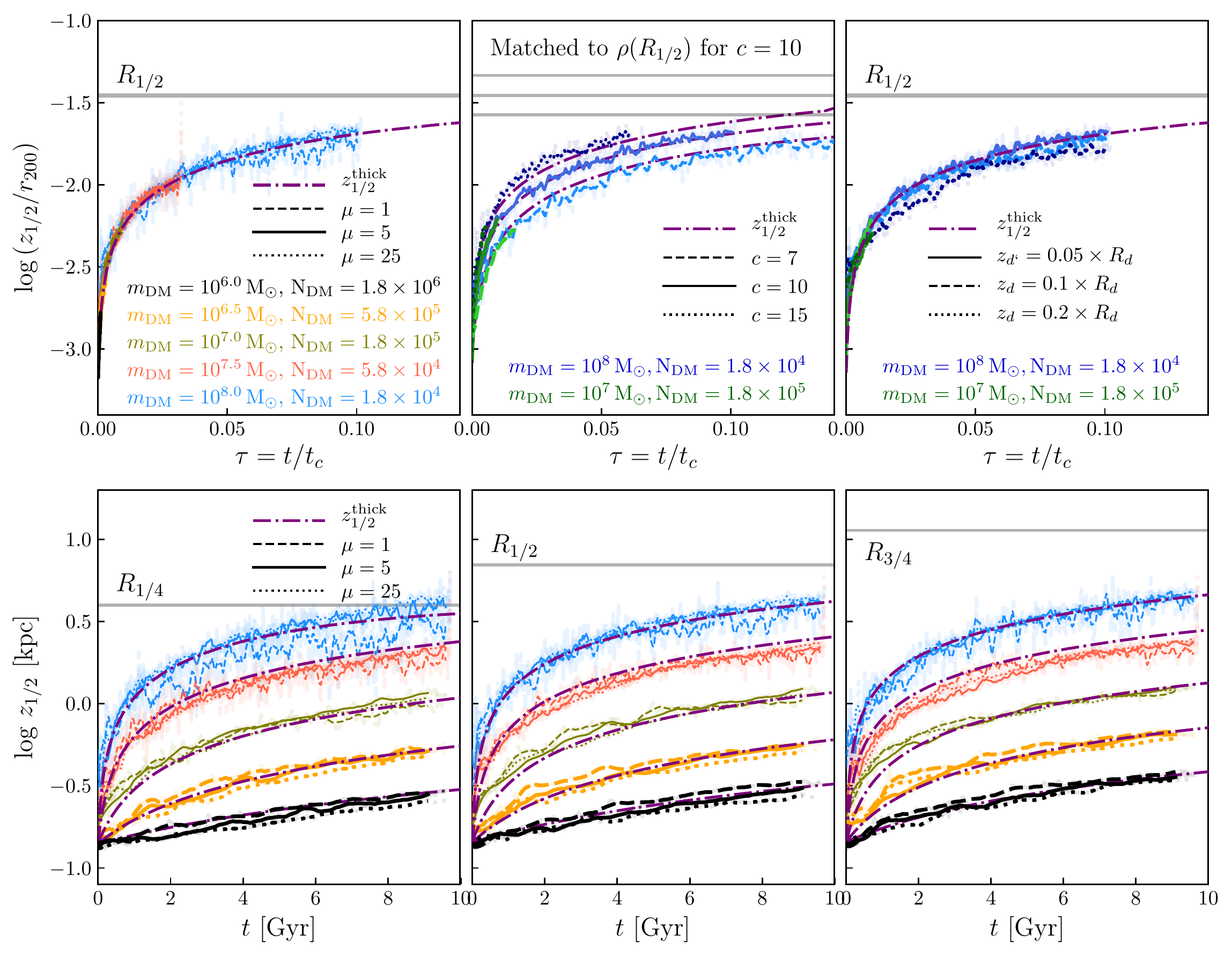}
  \caption{{\em Upper panels:} Same as Figure~\ref{fig4}, but for the evolution of the vertical half-mass height,
    $z_{1/2}$. {\em Lower panels:} Evolution of the half-mass height at the initial radii $R_{1/4}$ (left),
    $R_{1/2}$ (middle) and $R_{3/4}$ (right; these radii are indicated by horizontal grey lines in each panel)
    for our fiducial models. Different colours correspond to different dark matter particle masses and
    different line styles to different $\mu=m_{\rm DM}/m_\star$. In all panels, the purple dot-dashed curves correspond
    to $z_{1/2}^{\rm thick}$ (see eq.~\ref{eq:zthick} and Section~\ref{sSecScaleHeight} for details) calculated assuming $\sigma_z$
    evolves according to eq.~\ref{eq:empheat}.}
  \label{fig8}
\end{figure*}
%_________________________________________                                                                                                                                                                                 

\subsubsection{Evolution of the vertical half-mass height}
\label{ssSecThickThin}

The upper-panels of Figure~\ref{fig8} plot the evolution of the normalized vertical half-mass heights (i.e. $z_{1/2}/r_{200}$)
for the same models used for Figure~\ref{fig4}. The left-most panel
shows results for a range of $m_{\rm DM}$ and $\mu$ (see legend); in this case, scale heights are measured at the initial radius $R_{1/2}$.
The middle panel shows results for runs that vary the halo concentration (with the disk's spin parameter adjusted to ensure the stellar mass
profile remains the same between models). In this case, results are plotted at $R_{1/2}$ for our fiducial model ($c=10$), but at radii
corresponding to the same DM density for the other models (which correspond to different DM velocity dispersions and stellar densities).
In the right-most panel we plot results measured at $R_{1/2}$ but for disks with varying initial scale heights ($z_{\rm d}/R_{\rm d}=0.05$,
$0.1$ and $0.2$, as indicated). In all panels, simulations are shown as coloured lines; dot-dashed purple curves show the half-mass height,
$z_{1/2}^{\rm thick}$ (eq.~\ref{eq:zthick}), calculated as described above (note that the time-dependence of the
scale height is driven solely by the predicted evolution of the vertical velocity dispersion at $R_{1/2}$, i.e. by eq.~\ref{eq:empheat}).

The lower panels of Figure~\ref{fig8} plot the evolution of $z_{1/2}$ for the same models as in the upper-left panel, but at initial radii
corresponding to $R_{1/4}$ (left), $R_{1/2}$ (middle) and $R_{3/4}$ (right; each of these characteristic radii are shown as
horizontal dashed lines in the respective panel). In this case, quantities are expressed in physical units to better
emphasize the impact of collisional heating on the thickness of simulated disks. As with the upper panels, dot-dashed purple
lines show the predicted scale height evolution, $z_{1/2}^{\rm thick}$.
In all cases, the simple empirical model described above for the vertical scale heights of
disks -- combined with the evolution of the vertical velocity dispersion profiles predicted by eq.~\ref{eq:empheat} -- provides
an accurate description of our numerical results.

\section{Application to cosmological simulations of galaxy formation}
\label{SecCosmo}

The results presented above suggest that simulated disk galaxies in the presence of
coarse-grained dark matter haloes are susceptible to spurious collisional heating by dark matter particles.
This echos the findings of  \citet{Sellwood2013}, who showed that the relaxation of
stellar disks due to star-star encounters also poses a threat to simulations of galaxy formation.

Collisional heating affects the kinematics of disk particles and their vertical structure at essentially all galacto-centric radii
(unlike the collisional relaxation of DM haloes, which is primarily limited to their central regions). Heating rates depend primarily
on the mass, local density and characteristic velocity dispersion of DM particles, and can be accurately described by a simple semi-empirical
model which has its roots in the analytic theory developed by \citet{Chandrasekhar1960} and extended by \citet{LO1985}. Below we discuss
the implementation of our model and its implications for the interpretation of disk galaxy structure in cosmological simulations.

%_________________________________________
\begin{figure*}
  \includegraphics[width=0.9\textwidth]{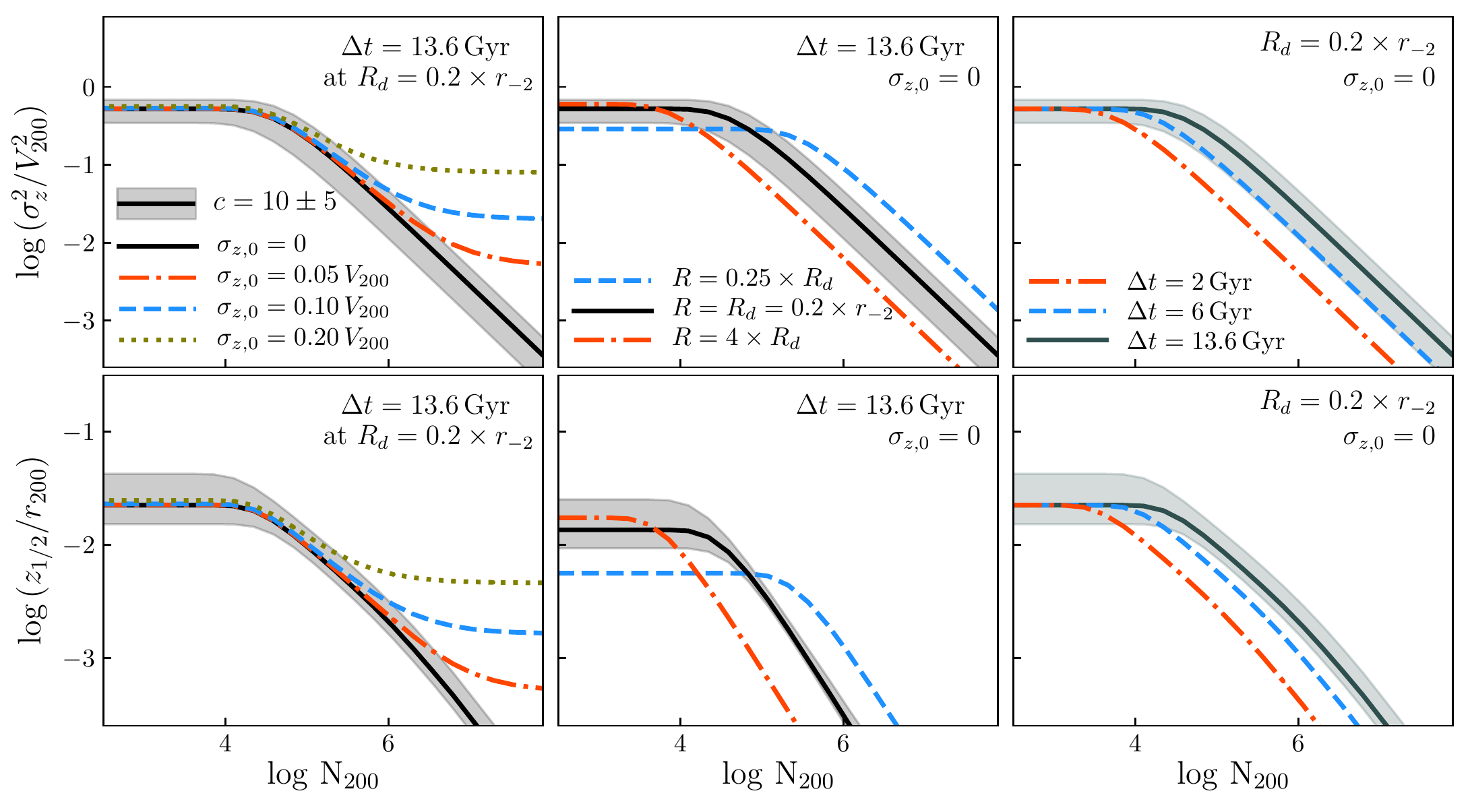}
  \caption{The vertical stellar velocity dispersion (squared, and normalized by $V_{200}^2$; top panels) and the vertical half-mass 
    height (normalized by $r_{200}$; bottom panels) as a function of the number of DM halo particles $N_{200}$ within the virial radius,
    $r_{200}$. Different columns show the effect of varying different model parameters. Panels on the left show the impact of
    varying $\sigma_{z,0}$, the initial stellar vertical velocity dispersion. The black line corresponds to the limiting case,
    $\sigma_{z,0}=0$; red (dot-dashed), blue (dashed) and green (dotted) lines to $\sigma_{z,0}/V_{200}=0.05$, 0.1 and 0.2, respectively.
    Each of these curves measure the integrated effect of collisional heating measured at the disk scale radius, $R_{\rm d}$ (approximated 
    as $R_{\rm d}=0.2\times r_{-2}$, where $r_{-2}$ is the scale radius of an NFW halo) after $\Delta t=13.6\,{\rm Gyr}$ in a DM halo with concentration
    parameter $c=10$. The shaded region shows the effect of varying the halo concentration between 5 and 15 (shown only for $\sigma_{z,0}=0$).
    The middle panels show the impact of collisional heating at different radii. Black lines correspond to $R=R_{\rm d}=0.2\times r_{-2}$ (the shaded
    region again shows the impact of varying $c$ from 5 to 15), whereas blue (dashed) and red (dot-dashed) lines correspond to $R=R_{\rm d}/4$ and
    $R=4\times R_{\rm d}$, respectively (each curve assumes $\sigma_{z,0}=0$ and $\Delta t=13.6\,{\rm Gyr}$). The right-most panels show $\sigma_z$ and
    $z_{1/2}$ measured at $R_{\rm d}$ for different integration times: $\Delta t=2,\,\,6,$ and 13.6 Gyr ($\sigma_{z,0}=0$).}
  \label{fig_model}
\end{figure*}
%_________________________________________

\subsection{Implementation of the empirical disk heating model}
\label{sSecImple}

\subsubsection{Velocity dispersions}
\label{ssSecImpleDisp}

Our empirical model for collisional disk heating is simple to implement. The time-dependence of the stellar velocity dispersion
at any cylindrical radius $R$ is given by eq.~\ref{eq:empheat}. In practice, heating rates depend on galacto-centric radius through
the radial dependence of the local density and velocity dispersion of DM particles (eq.~\ref{eq:rhohern} and \ref{eq:sighern} for the models discussed
in this paper, respectively), which can be determined from numerical simulations or calculated using empirical models that describe
the structure of DM haloes as a function of their mass \citep[e.g.][]{Ludlow2014,Correa2015c,Diemer2015,Ludlow2016,DiemerJoyce2019}.

%__________________________________________________________________
\begin{figure*}
\centering
\begin{minipage}[b]{.48\textwidth}
  \includegraphics[width=0.95\textwidth]{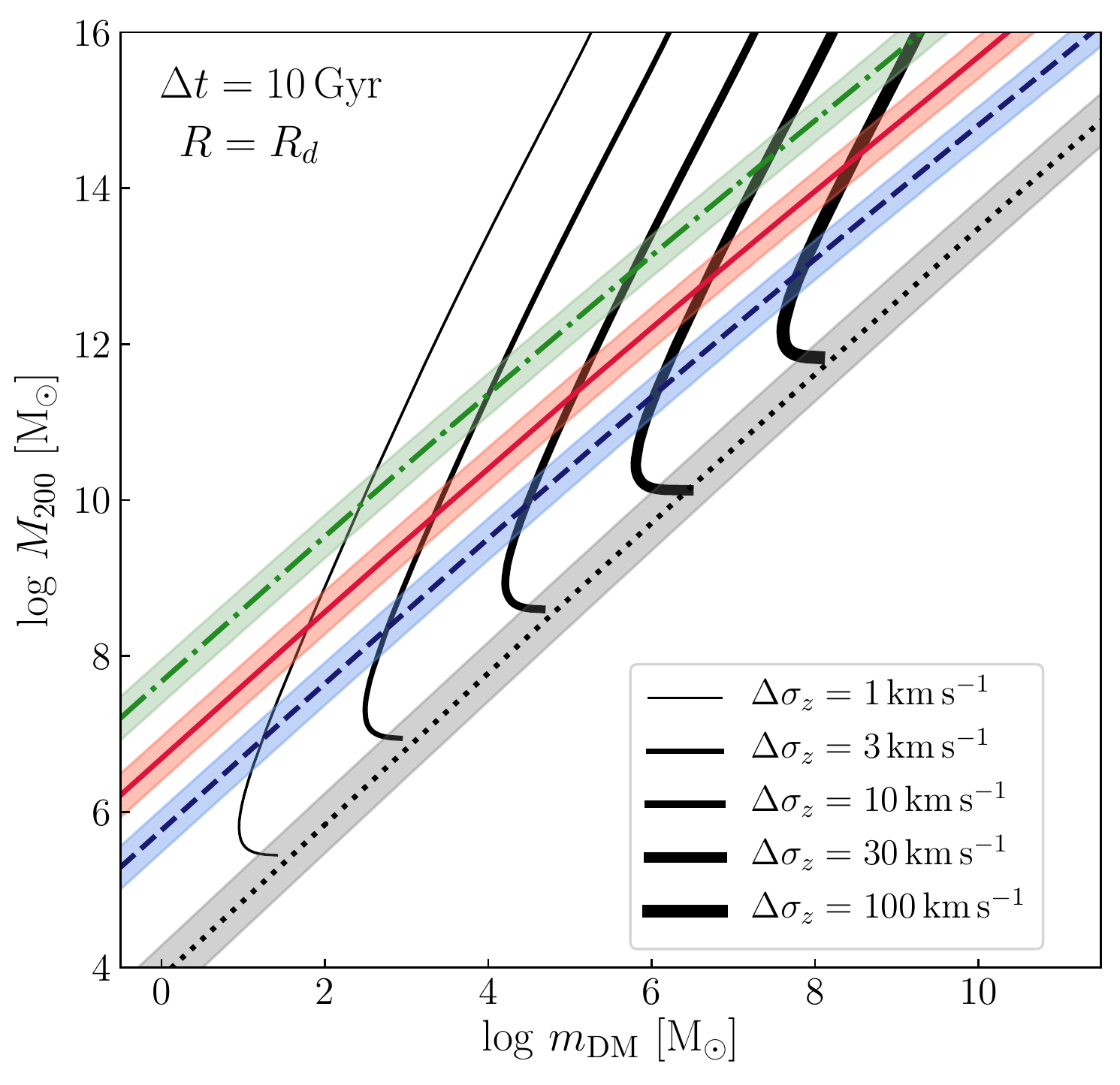}
\end{minipage}\qquad
\begin{minipage}[b]{.48\textwidth}
  \includegraphics[width=0.95\textwidth]{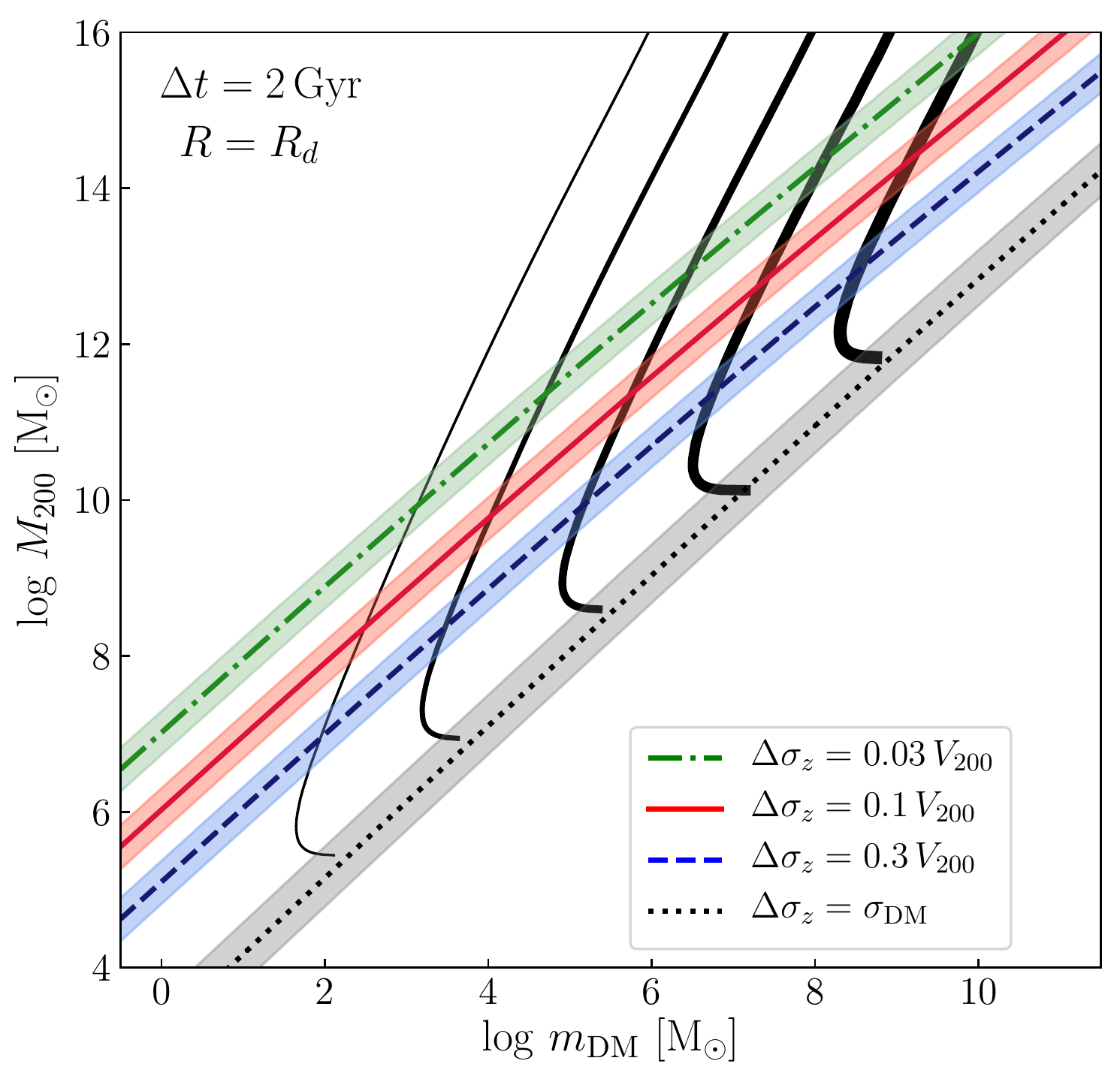}
\end{minipage}
\caption{The mass of haloes, $M_{200}$, below which simulated disk galaxies are expected to be adversely affected by
  collisional heating. Results are plotted as a function of the dark matter particle mass, $m_{\rm DM}$. The dot-dashed
  green lines correspond to a spurious increase in the vertical velocity dispersion of disk stars by
  $\Delta\sigma_z=0.03\times V_{200}$ over  $\Delta t=10\,{\rm Gyr}$ (left) and  $2\, {\rm Gyr}$ (right),
  the solid red lines to  $\Delta\sigma_z=0.1\times V_{200}$, 
  the dashed blue line to $\Delta\sigma_z=0.3\times V_{200}$, and the dotted black line to the maximum possible value
  $\Delta\sigma_z=\sigma_{\rm DM}$ (note that we assume $\sigma_{z,0}=0$ for each curve). All results are calculated at the disk's
  characteristic radius, $R_{\rm d}$, with shaded regions indicating the broader radial range $R=(0.5-2)\,R_{\rm d}$. In order of increasing
  thickness, solid black lines correspond to the halo masses below which collisional heating increases the vertical
  velocity dispersion of disk stars by at least $\Delta\sigma_z=1, \,\, 3,\,\,10,\,\,30$ and $100\,{\rm km\,s^{-1}}$ (in these cases, results are
  plotted only at $R=R_{\rm d}$ for clarity). We assume $R_{\rm d}=0.2\times r_{-2}$, where $r_{-2}=r_{200}/c$ is the characteristic
  radius of the NFW dark matter halo; the latter is prescribed using the analytic model of \citet{Ludlow2016}. Note that
  each legend applies to curves in both panels, but are plotted in separate panels to avoid cluttering the figure.} 
\label{fig9}
\end{figure*}
%__________________________________________________________________

The best-fit parameters for
eq.~\ref{eq:empheat} were determined in Section~\ref{sSecModel} and are provided in Table~\ref{TabBestFit}, for reference.
Only one parameter remains to be specified: $\sigma_{i,0}$, the {\em initial} components of the stellar velocity dispersion.
To gauge the potential importance of spurious collisional heating we can
set $\sigma_{i,0}$ to zero and compare the predicted values of $\sigma_i(R,t)$ to values of interest (note that stars in a real galaxy are
born with a small but non-zero velocity dispersion, $\sigma_{i,0}\lesssim 10\,{\rm km/s}$, which justifies this assumption for most
galaxies of interest). The solid black line in the
upper-left panel of Figure~\ref{fig_model}, for example, shows the vertical velocity dispersion (normalized by $V_{200}$) due to
collisional heating at the characteristic disk scale radius, $R_{\rm d}$, after $\Delta t=13.6\,{\rm Gyr}$ (results are similar
for $\sigma_R$). The results
are plotted as a function of the number of DM particles $N_{200}$ within the virial radius $r_{200}$. We assume a Hernquist profile for
the DM halo with an NFW-equivalent concentration of $c=10$, but show the impact of varying $c$ between 5 and 15 using a shaded region.
For the disk, we assume $R_{\rm d}=0.2\times r_{-2}$ where $r_{-2}=r_{200}/c$ is the scale radius of the corresponding NFW halo (this approximation
provides a reasonable description of the scale radii of may observed galaxies; see, e.g. \citealt{Navarro2017}). Colored lines
show results for the same set-up, but for various non-zero values of the initial stellar velocity dispersion, $\sigma_{z,0}$. 
For $c=10$, the vertical velocity dispersion at $R_{\rm d}$ reaches 90 per cent of its maximum value of
$\Delta\sigma_{z}\approx \sigma_{\rm DM}\approx 0.52\,V_{200}$
after $\Delta t=13.6\,{\rm Gyr}$ for $N_{200}\lesssim 3.26\times 10^4$, although the precise value depends on concentration. Note that this
is a limiting case, and collisional heating can be substantial for even larger $N_{200}$. For example,
$\Delta\sigma_z(R_{\rm d}) \approx 0.1\times V_{200}$ for $N_{200}\approx 2.96\times 10^6$ (assuming $c=10$). 

The upper-middle panel of Figure~\ref{fig_model} shows the vertical velocity dispersion of disk stars (normalized by $V_{200}$)
after $\Delta t=13.6\,{\rm Gyr}$ at $0.25\,R_{\rm d}$ (dashed blue line), $R_{\rm d}$ (black line) and $4\,R_{\rm d}$ (dot-dashed red
line) within DM haloes of varying $N_{200}$. As in the left-hand panel,
we assume $R_{\rm d}=0.2\times r_{-2}$ and $c=10$ (the grey shaded region again indicates the effect of varying $c$ from 5 to 15, but is
shown only for the $R=R_{\rm d}$ curve). Note that $\sigma_z$ reaches different maximum values at different radii and for different values of $N_{200}$.
This is because the maximum stellar velocity dispersion is determined by the local velocity dispersion of DM, which has a weak radial gradient.
Note too that, for a broad range of $N_{200}$, collisional heating is less problematic at large $R$ where the density of DM particles is lower.
This suggests that collisional heating may alter the morphologies of simulated galaxies, a possibility we will explore in future work.

The upper right-hand panel of Figure~\ref{fig_model} compares the vertical velocity dispersion at $R_{\rm d}$ after $\Delta t=2\,{\rm Gyr}$ (red dot-dashed line),
$6\,{\rm Gyr}$ (dashed blue line) and $13.6\,{\rm Gyr}$ (black line). As above, we assume a Hernquist DM halo with NFW-equivalent concentration
$c=10$ (the shaded region has the same meaning as in the other panels), and plot results as a function of $N_{200}$. Fewer particles
are required to suppress collisional heating over shorter timescales, as expected since it is an integrated effect. For example,
the vertical velocity dispersion (measured at $R_{\rm d}$) of an initially thin stellar disk will not exceed $\approx 10$ per cent of $V_{200}$
after $2\,{\rm Gyr}$ provided $N_{200}\gtrsim 4.29\times 10^5$, whereas the same degree of heating will be reached for $N_{200}\lesssim 1.28\times 10^6$
or $\lesssim 2.96\times 10^6$ for integration times of $6$ and $13.6\,{\rm Gyr}$, respectively.

The evolution we find for the radial velocity dispersion (not shown) is similar to that for the vertical velocity dispersion.

\subsubsection{Vertical scale height}
\label{ssSecImplezhalf}

As discussed in Section~\ref{sSecScaleHeight}, the evolution of $\sigma_z$ determines the evolution of the vertical scale height of simulated
disks. In our analysis, we characterized vertical scale heights by means of the vertical half-mass height, $z_{1/2}$. A useful
approximation for $z_{1/2}$ (at least over the range of radii and galaxy/halo properties studied in this paper) is given by
eq.~\ref{eq:zthick}, which provides a means of modelling $z_{1/2}$ in terms of $\sigma_z$, $R$ and the structural
properties of the disk and halo.

In the lower panels of Figure~\ref{fig_model} we plot $z_{1/2}^{\rm thick}$ (normalized by $r_{200}$) for the same set of models
displayed in the upper panels for $\sigma_z$. Note that, even in the most extreme cases of collisional heating considered here, the
half-mass of disks do not exceed a few per cent of their host's virial radius, $r_{200}$. Indeed, the maximum half-mass
heights are of order the half-mass {\em radii} of the disks. This is because scattering transfers energy
from DM to stellar particles, but (because the disk rotates more rapidly than the halo) transfers angular
momentum from the stellar to DM particles. This increases random motions in the disk while preventing it from expanding
to fill the halo's virial volume. Thus, for a given interval $\Delta t$,
$z_{1/2}$ increases considerably more than $R_{1/2}$ does, a result already apparent in Figure~\ref{fig1} (lower left-hand panel).

There are a few additional results in Figure~\ref{fig_model} worth highlighting. One: the
{\em minimum} vertical scale height of simulated disks are determined by the initial vertical velocity dispersion of their stars
(lower left-hand panel). Two: our model predicts a outward flaring of stellar disks as a result of spurious heating; i.e., if
allowed to reach its maximum thickness, the scale height will increase with $R$, at least over the radial range
$R_{\rm d}/4\leq R\leq 4\times R_{\rm d}$ (lower middle panel; see also the lower-left panel of Figure~\ref{fig1}).
And three: vertical scale heights grow monotonically with time, before plateauing at a maximum physical scale height that is determined
by the local velocity dispersion of halo particles and the half-mass radius of the disk (lower right panel). 

%_________________________________________
\begin{figure*}
  \includegraphics[width=0.9\textwidth]{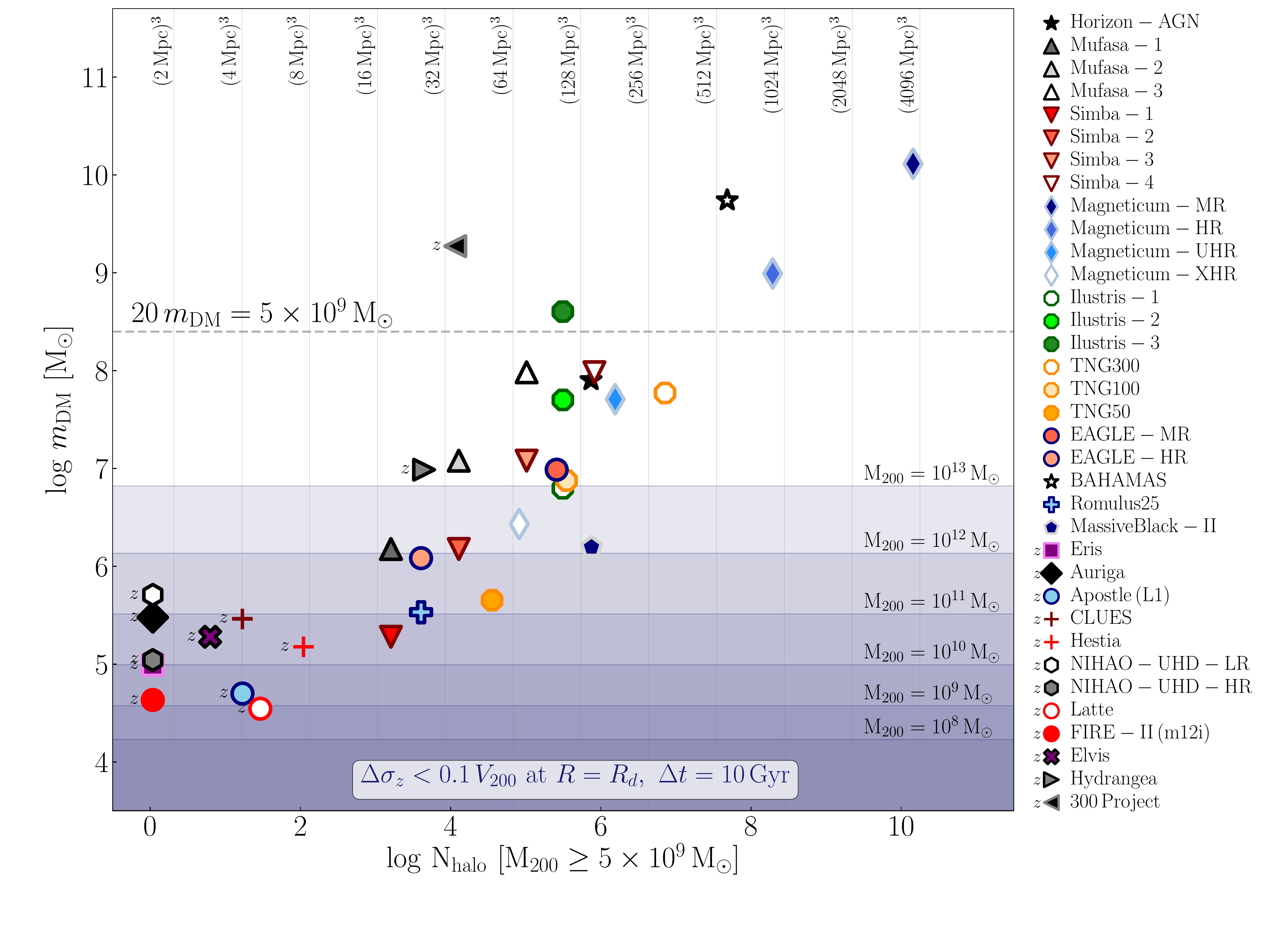}
  \caption{The DM particle mass plotted versus the expected number of DM haloes with virial mass $M_{200}>5\times 10^9\,{\rm M_\odot}$ for
    a number of recent cosmological and zoom hydrodynamical simulations. Halo abundances are estimated using the \citet{Tinker2008}
    mass function; the value $M_{200}=5\times 10^9\,{\rm M_\odot}$ roughly corresponds to the halo mass above which galaxy occupation reaches
    100 per cent \citep{Ale2020}. Vertical grey lines mark fixed volumes, increasing by factors of 8 from $(2\,{\rm Mpc})^3$ to
    $(4096\,{\rm Mpc})^3$; the horizontal dashed line indicates the minimum DM particle mass required to resolve
    $M_{200}=5\times 10^9$ haloes with at least 20 particles. The blue shaded regions highlight the DM particle masses below which
    spurious collisional heating (at $R=R_{\rm d}$, integrated over $\Delta t=10\,{\rm Gyr}$) remains below  $\Delta\sigma_z\lesssim 0.1\,V_{200}$;
    darker shades correspond to haloes of decreasing virial mass, ${\rm M_{200}}$, as labelled. As in previous figures, we assume
    $R_{\rm d}=0.2\times r_{-2}$, where $r_{-2}=r_{200}/c$ is the
    NFW-equivalent scale radius of the \citet{Hernquist1990} DM halo used in our model. The majority of
    state-of-the-art zoom simulations, and a few of the highest-resolution, smallest-volume cosmological runs, are in principle
    able to resolve cool stellar disks (e.g. $\Delta\sigma_z\lesssim 0.1\,{\rm V_{200}}$) in Milky Way-mass haloes, but most
    large volume cosmological runs cannot. The various simulations plotted here are labelled on the right-hand side of the
    figure (see Table~\ref{TabCosmoImp} for additional details). Symbols indicating cosmological zoom simulations have been prefixed with
    a ``$z$'' to distinguish them from uniform-resolution simulations of cosmological volumes.}
  \label{fig10}
\end{figure*}
%_________________________________________

\subsection{Implications for existing and future cosmological simulations}
\label{sSecImpli}

Our model can now be used to assess the importance of collisional disk heating for cosmological simulations
of galaxy formation. In Figure~\ref{fig9} we plot, as a function of the DM particle mass, the halo masses below which disk
heating is expected to become important. We assume a \citet{Hernquist1990} profile for the DM halo with mass-dependent NFW-equivalent
concentrations taken from the analytic model of \citet{Ludlow2016}. The various lines correspond to different amounts
of collisional heating affecting an initially perfectly-thin (i.e. $\sigma_{z,0}=0$) stellar disk, and are plotted at $R=R_{\rm d}$
(we assume $R_{\rm d}=0.2\times r_{-2}$; for several curves we show the impact of varying $R$ from $0.5$ to $2\times R_{\rm d}$ using shaded
regions). In both panels, the dot-dashed green lines correspond to $\Delta\sigma_z=0.03\times V_{200}$, the solid red lines to
$\Delta\sigma_z=0.1\times V_{200}$, and the dashed blue lines to $\Delta\sigma_z=0.3\times V_{200}$;
dotted black lines indicate a maximally-heated disk, $\Delta\sigma_z=\sigma_{\rm DM}$. Solid black lines in both panels
(in order of increasing thickness) correspond to fixed physical degrees of heating: $\Delta\sigma_z=1\,{\rm km \,s^{-1}}$, $3\,{\rm km \,s^{-1}}$, 
$10\,{\rm km \,s^{-1}}$, $30\,{\rm km \,s^{-1}}$, and $100\,{\rm km \,s^{-1}}$ respectively. The left- and right-hand panels of Figure~\ref{fig9}
correspond to integration times of $\Delta t=10\,{\rm Gyr}$ and $2\,{\rm Gyr}$ respectively. 

For the dark matter particle masses adopted for many recent cosmological simulations, $m_{\rm DM}\approx 10^4\,{\rm M_\odot}$
to $\approx 10^8\,{\rm M_\odot}$, collisional heating is expected to be important over a broad range of halo masses.
More specifically, the locations at which a vertical line in Figure~\ref{fig9} intersects the curves corresponding 
to a fixed (fractional or absolute) amount of heating mark
the halo masses below which we expect collisional heating to be important for that particle mass. Clearly the properties of galaxies
in many simulations may be unduly affected by collisional heating, including their structure, morphology and
kinematics.

For example, all recent large-volume cosmological hydrodynamical simulations (i.e. $V\gtrsim 100^3\,{\rm Mpc}^3$) used DM particle
masses $m_{\rm DM}\gtrsim 10^6\,{\rm M_\odot}$ (this includes the \eagle simulations, Ilustris, TNG300, TNG100, TNG50, Horizon-AGN,
New Horizon and Magneticum, but also Mufasa and Simba; see Table~\ref{TabCosmoImp} for simulation details and 
references). We therefore expect collisional
heating in these runs, integrated over $\approx 10\, {\rm Gyr}$, to increase the vertical velocity dispersion of disk stars at $R=R_{\rm d}$ to
$\gtrsim 0.1\,V_{200}$ ($0.03\,V_{200}$) for halo masses $M_{200}\lesssim 1.6\times 10^{12}\,{\rm M_\odot}$
($1.4\times 10^{13}\,{\rm M_\odot}$), corresponding to haloes resolved with fewer than $N_{200}=1.6\times 10^6$ ($1.4\times 10^7$)
DM particles. These mass scales are reduced somewhat for shorter integration times, but remain problematic. For example, for
$\Delta t=2\,{\rm Gyr}$ we find $\Delta\sigma_z \gtrsim 0.1\,V_{200}$ ($0.03\,V_{200}$) for $M_{200}\lesssim 3.8\times 10^{11}\,{\rm M_\odot}$
($3.3\times 10^{12}\,{\rm M_\odot}$), corresponding to $N_{200}=3.8\times 10^5$ ($3.3\times 10^6$). We therefore expect young,
massive galaxies (corresponding to shorter $\Delta t$) to experience less heating than older ones, but because their haloes grow
in part through the merger and accretion of low-mass, poorly-resolved progenitors, the effects of spurious heating may still be
important.

Results become substantially worse as the DM particle mass increases, and a few of the runs mentioned above are therefore
{\em more} vulnerable to collisional heating than just mentioned. For $m_{\rm DM}=10^7\,{\rm M_\odot}$ -- which is comparable to, or even better than
the mass resolution of TNG300, TNG100, Horizon-AGN, Magneticum, and the intermediate-resolution \eagle simulation -- collisional
heating is likely to strongly affect the structure and kinematics of simulated disks (i.e. $\Delta\sigma_z\gtrsim 0.1\,V_{200}$)
in haloes with masses less than about 
$M_{200}\lesssim 1.2\times 10^{13} {\rm M_\odot}$ (or equivalently, $N_{200}\lesssim 1.2\times 10^6$).
Thus, virtually all star-forming main sequence galaxies will be spuriously heated in these runs. 
More specifically, galaxies hosted by haloes with $M_{200}\approx 10^{13}\,{\rm M_\odot}$ ($\approx 10^{12}\,{\rm M_\odot}$)
will experience a spurious increase in their vertical velocity dispersion at $R\approx R_{\rm d}$ of
$\approx 40\,{\rm km\,s^{-1}}$ ($\approx 62\,{\rm km\,s^{-1}}$) over roughly $\Delta t=10\,{\rm Gyr}$;
their radial velocity dispersion will increase by $\Delta\sigma_R\gtrsim 58\,{\rm km\, s^{-1}}$ ($\gtrsim 84\,{\rm km\, s^{-1}}$)
and their half-mass heights by $\Delta z_{1/2}\gtrsim 750\,{\rm pc}$ ($\gtrsim 1.5\,{\rm kpc}$) over the same timescale.
Collisional heating is also problematic over much shorter timescales: For example, for the same halo masses $\Delta\sigma_z\gtrsim 18\,{\rm km\,s^{-1}}$
($\gtrsim 30\,{\rm km\,s^{-1}}$), $\Delta\sigma_R\gtrsim 26\,{\rm km\,s^{-1}}$ ($\gtrsim 43\,{\rm km\,s^{-1}}$) and 
$\Delta z_{1/2}\gtrsim 196\,{\rm pc}$ ($\gtrsim 515\,{\rm pc}$) for $\Delta t= 2\,{\rm Gyr}$.

Increasing the mass resolution of DM particles in cosmological simulations is the best way to suppress the spurious collisional
heating of galaxy disks. For DM particle masses $m_{\rm DM}\approx 10^5\,{\rm M_\odot}$ (which is comparable to the
highest uniform-resolution cosmological simulation published to date, i.e. the $V=(12.5\, h^{-1}{\rm Mpc})^3$ Simba run), spurious disk
heating over a timescale of $\Delta t=10\,{\rm Gyr}$ is below the level of $\Delta\sigma_z\approx 0.03\times V_{200}$ ($\approx 6\,{\rm km\,s^{-1}}$) for
$M_{200}\gtrsim 1.8\times 10^{12}\,{\rm M_\odot}$ ($N_{200}\approx 1.8\times 10^7$) and below $\approx 0.1\times V_{200}$
($\approx 10\,{\rm km\,s^{-1}}$) for  $M_{200}\gtrsim 2.1\times 10^{11}\,{\rm M_\odot}$ ($N_{200}\gtrsim 2.1\times 10^6$). For a
DM particle mass of $m_{\rm DM}\approx 5\times 10^5$ (comparable to the value adopted for the TNG50 simulation), we expect,
over $\Delta t \approx 10\,{\rm Gyr}$, $\Delta\sigma_z\approx 0.03\times V_{200}$ ($\approx 10\,{\rm km\,s^{-1}}$) for
$M_{200}\approx 7.4\times 10^{12}\,{\rm M_\odot}$ ($N_{200}\approx 1.5\times 10^7$) and $\Delta\sigma_z\approx 0.1\times V_{200}$
($\approx 16\,{\rm km\,s^{-1}}$) for  $M_{200}\approx 8.7\times 10^{11}\,{\rm M_\odot}$ ($N_{200}\approx 1.7\times 10^6$). This may
explain why \citet{Pillepich2019} report poor convergence in the vertical scale heights of disks in TNG50 with increasing mass resolution.

The most recent generation of zoom simulations, which sacrifice galaxy statistics in favour of resolution, typically adopt
DM particle masses in the range $10^4 \lesssim m_{\rm DM}/{\rm M_\odot}\lesssim {\rm a\, few}\times 10^5$ (see Table~\ref{TabCosmoImp}
for several examples). These runs often target Milky Way-mass galaxies ($M\approx 10^{12}\,{\rm M_\odot}$) and, although it remains
challenging to suppress collisional heating below $\Delta\sigma_z\lesssim 0.03\,V_{200}$ ($\approx 4.9\,{\rm km\,s^{-1}}$) over 10 Gyr
in these halos (which would require $m_{\rm DM}\lesssim 5.2\times 10^4\,{\rm M_\odot}$), they typically succeed in suppressing it below
$\Delta\sigma_z\lesssim 0.1\,V_{200}$ ($\approx 16\,{\rm km\,s^{-1}}$), which requires $m_{\rm DM}\lesssim 5.9\times 10^5\,{\rm M_\odot}$.

These are, of course, lower limits on disk heating in simulations, as galaxies may not evolve in isolation at all times. But they
will experience spurious DM heating at all times, even if they also occasionally merge with other galaxies, or experience disk
heating as a result of other physical processes, such as orbital scattering by giant molecular clouds or instabilities, or accretion-driven
heating.

Our results can therefore be used to determine the range of DM particle masses for which spurious collisional heating 
will be unimportant in DM haloes of a given virial mass, and thus can be used as a guide for the selection of numerical
parameters for simulations of galaxy formation and evolution. As discussed above, suppressing spurious heating in
simulations will always favor higher (DM) mass resolution.
The optimal particle mass for a particular simulation will therefore represent a compromise between the simulation's
resolution and volume (or equivalently the number of resolved objects). Figure~\ref{fig10} helps visualize the situation. Here
we plot, for a number of recent state-of-the-art cosmological and zoom simulations, the DM particle mass versus the expected number
of DM haloes with virial masses above ${\rm M_{200}}\geq 5\times 10^9\,{\rm M_\odot}$ (which roughly corresponds to the halo mass
above which galaxy occupation reaches $\approx 100$ per cent; \citealt{Ale2020}). Vertical grey lines mark fixed volume increments,
and the horizontal dashed line corresponds to particle masses for which $5\times 10^9\,{\rm M_\odot}$ haloes are resolved with 
20 particles. 

The blue shaded regions in Figure~\ref{fig10} indicate the DM particle masses below which collisional heating (integrated over
$\Delta t=10\,{\rm Gyr}$) does not exceed $0.1\times V_{200}$ in haloes of different mass. Darker shades correspond to decreasing halo masses, with values ranging
from ${\rm M_{200}}=10^{13}\,{\rm M_\odot}$ to $10^{8}\,{\rm M_\odot}$ in steps of 1 dex, as indicated. Note that the majority of zoom
simulations targeting Milky Way-mass galaxies or local group analogues (e.g. Auriga, Eris, Apostle, FIRE, Latte, Elvis, CLUES, Hestia)
are unlikely to be strongly impacted by spurious collisional heating, although both the progenitors of the central galaxies that these
simulations target and their present-day satellites will be impacted. However,
most simulations of large volumes, e.g. $\gtrsim (100^3\,{\rm Mpc}^3)$, which typically adopt DM particle masses 
$m_{\rm DM}\gtrsim 10^6\,{\rm M_\odot}$, are likely to fall victim to the numerical heating effects discussed in this paper.

Finally, we stress that the magnitude of the heating one may wish to suppress will depend on the subgrid physics adopted for the run,
and on the scientific goals of the simulation programme.
Simulations that properly resolve the cold phase of the inter-stellar medium, for example, are expected to result in some
galaxies having cold, molecular gas disks with $\sigma_z\ll 10\,{\rm km\,s^{-1}}$, giving rise to stellar disks that, at least
initially, have very low vertical velocity dispersions and small scale heights. In these cases, DM particle masses should be
chosen so that $\Delta\sigma_z\approx$ a few ${\rm km\,s^{-1}}$ in order to ensure that spurious collisional heating does not
alter the kinematics of stellar particles. Simulations that oppose artificial fragmentation by preventing gas cooling
below $\approx 10^4\,{\rm K}$ have a natural kinematic floor of $\approx 10\,{\rm km\,s^{-1}}$, which provides a sensible 
target above which collisional heating should be suppressed. The solid black
curves in Figure~\ref{fig9} can be used to guide the selection of particle masses that adhere to these restrictions. For
example, the vertical heating of disk stars will not exceed $\Delta\sigma_z\approx 3\,{\rm km\,s^{-1}}$ over 10 Gyr in a
simulated dwarf galaxy of halo mass $M_{200}\approx 10^{11}\,{\rm M_\odot}$ provided $m_{\rm DM}\lesssim 7.1\times 10^3\,{\rm M_\odot}$
(corresponding to $N_{200}\gtrsim 1.4\times 10^7$); the
heating will not exceed $\Delta\sigma_z\approx 10\,{\rm km\,s^{-1}}$ over 10 Gyr provided $m_{\rm DM}\lesssim 8.0\times 10^4\,{\rm M_\odot}$
(or $N_{200}\gtrsim 1.3\times 10^6$).

Some large-volume simulations are of course not intended for studies of galaxy structure, but rather for studies of the structure of the inter- or
circum-galactic medium, the role of environmental effects in galaxy evolution, the properties of galaxy groups or clusters, or the imprint of
baryons on cosmology and large-scale structure. It seems unlikely that spurious collisional disk heating is detrimental to simulations aimed at
addressing these topics, but the effect should nonetheless be recognized as one of their important limitations.

In Table~\ref{TabCosmoImp}, we tabulate the halo masses below which collisional heating (specifically, $\Delta\sigma_z/V_{200}=0.1$
over $\Delta t=10\,{\rm Gyr}$) may become important for the cosmological hydrodynamical simulations plotted in Figure~\ref{fig10}. 

\section{Summary and conclusions}
\label{SecSummary}

Spurious collisional heating of simulated stellar disks by coarse-grained DM particles affects the vertical and radial
kinematics of disk stars, and as a consequence, the structure of their stellar distribution. The effects are irreversible,
and depend primarily on the number of DM particles and the structural properties of the disk's DM halo. A
rigorous theoretical description of disk heating was outlined by \citet{LO1985}, who calculated (among other
quantities) the rate at which the vertical, radial and azimuthal components of the stellar velocity dispersion of disk stars increases with
time as a result of scattering off point-mass perturbers (black holes in their case, which were at the time viable
candidates for DM). Their analytic model provides an accurate description of the early stages of our simulations results (see
Figure~\ref{fig2}), but cannot account for the strong radial gradients in the dark matter density across stellar
disks that are embedded within realistic DM haloes, nor the asymptotic velocity dispersion of stellar particles
set by either energy equipartition or the local escape velocity of the halo (eq.~\ref{eq:asympv}). Here we provide a
simple empirical amendment to their equations that accurately describes the results of our simulations (see
eq.~\ref{eq:empheat}).

Collisional heating is largely independent of stellar particle mass, but may be exacerbated in simulations
that adopt stellar-to-DM particle mass ratios that differ substantially from unity (see Appendix~\ref{sec:A2} and
\citealt{Ludlow2019}). As a result, attempts to artificially increase the baryon-to-DM mass resolution (i.e. to
decrease the baryon-to-DM particle mass ratio) in simulations will only benefit runs that reach sufficiently high
DM-mass resolution to ensure that collisional heating due to DM particles can be safely ignored. If it cannot,
the effect can only be suppressed by increasing the resolution of the DM component relative to that of the stars.

Our results have implications for the reliability with which simulations can be used to interpret the vertical, radial and kinematic
gradients in the ages and metallicities of stars. For example, in the solar
neighbourhood the velocity dispersion of stars increases with age and with decreasing metallicity
\citep[e.g.][]{Wielen1977,Lacey1984,Quillen2001}; at fixed cylindrical radius $R$, the scale height of stars increases with age
\citep[e.g.][]{Casagrande2016}. It is plausible that such trends arise due to the cumulative effects
of scattering (due to, for example, molecular clouds or globular clusters), or dynamical heating due to accretion
events or disk instabilities. Disks in cosmological simulations, if not sufficiently well-resolved, may however exhibit
similar age- or metallicity-dependent gradients as a result of {\em spurious} collisional heating (although see \citealt{Navarro2018}
and \citealt{Grand2016} for examples of simulations in which they are not spurious). Studies of metallicity and age gradients based on simulated disk galaxies should suppress these numerical effects.

%___________________________________________________________________________________________
\begin{center}
  \begin{table*}
    \caption{For a number of recent state-of-the-art hydrodynamical simulations, we here provide a summary of the halo virial masses,
      ${\rm M_{200}}$, below which we expect
      collisional vertical disk heating to be of order $\Delta\sigma_z=0.1\times V_{200}$ over a timescale of $\Delta t=10^{10}\,{\rm yr}$
      (see Figure~\ref{fig9}). The first three columns provide the simulation name, the journal reference and the type of simulation
      (i.e. whether a uniform mass-resolution cosmological simulation or zoom). Next we list the simulation's DM particle mass resolution
      ($m_{\rm DM}$), and the virial mass and corresponding DM particle number ($M_{200}$ and $N_{200}$, respectively) below which spurious
      heating of an initially-cold stellar disk results in $\Delta\sigma_z/V_{200}\geq 0.1$; the corresponding magnitude of {\em physical} heating
      (i.e. in units of ${\rm km\,s^{-1}}$) is also provided, as are the corresponding changes in vertical disk scale height,
      $\Delta z_{1/2}$, calculated assuming $f_\star=0.01$ (both in physical units and as a fraction of $r_{200}$). For zoom simulations
      that vary the mass resolution (either for an individual galaxy, or a sample of galaxies), we quote values for the highest-resolution
      run published.}
    \begin{tabular}{l l l r r r r r r r r r}\hline \hline
      & ${\rm Simulation}$ & Reference & Type&$m_{\rm DM}$  & ${\rm M_{200}}$  & ${\rm N_{200}}$ & $\Delta\sigma_z$ &  $\Delta\sigma_z$    & $\Delta z_{1/2}$ & $\Delta z_{1/2}$ &\\
      &                    &      &     &$[M_\odot]$   &  $[M_\odot]$     &  $[10^6]$       & $[V_{200}]$      & $[{\rm km\,s^{-1}}]$ & $[r_{200}]$      & $[{\rm kpc}]$    &\\\hline

      &   Horizon-AGN      & \citet{Dubois2014} & Cosmo. & $8\times 10^7$    & $7.5\times 10^{13}$ & 0.9   & 0.10 & 69.0  & $2.6\times 10^{-3}$ & 1.76    &            \\
      &   New Horizon      & \citet{Dubois2020} & Cosmo. & $1.2\times 10^6$  & $1.9\times 10^{12}$ & 1.6   & 0.10 & 20.2  & $1.4\times 10^{-3}$ & 0.27    &            \\
      &   Mufasa-1         & \citet{Dave2016}   & Cosmo. & $9.6\times 10^7$  & $8.8\times 10^{13}$ & 0.9   & 0.10 & 72.6  & $2.8\times 10^{-3}$ & 1.99    &            \\
      &   Mufasa-2         &  \dittostraight    & Cosmo. & $1.2\times 10^7$  & $1.4\times 10^{13}$ & 1.2   & 0.10 & 39.7  & $1.9\times 10^{-3}$ & 0.75    &            \\
      &   Mufasa-3         &  \dittostraight    & Cosmo. & $1.5\times 10^6$  & $2.3\times 10^{12}$ & 1.5   & 0.10 & 21.6  & $1.3\times 10^{-3}$ & 0.30    &            \\
      &   Simba-1          & \citet{Dave2019}   & Cosmo. & $9.6\times 10^7$  & $8.8\times 10^{13}$ & 0.9   & 0.10 & 72.6  & $2.8\times 10^{-3}$ & 1.99    &            \\
      &   Simba-2          &   \dittostraight   & Cosmo. & $1.2\times 10^7$  & $1.4\times 10^{13}$ & 1.2   & 0.10 & 39.7  & $1.9\times 10^{-3}$ & 0.75    &            \\
      &   Simba-3          &   \dittostraight   & Cosmo. & $1.5\times 10^6$  & $2.3\times 10^{12}$ & 1.5   & 0.10 & 21.6  & $1.4\times 10^{-3}$ & 0.30    &            \\
      &   Simba-4          &   \dittostraight   & Cosmo. & $1.9\times 10^5$  & $3.6\times 10^{11}$ & 1.9   & 0.10 & 11.7  & $1.1\times 10^{-2}$ & 0.12    &            \\
      &   Magneticum       & \citet{Dolag2016}  & Cosmo.&$1.3\times 10^{10}$ & $6.1\times 10^{15}$ & 0.5   & 0.10 & 298.1 & $5.9\times 10^{-3}$ & 17.5    &            \\
      &   Ilustris-1 & \citet{Vogelsberger2014} & Cosmo. & $6.3\times 10^6$  & $8.1\times 10^{12}$ & 1.3   & 0.10 & 32.8  & $1.7\times 10^{-3}$ & 0.56    &            \\
      &   Ilustris-2       &   \dittostraight   & Cosmo. & $5.0\times 10^7$  & $5.0\times 10^{13}$ & 1.0   & 0.10 & 60.1  & $2.4\times 10^{-3}$ & 1.42    &            \\
      &   Ilustris-3       &   \dittostraight   & Cosmo. & $4.0\times 10^8$  & $3.0\times 10^{14}$ & 0.8   & 0.10 & 109.4 & $3.4\times 10^{-3}$ & 3.72    &            \\
      &   TNG300         & \citet{Springel2018} & Cosmo. & $5.9\times 10^7$  & $5.8\times 10^{13}$ & 1.0   & 0.10 & 62.9  & $2.5\times 10^{-3}$ & 1.58    &            \\
      &   TNG100           &  \dittostraight    & Cosmo. & $7.5\times 10^6$  & $9.5\times 10^{12}$ & 1.3   & 0.10 & 34.5  & $1.8\times 10^{-3}$ & 0.63    &            \\
      &   TNG50         & \citet{Pillepich2019} & Cosmo. & $4.5\times 10^5$  & $7.9\times 10^{11}$ & 1.8   & 0.10 & 15.0  & $1.2\times 10^{-3}$ & 0.18    &            \\
      & \textsc{eagle-MR}  & \citet{Schaye2015} & Cosmo. & $9.7\times 10^6$  & $1.2\times 10^{13}$ & 1.2   & 0.10 & 37.3  & $1.9\times 10^{-3}$ & 0.69    &            \\
      & \textsc{eagle-HR}  &  \dittostraight    & Cosmo. & $1.2\times 10^6$  & $1.9\times 10^{12}$ & 1.6   & 0.10 & 20.3  & $1.4\times 10^{-3}$ & 0.27    &            \\
      &  BAHAMAS         & \citet{McCarthy2017} & Cosmo. &  $5.5\times 10^9$ & $2.9\times 10^{15}$ & 0.5   & 0.10 & 231.2 & $5.1\times 10^{-3}$ & 11.82   &            \\
      &  Massive Black-II & \citet{Khandai2015} & Cosmo. &  $1.6\times 10^6$ & $2.4\times 10^{12}$ & 1.5   & 0.10 & 21.8  & $1.4\times 10^{-3}$ & 0.30    &            \\
      &  Romulus25        & \citet{Tremmel2017} & Cosmo. &  $3.4\times 10^5$ & $6.1\times 10^{11}$ & 1.8   & 0.10 & 13.8  & $1.1\times 10^{-3}$ & 0.15    &            \\

      &   Eris             & \citet{Eris2011}   & zoom & $9.8\times 10^4$  &  $2.0\times 10^{11}$  & 2.0  & 0.10 & 9.6   & $9.7\times 10^{-4}$ & 0.09  &            \\
      &   Auriga           & \citet{Grand2017}  & zoom & $3.0\times 10^5$  &  $5.5\times 10^{11}$  & 1.8  & 0.10 & 13.4  & $1.1\times 10^{-3}$ & 0.15  &            \\
      &   Apostle (L1)     & \citet{Sawala2016} & zoom & $5.0\times 10^4$  &  $1.1\times 10^{11}$  & 2.2  & 0.10 & 7.8   & $8.8\times 10^{-4}$ & 0.07  &            \\
      &   CLUES         & \citet{Libeskind2010} & zoom & $2.9\times 10^5$  &  $5.3\times 10^{11}$  & 1.8  & 0.10 & 13.2  & $1.1\times 10^{-3}$ & 0.15  &            \\
      &   HESTIA           & \citet{HESTIA2020} & zoom & $1.5\times 10^5$  &  $2.9\times 10^{11}$  & 2.0  & 0.10 & 10.9  & $9.9\times 10^{-4}$ & 0.11  &            \\
      &   NIHAO-UHD        &  \citet{Buck2020}  & zoom & $5.1\times 10^5$  &  $8.8\times 10^{11}$  & 1.7  & 0.10 & 15.7  & $1.2\times 10^{-3}$ & 0.19  &            \\
      &   Latte            & \citet{Wetzel2016} & zoom & $3.5\times 10^4$  &  $7.9\times 10^{10}$  & 2.3  & 0.10 & 7.0   & $8.6\times 10^{-4}$ & 0.06  &            \\
      &   FIRE-II (m12i)  & \citet{Hopkins2018} & zoom & $4.3\times 10^4$  &  $9.6\times 10^{10}$  & 2.2  & 0.10 & 7.4   & $8.7\times 10^{-4}$ & 0.07  &            \\
      &   Elvis            &  \citet{Elvis2014} & zoom & $1.9\times 10^5$  &  $3.6\times 10^{11}$  & 1.9  & 0.10 & 11.7  & $1.1\times 10^{-3}$ & 0.13  &            \\
      &   Hydrangea        & \citet{Bahe2017}   & zoom & $9.7\times 10^6$  &  $1.2\times 10^{13}$  & 1.2  & 0.10 & 37.3  & $1.9\times 10^{-3}$ & 0.69  &            \\
      &   The 300 Project  & \citet{Cui2018}    & zoom & $1.9\times 10^9$  &  $1.1\times 10^{15}$  & 0.6  & 0.10 & 169.5 & $4.5\times 10^{-3}$ & 7.58  &            \\
   \end{tabular}
    \label{TabCosmoImp}
  \end{table*}
\end{center}
%___________________________________________________________________________________________

In simulations, collisions with DM particles randomly perturb stellar particle orbits, which may disturb ordered flows and disrupt the
coherent motions reminiscent of rotationally-supported disks or tidal streams. The spatial distribution and kinematics of stars in simulated
galaxies serve as proxies for their morphology, and are often used to decompose them into distinct kinematic
components. Collisional heating is likely to have important consequences for the structure and morphology of simulated galaxies,
particularly in studies based on uniform-resolution cosmological simulations, in
which galaxies with different stellar masses are resolved with different numbers of dark matter particles.
We will address the implications of collisional heating for the morphological transformation of galaxies in a
follow-up paper.

Because galaxy disks are dynamically cold relative to their surrounding dark matter haloes, collisions between
dark matter and stellar particles will, over time, increase the velocity dispersion of stars in an attempt to
establish energy equipartition between the two components. This is true regardless of the DM-to-stellar particle
mass ratio, $\mu$ (recall that equipartition is satisfied when $\mu=[\sigma_\star/\sigma_{\rm DM}]^2$), although it 
will be accelerated when $\mu > 1$. Spurious collisional heating of galaxies will therefore artificially increase
their sizes, as the increased kinetic energy of stellar particles drives them onto higher energy orbits within the
halo potential \citep[see also][]{Ludlow2019}.

\vspace{0.25cm}
The highlights of our work can be summarized as follows:

\begin{enumerate}[i]
\item Spurious collisional heating due to DM halo particles affects the velocity dispersion and scale heights of disk
  stars at all galacto-centric radii (Figure~\ref{fig1}). The magnitude of the effect depends on the mass of DM particles (or on the
  number of particles at fixed halo mass), but also
  on their local density and velocity dispersion (Figure~\ref{fig2}; the former dominates due to its much stronger radial gradients). As a
  result, heating is often more substantial in the denser, colder centres of DM haloes than in their outskirts, an effect that may
  drive artificial morphological changes in disk galaxies. The effects of collisional heating on the (radial and vertical)
  velocity dispersion of disk stars, and on their scale heights (Figure~\ref{fig4}), can be described by
  our eqs.~\ref{eq:empheat} and \ref{eq:zthick}, respectively.

\item In haloes resolved with a given number of DM particles (or equivalently, for fixed $m_{\rm DM}$), the effects of collisional
  heating are approximately independent of the mass of stellar particles, at least
  over the range $1\leq \mu \leq 25$. This suggests that the spurious heating of stellar particle orbits 
  is primarily due to incoherent fluctuations in the DM particle distribution (which, for fixed $m_{\rm DM}$ and $\mu\gtrsim 1$,
  deflect stellar particle orbits by approximately the same angle) rather than the more gradual diffusion of energy between DM
  and stellar particles of unequal mass (see e.g. Figure~\ref{fig1} or \ref{fig4}, but also the discussion in Appendix~\ref{sec:A2} and
  \citealt{Ludlow2019}). Because the total
  mass of the DM halo greatly exceeds that of the stars, other galaxy components, such as bulges or stellar haloes, contribute little
  to the heating of disk stars (see Appendix~\ref{sec:A4}). Collisional heating is nearly independent of the gravitational
  softening length ($\epsilon$) provided the vertical scale height ($z_{\rm d}$) of a simulated galaxy is resolved with at least one
  softening length, i.e. $\epsilon\lesssim z_{\rm d}$ (Appendix~\ref{sec:A5}). 
  
\item Our empirical model can be used to help interpret results from current cosmological simulations (see~Figures~\ref{fig9} and \ref{fig10},
  Table~\ref{TabCosmoImp}, and the discussion in section~\ref{sSecImpli}), and to guide the selection of DM particle masses and gravitational
  softening lengths for future high-resolution
  simulations of galaxy formation. For example, to ensure that the vertical collisional heating of stars in the disk of a typical
  Milky Way-mass galaxy (i.e. $M\approx 10^{12}\,{\rm M_\odot}$) does not exceed $\Delta\sigma_z\approx 5 \, {\rm km\,s^{-1}}$
  (roughly the minimum scale set by a gas particle temperature floor of $10^4\,{\rm K}$) over $\Delta t=10\,{\rm Gyr}$, the DM
  particle mass must be $m_{\rm DM}\lesssim 5.5\times 10^4\,{\rm M_\odot}$ ($N_{200}\gtrsim 1.8\times 10^7$). Maintaining
  $\Delta\sigma_z\lesssim 0.1\,V_{200}\approx 16 \, {\rm km\,s^{-1}}$ (again for $\Delta t=10\,{\rm Gyr}$) requires
  $m_{\rm DM}\lesssim 5.9\times 10^5\,{\rm M_\odot}$ ($N_{200}\gtrsim 1.7\times 10^6$). For a typical dwarf galaxy,
  with $M_{200}\approx 10^{11}\,{\rm M_\odot}$,
  DM particle masses of $m_{\rm DM}\lesssim 2.0\times 10^4\,{\rm M_\odot}$ and $4.5\times 10^4\,{\rm M_\odot}$ (or equivalently,
  $N_{200}\gtrsim 5\times 10^6$ and $N_{200}\gtrsim 2.2\times 10^6$, respectively) ensure collisional heating does not
  exceed $\Delta\sigma_z=5\,{\rm km\,s^{-1}}$ and $0.1\,V_{200}$, respectively. The results are based on heating rates measured
  at the characteristic disk scale length, which we take to be $\approx 1.4\,{\rm kpc}$ and $3.8\,{\rm kpc}$ for the dwarf
  and Milky Way-mass galaxies, respectively (i.e. $0.2\times r_{-2}$ for typical values of the DM halo's concentrations). Softening lengths
  should be smaller than the vertical scale height of a thin, molecular disk,  i.e. $\epsilon\lesssim 100\,{\rm pc}$.
  
\end{enumerate}

In closing, we note a number of limitations of our study that could be improved upon in future work.
One is that by design our controlled simulations of equilibrium disks do not account for a number of important
physical processes that may also give rise to secular disk heating. Among them are accretion-driven heating,
or orbital scattering by instabilities, globular clusters, giant molecular clouds or spiral density waves; spurious heating of disks
by star particles in stellar haloes or bulges are also ignored (but see Appendix~\ref{sec:A4}). Nevertheless, collisional
heating is an integrated effect, and we expect heating rates estimated from eq.~\ref{eq:empheat} to yield sensible
{\em lower limits} to the spurious heating due to coarse-grained DM particles. Perhaps most importantly, our study
neglects the time-dependent growth of the DM potential (which introduces a time-dependent DM density and characteristic
velocity), as well as the fact that most simulated disk galaxies actively form stars, thus replenishing a thin disk with
a low velocity dispersion stellar component. These issues should be the focus of future work.

\section*{Acknowledgements}
We would like to thank our referee, Cedric Lacey, for a prompt and useful report on our paper.
ADL and DO acknowledge financial support from the Australian Research Council through their Future Fellowship
scheme (project numbers FT160100250, FT190100083, respectively). This work is partly funded by Vici grant 639.043.409
from the Dutch Research Council (NWO). This work has benefited from the following public {\textsc{python}} packages:
{\textsc{scipy}} \citep{scipy}, {\textsc{numpy}} \citep{numpy}, {\textsc{matplotlib}} \citep{matplotlib} and
{\textsc{ipython}} \citep{ipython}.

%%%%%%%%%%%%%%%%%%%%%%%%%%%%%%%%%%%%%%%%%%%%%%%%%%
\section*{Data Availability}
The simulations used in this paper can be made available upon request. Theoretical results are reproducible using
the equations provided in the paper.

%%%%%%%%%%%%%%%%%%%% REFERENCES %%%%%%%%%%%%%%%%%%
\bibliographystyle{mnras}
\bibliography{paper} % if your bibtex file is called mybibs.bib

%%%%%%%%%%%%%%%%% APPENDICES %%%%%%%%%%%%%%%%%%%%%

\appendix

\section{Justification for several assumptions and choices used in this paper}
\label{sec:A1}

%_________________________________________                                                                                                                                                                                 
\begin{figure*}
  \includegraphics[width=0.7\textwidth]{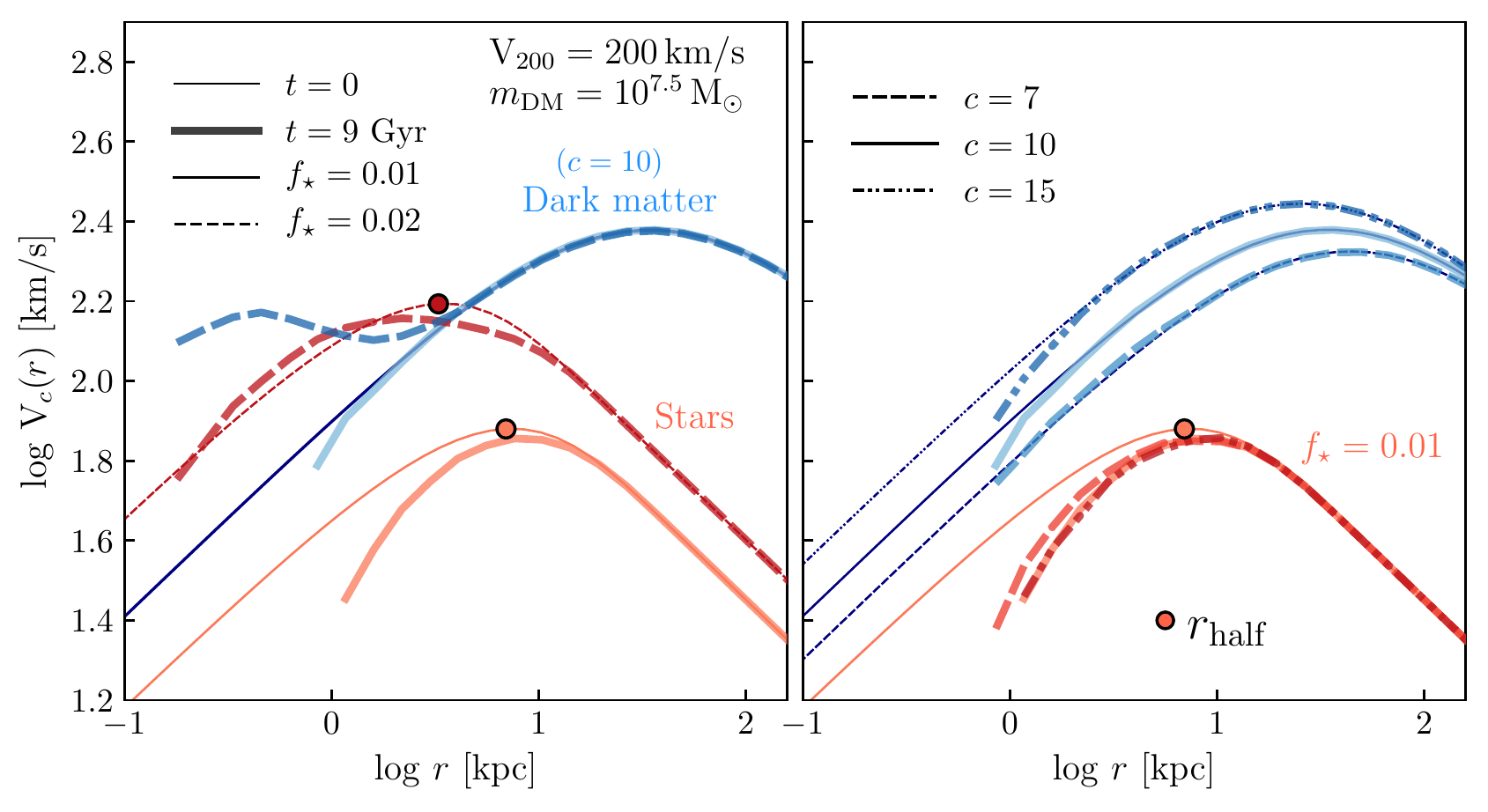}
  \caption{Circular velocity profiles due to DM (blue curves) and stars (red) for models
    adopting  ${\rm V_{200}=200\, km\,s^{-1}}$ and $m_{\rm DM}=10^{7.5}\,{\rm M_\odot}$. The left-hand panel
    corresponds to a DM halo with concentration $c=10$ and spin parameter $\lambda_{\rm DM}=0.03$. Two sets of red curves correspond
    to circular velocity profiles of disk stars (both disks have $f_j=j_{\rm d}/j_{\rm DM}=1$) for stellar mass fractions of $f_\star=0.01$
    (lighter colour, solid lines) or $f_\star=0.02$ (darker colour, dashed lines). Thin lines show the initial
    profiles for each component and thick lines the corresponding profiles after $\Delta t=9\,{\rm Gyr}$. Note that the galaxy with
    $f_\star=0.02$ leads to a significant contraction of the DM halo within the initial stellar half-mass radius (shown as outsized
    circles along the initial stellar $V_c(r)$ profiles), whereas the one with $f_\star=0.01$ does not.
    The right-hand panel, similarly, shows the circular velocity profiles due to stars ($f_\star=0.01$; red
    curves) and DM (blue curves) for varying concentration parameters ($c=7$, 10 and 15; note that for
    each value of $c$ we have adjusted $f_j$ to ensure that the stellar mass profile remains fixed).
    As on the left, the initial profiles are plotted using thin lines and the profiles after $\Delta t=9\,{\rm Gyr}$ using thick lines.
    Regardless of the DM halo's concentration, a stellar disk with $f_\star=0.01$ does not affect its mass profile. }
  \label{figA1}
\end{figure*}
%_________________________________________                                                                                                                                                                                 

%_________________________________________                                                                                                                                                                                 
\begin{figure}
  \includegraphics[width=0.45\textwidth]{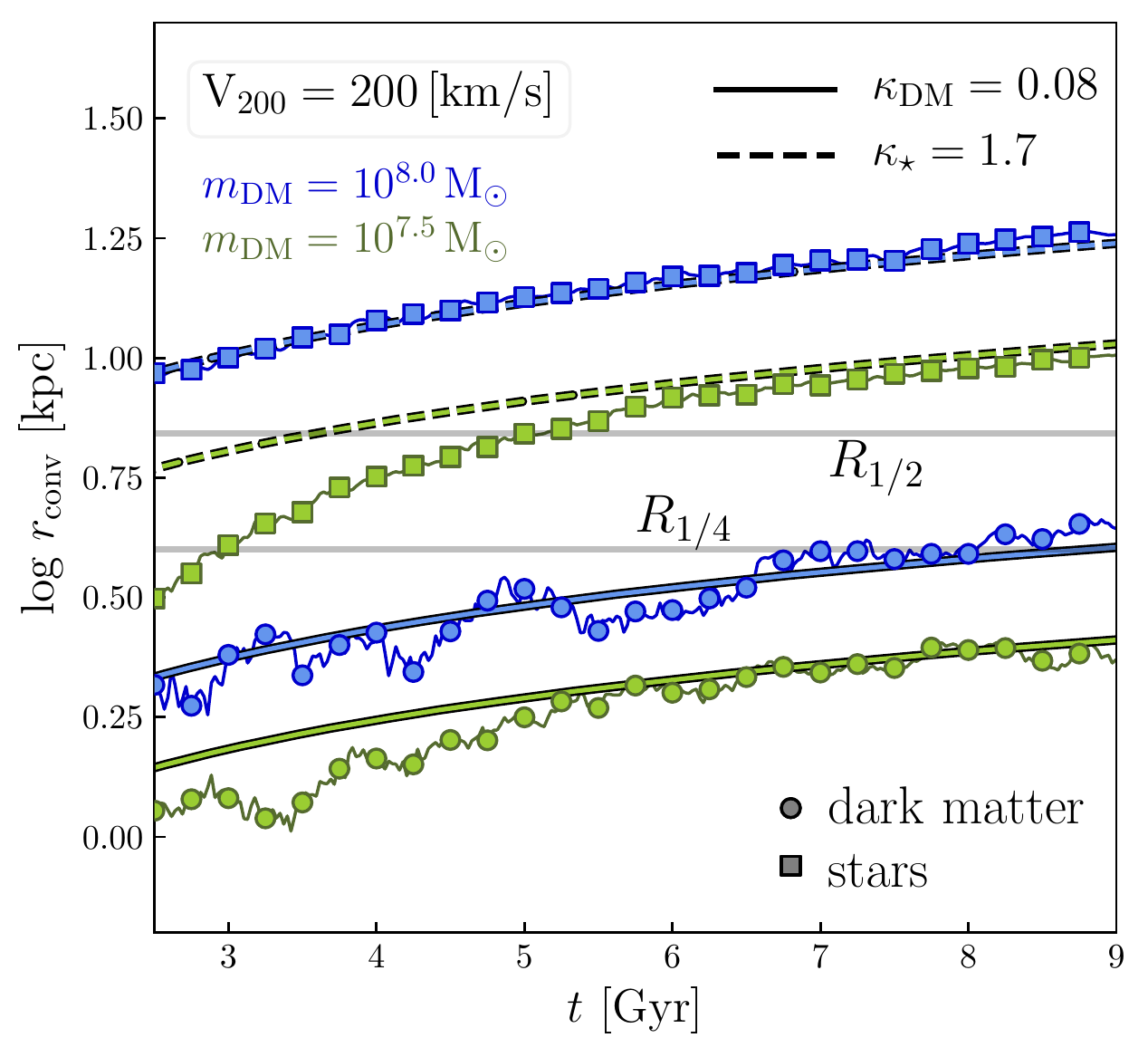}
  \caption{Convergence radii ($r_{\rm{conv}}$; i.e. the radius at which simulated $V_{\rm c}(r)$ profiles first deviate by
    $\approx 10$ per cent
    from their initial profiles) for our fiducial model (${\rm V_{200}}=200\,{\rm km\,s^{-1}}$, $f_\star=0.01$, $\mu=5$) run
    with DM particle masses $m_{\rm DM}=10^{7.5}\,{\rm M_\odot}$ (green) and $m_{\rm DM}=10^8\,{\rm M_\odot}$ (blue). Results are shown separately
    for the stellar disk (connected squares) and for the DM halo (circles). Dashed and solid lines show the radii at
    which the relaxation time of the stars or DM halo (in units of the current age of the universe) are $\kappa_\star=1.7$ and
    $\kappa_{\rm DM}=0.08$, respectively (see text for details). The horizontal dotted lines mark the initial characteristic
    radii $R_{1/4}$ and $R_{1/2}$. Note that the convergence radius of the DM halo does not exceed $R_{1/4}$, even in
    our lowest-resolution run (corresponding to $m_{\rm DM}=10^8\,{\rm M_\odot}$ for the fiducial runs shown here).}
  \label{figA2}
\end{figure}
%_________________________________________

\subsection{$f_{\star}=0.01$}

Because the collisional heating rate anticipated by eq.~\ref{eq:empheat} depends only on properties of the DM
halo, it should be relatively insensitive to the stellar distribution, and in particular (within reason) to
the stellar mass fraction. All of the simulations analyzed in this paper adopted a
stellar to {\em total} mass fraction of $f_\star\equiv {\rm M_\star/M_{200}}=0.01$. This value is relatively
low, particularly for the interesting case of haloes with masses
$\approx 10^{12}\,{\rm M_\odot}$, for which $f_\star\gtrsim 0.03$ \citep{Posti2021}. We nevertheless adopt $f_\star=0.01$
since, for each of the halo models we consider, it ensures that the disk is gravitationally sub-dominant
at all radii, and has a negligible impact on the dynamics and spatial distribution of halo particles.
For studies such as ours, larger values of $f_\star$ are undesirable for a couple of reasons: massive disks
1) may lead to a contraction of the inner halo which modifies the DM distribution in non-trivial ways, and
2) are vulnerable to instabilities, which rapidly alter the
disk structure. Both of these effects complicate the interpretation of the simulation results and the development
of the empirical model described in Section~\ref{sSecModel}. Collisional heating of more massive disks will be studied
in future work.

The contraction of the inner DM halo due to a massive disk is shown explicitly in Figure~\ref{figA1}, where we plot
the circular velocity
profiles due to DM (blue) and stars (red) for a few example halo/galaxy models. The left-hand panel shows a
DM halo with concentration $c=10$ (our fiducial choice) and stellar disks with $f_\star=0.01$ (also our
fiducial choice, shown using solid red lines) and $f_\star=0.02$ (dashed red lines). In both cases, thin
and thick lines, respectively, distinguish the initial profiles (i.e. the analytic profiles used to construct each model)
from those measured in the simulations after $\Delta t=9\,{\rm Gyr}$. Note that for $f_\star=0.02$ the more massive disk
significantly alters the mass profile of the DM halo on spatial scales $\lesssim R_{1/2}$ (outsized circles
mark the location of the {\em initial} disk half-mass radius). This may lead to difficulties interpreting the underlying
heating rates anticipated by eq.~\ref{eq:empheat} since the local density of dark matter is no longer independent of the
properties of the disk.

The right-hand panel shows analogous results for our fiducial disk (i.e. $f_\star=0.01$ and $f_j$ adjusted
to ensure a fixed stellar mass profile)
but for DM haloes with $c=7$ (dashed lines), 10 (solid) and 15 (dot-dashed). In none of these cases does
the presence of the stellar disk alter the distribution of dark matter.

\subsection{Limitations due to collisional relaxation of the DM halo}
\label{A1a2}

In cosmological DM-only simulations, the radius at which the local collisional relaxation time is of order the age of
the universe, $t_{\rm H}$, i.e. where $\kappa(r)\equiv t_{\rm relax}(r)/t_{\rm H}\approx 1$, marks the well-known ``convergence radius''
of a simulated halo. Specifically, circular velocity profiles of haloes converge (to better than $\lesssim 10$ per cent)
to those of arbitrarily-high mass and force resolution provided $\kappa\simeq 0.6$ (for {\em individual} haloes;
see \citealt{Power2003}) or $\kappa\simeq 0.18$ (for {\em median} mass profiles; see \citealt{LSB2019}).
For radii $r\lesssim r_{\rm conv}$, the collisional relaxation of dark matter particles thermalizes their velocity distribution
and suppresses the internal density. It is desirable to test the validity of eq.~\ref{eq:empheat} at radii
deemed ``converged'' by similar arguments in order to ensure that the local DM density does not develop an (unknown and unwanted)
time dependence. But what value of $\kappa$ is relevant for our idealized runs? What determines the innermost resolved radius?

Figure~\ref{figA2} provides some clues. Here we plot the measured convergence radii for stars (connected squares)
and DM
(circles) in the two lowest-resolution realisations of our fiducial galaxy/halo model (i.e. ${\rm V_{200}=200\, km\,s^{-1}}$,
$f_\star=0.01$, $\mu=5$). We calculate $r_{\rm conv}$ by locating the radius beyond which the ${\rm V_{\rm c}}(r)$
profiles converge to $\lesssim 10$ per cent relative to the {\em initial} profiles. The
convergence radius grows with time, as expected, and can be described reasonably well assuming $\kappa_\star=1.7$ (for stars)
and $\kappa_{\rm DM}=0.08$ for DM (note that we have assumed $t_{\rm relax}\propto N_{\rm DM}/(8\ln N_{\rm DM})$, such that
relaxation is driven by the DM particles; $t_{\rm H}$ is the current age of the universe). These results depend primarily on
$N_{\rm DM}$, but only slightly on $\mu$ (not shown). For the purposes of this paper it suffices to say that, for DM,
$r_{\rm conv}\lesssim R_{1/4}$ even at the lowest mass-resolution studied (horizontal dotted lines mark the initial values
of $R_{1/4}$ and $R_{1/2}$). We therefore expect the impact of collisional relaxation on the structure of the DM halo to be
negligible at $r \gtrsim R_{1/4}$ in all models, which is the {\em minimum} radius we considered in our analysis. Our
analysis is therefore robust to collisional relaxation in the DM component.

Figure~\ref{figA2} also reveals that the convergence radius of stars exceeds that of the DM by nearly a factor of 5,
but it is premature (and unlikely correct) to assume that this is a generally-valid result. For
example, it is {\em not} due to mass segregation: it is approximately valid for $\mu=1$ as well.
It is more likely is that this is a general manifestation of energy diffusion associated with equipartition:
in our model, stars form a dynamically-{\em cold} disk in the centre of a comparatively {\em hot} halo:
stars, on average, {\em gain} energy through collisions with
DM particles, causing them to diffuse from the central regions of haloes outward. It is therefore likely that the
convergence radii of simulated galaxies depend not only on the distribution and number density of DM particles,
but also on $\mu$, and on the initial phase-space distribution of stellar particles (i.e. on their initial
segregation with respect to the DM, which will depend on the initial scale height, angular momentum, mass fraction, morphology, etc).
We conclude that the convergence radii of galaxies generally exceed those of their DM haloes, but the topic
requires further investigation.

%_________________________________________                                                                                                                                                                                 
\begin{figure}
  \includegraphics[width=0.4\textwidth]{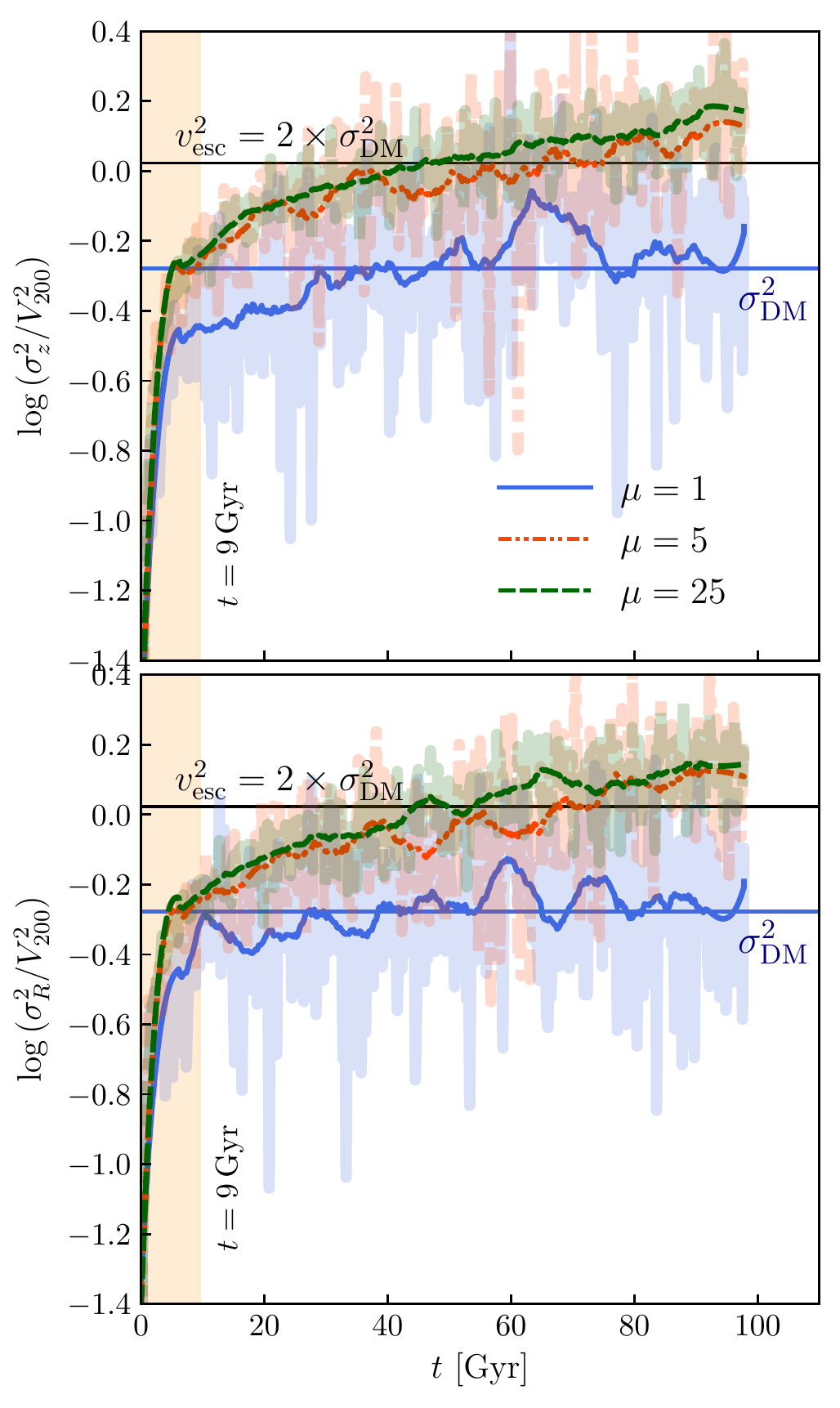}
  \caption{The evolution of $\sigma_z^2$ (upper panel) and $\sigma_R^2$ (lower panel) measured at $R_{1/4}$
    for our fiducial model (see Section~\ref{sSecICs}) with $m_{\rm DM}=10^8\,{\rm M_\odot}$ and
    for three values of $\mu$ (different line styles and colors). Results are shown for $\Delta t=100\,{\rm Gyr}$ to emphasize the asymptotic behaviour
    of the velocity dispersion of stellar particles. Horizontal lines mark two characteristic velocity dispersions:
    $\sigma^2_{\rm DM}$, the 1-dimensional dark matter velocity dispersion at $R_{1/4}$ (squared); and
    $v_{\rm esc}^2=2\times\sigma_{\rm DM}^2$, the local escape velocity for a system in virial equilibrium. Note
    that a simulated stellar system in energy equipartition with its dark matter halo would have
    $\sigma_\star^2=\mu\times\sigma_{\rm DM}^2$, which can only be achieved
    for $\mu\leq 2$; for larger $\mu$, the transfer of energy from dark matter to stellar particles leads to their slow
    evaporation from the system, such that, regardless of $\mu$, the maximum local velocity dispersion of
    stars is approximately $\sqrt{2}\times\sigma_{\rm DM}$.}
  \label{figA3}
\end{figure}
%_________________________________________                                                                                                                                                                                 

\section{Asymptotic dependence of vertical and radial heating rates on dark matter-to-stellar particle mass ratio, $\mu$}
\label{sec:A2}

The collisional heating of stellar disks reported in this paper is largely insensitive to the
dark matter-to-stellar particle mass ratio, $\mu=m_{\rm DM}/m_\star$: at fixed $m_{\rm DM}$, similar results
were obtained for all values of $m_\star$ considered. As discussed in
Section~\ref{sSec_prelim2}, complete equipartition of energy should drive the stellar velocity dispersion
asymptotically to $\sigma_i\approx \sqrt{\mu}\,\sigma_{\rm DM}$; in practice, the virial theorem imposes
an asymptotic limit of $\sigma_i\approx \sqrt{2}\,\sigma_{\rm DM}$. The latter limit is
applicable to runs with $\mu\geq 2$, the former for those with $\mu < 2$ (see eq.~\ref{eq:asympv}).
Whether full energy equipartition will be reached, however, depends on a number of factors, including $\mu$, the {\em total}
relative masses of each component (i.e. $f_\star^{-1}-1$) as well as the number of particles in each, their initial
kinematic and spatial segregation, and the duration of the simulation, $\Delta t$.

The main results of our study, presented in Section~\ref{SecResults}, were based on simulations carried out
for $\Delta t=9.6\,{\rm Gyr}$. In Figure~\ref{figA3}, we present results for three of our fiducial
runs (i.e. $V_{200}=200\,{\rm km\,s^{-1}}$, $\lambda_{\rm DM}=0.03$, $f_\star=0.01$, $c=10$), extended to
$\Delta t=100\,{\rm Gyr}$. Upper and lower panels plot $\sigma_z^2$ and $\sigma_R^2$, respectively
(measured at the radius $R_{1/4}$ that encloses a quarter of the initial stellar mass).
The DM particle mass is $m_{\rm DM}=10^8\,{\rm M_\odot}$ and $\mu=1$ (blue solid lines), 5
(red dot-dashed) and 25 (green dashed). The relatively high DM particle mass was chosen in order to hasten
collisional heating.

The kinematics of disk stars undergo an initial phase of very rapid heating, lasting roughly
$\Delta t\approx 10\,{\rm Gyr}$. At later times, the effects of collisional heating taper off, and the
stellar velocity dispersion obtained from the various models approach the asymptotic values expected from
eq.~\ref{eq:asympv}, namely $\sigma_i\approx \sigma_{\rm DM}$ for $\mu=1$ and $\sigma_i\approx \sqrt{2}\,\sigma_{\rm DM}$
for $\mu=5$ and 25. These results confirm our expectations for the asymptotic velocity dispersion based on energy
equipartition and the virial theorem.

\section{Collisional heating due to stellar haloes and bulges}
\label{sec:A4}

%_________________________________________
\begin{figure*}
  \includegraphics[width=0.8\textwidth]{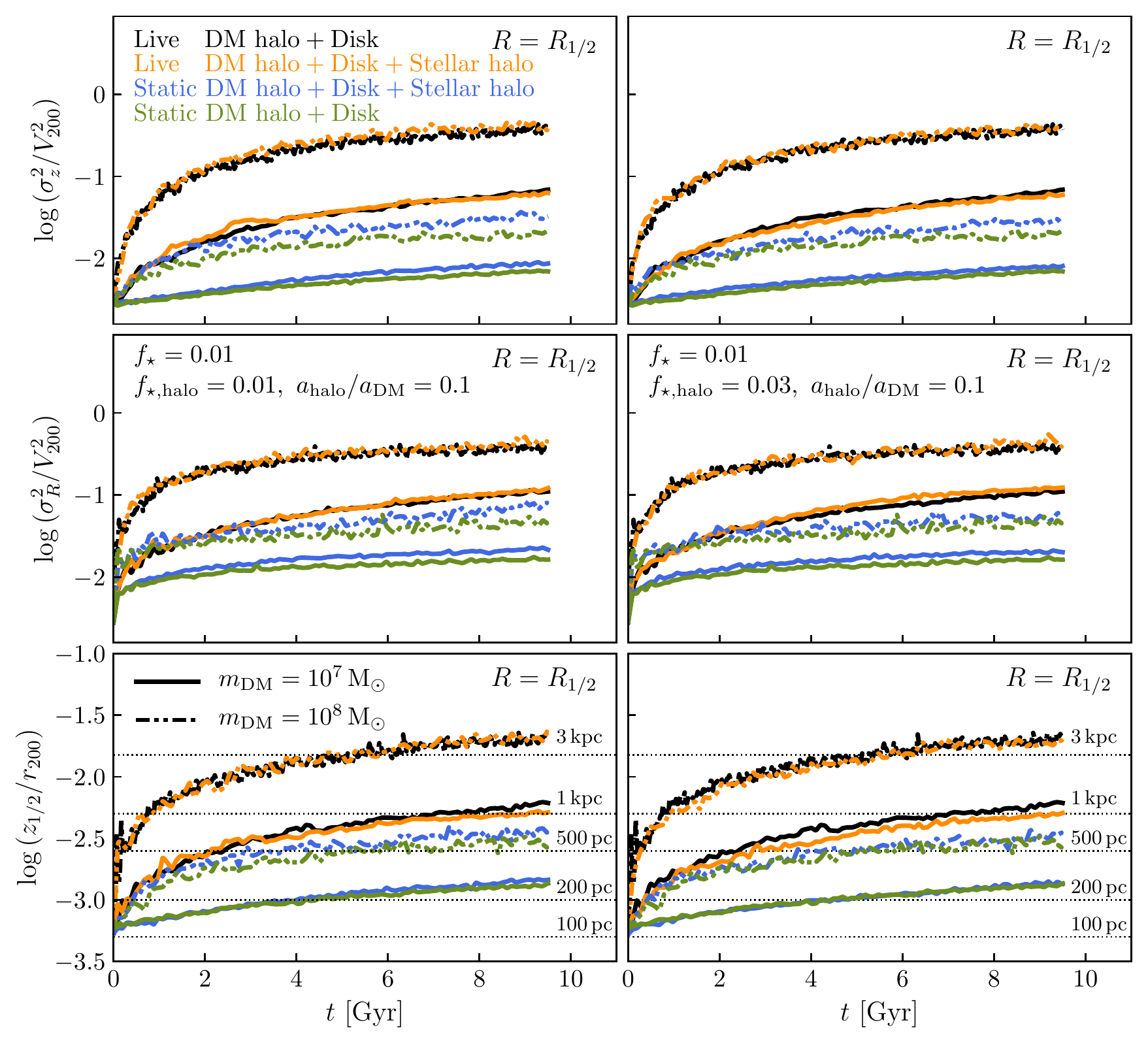}
  \caption{Contribution to collisional heating due to dark matter and stellar haloes. Top panels plot the evolution of the
    vertical velocity dispersion, $\sigma_z^2/{\rm V}_{200}^2$, middle panels plot the radial velocity dispersion,
    $\sigma_R^2/V_{200}^2$, and bottom panels the vertical half-mass height, $z_{1/2}/r_{200}$. All quantities are measured
    at the {\em initial} stellar half-mass radius of the disk, $R_{1/2}$. Different colour lines correspond to different
    galaxy/halo configurations: black to a live dark matter halo and a live thin disk (equivalent to the set-up used throughout
    the paper); yellow to a live DM halo, live disk and live stellar halo; blue to a static DM halo, live disk and live stellar
    halo; green to a static dark matter halo and live disk (with no stellar halo). All models correspond to our fiducial halo
    (${\rm V_{200}=200\, km\,s^{-1}}$, $\lambda_{\rm DM}=0.03$) and disk
    ($f_\star=0.01$, $f_j=1$), but for the {\em stellar} halo we consider mass fractions of $f_{\star,\rm halo}=0.01$ (left panels) and
    $f_{\star,\rm halo}=0.03$ (right panels). Different line-styles correspond to DM haloes sampled with particles of mass
    $m_{\rm DM}=10^7\,{\rm M_\odot}$ (solid) and $10^8\,{\rm M_\odot}$ (dash-dotted). Note that, in all cases, the
    heating and structural evolution of the disk is dominated by the presence of a coarse-grained DM halo, with a small
    but non-negligible contribution resulting from collisions between stellar particles.}
  \label{figA4}
\end{figure*}
%_________________________________________

In Section~\ref{SecResults} we established that spurious collisional heating of thin stellar disks by coarse-grained DM haloes
is cause for concern in simulations of galaxy formation. Realistic simulations, however, will produce galaxies with a wide range
of morphologies, even when limited to the disk galaxy population: disks are often partnered with bulges
or stellar haloes, for example. Are star particles in these components also able to collisionally excite
stellar motions in disks? What about collisions between disk stars?

We investigate this in
Figure~\ref{figA4}, where we plot the evolution of the vertical and radial velocity dispersion (upper and middle panels,
respectively) and the half-mass height (lower panels) obtained from a number of supplementary simulations. All models
adopt a Hernquist profile for the DM (the NFW-equivalent halo has a concentration $c=10$ and
$V_{200}=200\,{\rm km\,s^{-1}}$) and $\lambda_{\rm DM}=0.03$. As for the other models considered
in this paper, the disk contains a fraction $f_\star=0.01$ of the system's total mass, and the disk-to-halo specific
angular momentum ratio is $f_j=1$. Different line styles correspond to different DM particle masses, as
indicated; the stellar particle masses are chosen so that $\mu=5$ in all cases. Lines of different colour
denote different simulations, as follows. Black lines correspond to our fiducial model: a coarse-grained DM halo and a thin,
rotationally-supported stellar disk with parameters described above. Yellow lines are the same, but with an additional stellar bulge/halo
component with mass fractions equal to $f_{\star, {\rm halo}}=0.01$ (left) and 0.03 (right). Note that the spheroidal stellar component is also
modelled as a Hernquist sphere with a characteristic scale radius equal to one-tenth that of the DM halo's (i.e. $a_{\rm halo}/a_{\rm DM}=0.1$).
Comparing these curves, we conclude that collisional heating by stellar haloes or bulges is negligible compared to that due to the DM halo.

In order to verify these results, and to test the impact of collisions between stars {\em in the disk}, we have repeated
these two sets of simulations, but after replacing the ``live'' DM halo with a fixed Hernquist potential of
equivalent mass and characteristic size. The results are plotted as blue and green lines for the cases with and without
a ``live'' stellar halo, respectively. In this case, the heating rate drops considerably, and remains largely independent
of the presence or absence of a coarse-grained stellar halo/bulge. We have verified that these results apply to a relatively broad range
of stellar spheroidal structures, ranging from compact bulges to extended stellar haloes.

Finally, note that spurious
collisional heating is still present, albeit much reduced, even when the DM halo is modelled using a fixed analytic potential.
This suggests that collisional heating due to star-star encounters also affects simulations of galaxy structure (in
agreement with \citealt{Sellwood2013}).

\section{The impact of the gravitational softening length}
\label{sec:A5}

%_________________________________________                                            
\begin{figure*}
  \includegraphics[width=0.8\textwidth]{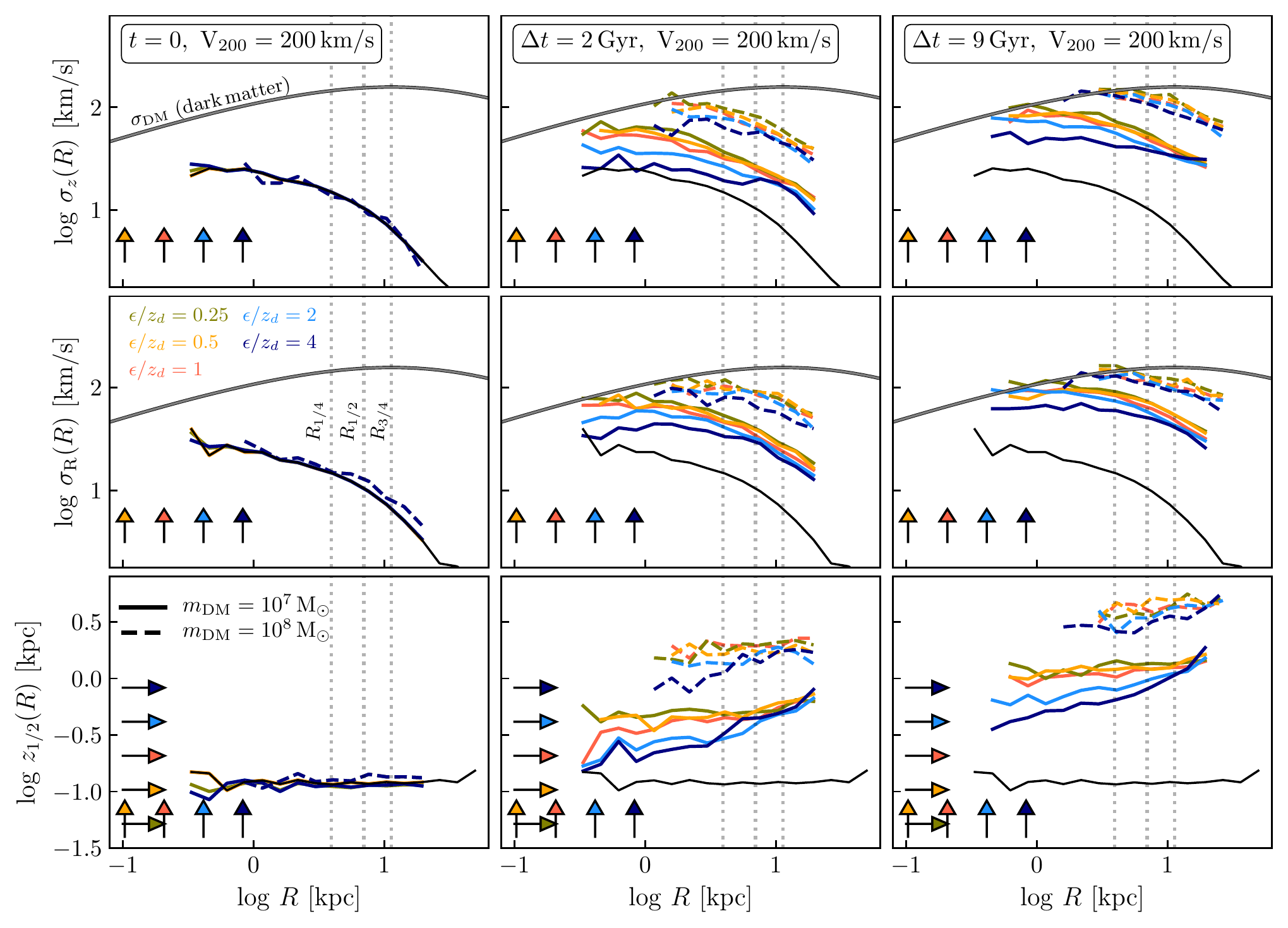}
  \caption{The effect of gravitational softening on the vertical (top panels) and radial (middle panels) velocity dispersion profiles
    of stellar particles for our fiducial galaxy/halo models. Bottom panels show the radial dependence of the
    vertical half-mass height, $z_{1/2}$. Panels on the left corresponds to $t=0$, i.e. the initial conditions of our simulations,
    middle panels to $\Delta t=2\,{\rm Gyr}$, and the right-most panels to $\Delta t=9\,{\rm Gyr}$ (the initial
    profiles in the highest-resolution run are shown for comparison using a thin black line in all panels). In all cases,
    profiles are plotted to the innermost radius that encloses at least 10 stellar particles. All
    runs adopt ${\rm V_{200}=200\,{\rm km\,s^{-1}}}$, $f_j=1$, and $f_\star=0.01$, but were repeated
    for $m_{\rm DM}=10^7\,{\rm M_\odot}$ (solid curves) and $m_{\rm DM}=10^8\,{\rm M_\odot}$ (dashed curves). Different colours
    correspond to different softening lengths $\epsilon$ (indicated by arrows), which vary from $\epsilon=z_{\rm d}/4$ to
    $\epsilon=4\times z_{\rm d}$ (our fiducial value is $\epsilon=z_{\rm d}$, where $z_{\rm d}=(2/\ln 3)\,z_{1/2}$ is 
    the disk's initial characteristic scale height; see eq.~\ref{eq:rhodisk}). Vertical dotted lines mark the initial characteristic radii
    $R_{1/4}$, $R_{1/2}$ and $R_{3/4}$, from left to right, respectively. Collisional heating rates are suppressed when
    $\epsilon$ is large, but are independent of softening provided $\epsilon/z_{\rm d}\lesssim 1$. }
  \label{figA5}
\end{figure*}
%_________________________________________

%_________________________________________                                            
\begin{figure*}
  \includegraphics[width=0.9\textwidth]{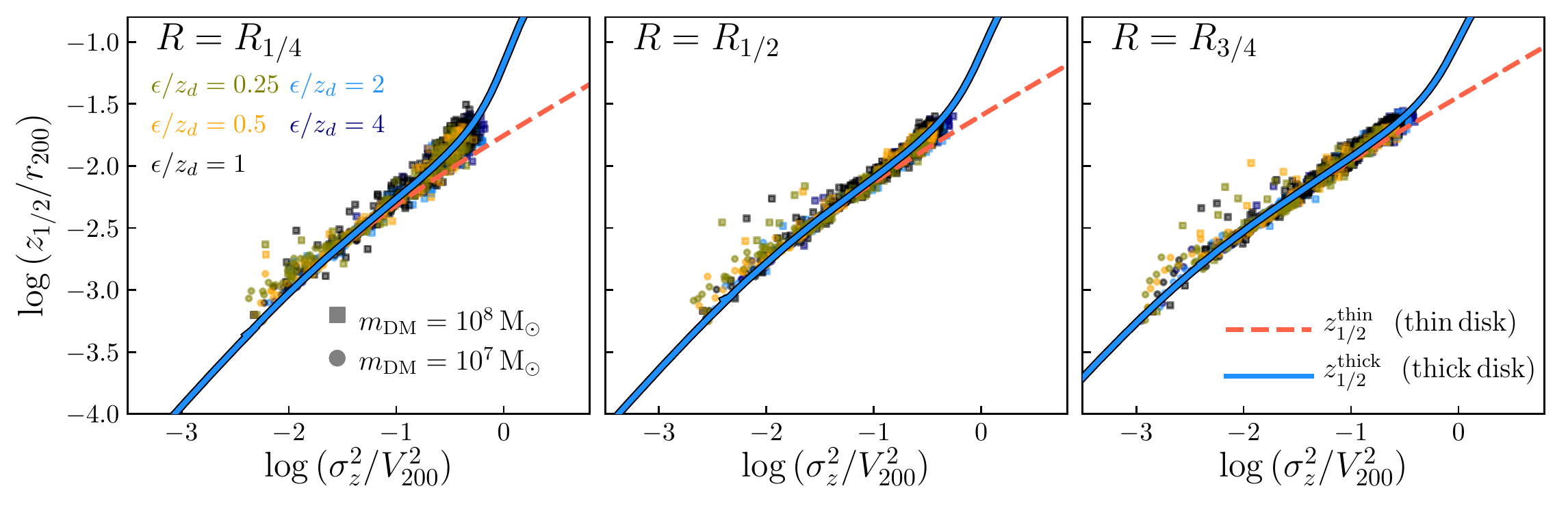}
  \caption{Half-mass scale height of stellar particles (normalized by $r_{200}$) plotted as a function of their
    vertical velocity dispersion (squared, and normalized by $V_{200}^2$). From left to right, different panels correspond to 
    different cylindrical radii, $R_{1/4}$, $R_{1/2}$ and $R_{3/4}$, respectively, where $R_f$ is the radius enclosing
    a fraction $f$ of the galaxy's initial stellar mass. Results from our simulations are plotted as individual points for the same fiducial models
    shown in Figure~\ref{figA5}; points are color-coded by gravitational softening length, $\epsilon$, with different symbols used
    for runs with different DM particle mass, as indicated. The dashed orange line shows the the vertical
    scale heights expected for thin disks, i.e. $z_{1/2}^{\rm thin}$, eq.~\ref{eq:z50thin}; the
    solid blue line shows $z_{1/2}^{\rm thick}$ for thick disks, i.e. eq.~\ref{eq:zthick}. Note that although both $z_{1/2}$
    and $\sigma_z$ depend on $\epsilon$ (most noticeably when $\epsilon \gtrsim z_{\rm d}$; Figure~\ref{figA5}), the relation between
    them does not. The vertical scale heights of simulated disks can therefore be estimated from their vertical velocity dispersion
    profiles using eq.~\ref{eq:zthick} independently of the gravitational softening length.}
  \label{figA6}
\end{figure*}
%_________________________________________

The simulations presented in this paper and all previous appendices used a gravitational softening length equal
to the initial scale height of the stellar disk, which is a fixed fraction of the disk scale radius, i.e.
$\epsilon=z_{\rm d}=0.05\,R_{\rm d}$. As discussed in Section~\ref{sSecICs}, this has the advantage of preserving the scale-invariance of
our runs, since the softening length scales self-consistently with the size of the disk in models with widely
varying ${\rm V_{200}}$ and $R_{\rm d}$. However, as discussed by \citet{Ludlow2020}, the softening length adopted in
hydrodynamical simulations of galaxy formation has a non-trivial relationship to galaxy structure: if it is too
large, it suppresses the small-scale clustering of stellar and DM particles, resulting in lower central densities,
larger galaxies, and DM haloes with softened ``cores''; if it is too {\em small}, is exacerbates the collisional
heating of stellar structures, escalating the spurious growth of galaxies sizes. Similar softening-dependent
effects may infiltrate the numerical results present in this work.

We explore the softening dependence of collisional disk heating for a couple of our fiducial models in
Figure~\ref{figA5} (${\rm V_{200}=200\,km\,s^{-1}}$, $c=10$, $\lambda_{\rm DM}=0.03$; $f_\star=0.01$, $f_j=1$,
$\mu=5$). We focus on the radial profiles of the vertical velocity dispersions ($\sigma_z$, upper panels), radial velocity
dispersion ($\sigma_R$, middle) and vertical half-mass height ($z_{1/2}$, bottom). Results are plotted at $t=0$
(left columns) and after $\Delta t=2\,{\rm Gyr}$ (middle) and $\Delta t=9\,{\rm Gyr}$ (right). Line styles indicate the
mass of  DM particles (solid for $m_{\rm DM}=10^7{\rm M_\odot}$; dashed for $m_{\rm DM}=10^8{\rm M_\odot}$), and different
coloured lines correspond to runs carried out with different softening lengths, ranging from $\epsilon/z_{\rm d}=0.25$ to 4
 (physical values of $\epsilon$ are marked using arrows in each panel).

The plot elicits a few comments. First, despite the wide range of softening lengths considered, the consequences of
collisional heating are apparent in all simulations, and at all radii. Nevertheless, the integrated effects are suppressed in runs
in which the softening is largest, particularly in the galaxy's central regions in runs
with $\epsilon > z_{\rm d}$ (see \citealt{Ludlow2020} and \citealt{Pillepich2019} for
similar results based on cosmological simulations). 

Importantly, however, the results are
largely independent of $\epsilon$  provided $\epsilon\lesssim z_{\rm d}$, i.e. provided softening length does
not exceed the scale height of the disk. This implies that -- in cosmological simulations, which typically
adopt softening lengths that are fixed in physical or comoving coordinates -- collisional disk heating may be
suppressed in low-mass, poorly-resolved galaxies, whose characteristic sizes can be comparable to the softening length.
Clearly, however, such systems cannot be considered spatially resolved in the first place.

Despite the tendency for collisional heating to be suppressed when $\epsilon\gtrsim z_{\rm d}$, the relation between
$z_{1/2}$ and $\sigma_z$ appears largely independent of the softening length. As a result, the scale heights of
simulated disks can still be inferred from their vertical velocity dispersion profiles, regardless of $\epsilon$. We show this explicitly in
Figure~\ref{figA6} (which is analogous to Figure~\ref{fig6} in the main body of the paper), where we plot the
vertical scale height of stellar particles (normalized by $r_{200}$) versus their vertical velocity dispersion
(squared and normalized by $V_{200}^2$) at different galacto-centric radii ($R_{1/4}$, $R_{1/2}$ and $R_{3/4}$ from left to
right, respectively; where $R_f$ is the cylindrical radius enclosing a fraction $f$ of the galaxy's initial stellar mass).
The results, which correspond to the same set of runs plotted in Figure~\ref{figA5}, clearly show that, despite
the softening dependence of $z_{1/2}$ and $\sigma_z$, the relation between these two quantities (at fixed $R_f$)
is independent of softening.

%%%%%%%%%%%%%%%%%%%%%%%%%%%%%%%%%%%%%%%%%%%%%%%%%%
% Don't change these lines
\bsp	% typesetting comment
\label{lastpage}
\end{document}